\documentclass[11pt,letterpaper]{article}
\pdfoutput=1
\usepackage{jheppub}
\usepackage{amsthm}
\usepackage{natbib}
\usepackage{tikz}
\usetikzlibrary{calc}

\usepackage{hyperref}
\usepackage{amsmath}
\usepackage{amsfonts}
\usepackage{graphicx}
\usepackage{caption}
\usepackage{placeins}
\usepackage{tqft}
\usepackage{comment}
\usepackage{array}
\usepackage[utf8]{inputenc} 
\def\be{\begin{equation}}
\def\ee{\end{equation}}
\def\ba{\begin{eqnarray}}
\def\ea{\end{eqnarray}}
\def\C{{\mathbb C}}
\def\CC{{\mathcal C}}
\def\R{{\mathbb R}}
\def\C{{\mathbb C}}
\def\Z{{\mathbb Z}}
\def\G{{\mathrm{G}}}
\def\B{{\mathcal{B}}}
\def\z{{\bar z}}
\def\F{{\mathcal{F}}}
\def\Vol{{\rm Vol}}
\def\Int{{\rm Int}}
\def\Vert{{\rm Vert}}
\def\Conv{{\rm Conv}}

\def\Spec{{\rm Spec}}
\def\Trop{{\rm Trop}}
\def\New{{\mathbf{N}}}
\def\Re{{\rm Re}}
\def\tA{{\tilde A}}
\def\A{{\cal A}}
\def\X{{\mathbf{X}}}
\def\x{{\mathbf{x}}}
\def\v{{\mathbf{v}}}
\def\uu{{\mathbf{u}}}
\def\y{{\mathbf{y}}}
\def\w{{\mathbf{w}}}
\def\n{{\mathbf{n}}}
\def\M{{\mathcal{M}}}
\def\oM{{\overline{\M}}}
\def\N{{\mathcal{N}}}
\def\I{{\mathcal{I}}}
\def\PP{{\mathcal{P}}}
\def\Q{{\mathcal{Q}}}
\def\PPP{{\mathbb P}}
\def\zz{{\mathbf{z}}}
\def\bI{{\bf I}}

\def\hC{{\hat C}}
\def\hT{{\hat T}}
\def\SL{{\rm SL}}
\def\bmu{{{\boldsymbol \mu}}}
\def\bla{{{\boldsymbol \lambda}}}
\newtheorem{claim}{Claim}
\theoremstyle{definition}
\newtheorem{example}{Example}
\numberwithin{example}{section}
\theoremstyle{remark}
\newtheorem{remark}{Remark}
\numberwithin{remark}{section}

\title{Stringy Canonical Forms}

\author[a,b]{Nima Arkani-Hamed,}
\author[c,d,e]{Song He,}
\author[f,g]{Thomas Lam}

\abstract{%The idea of positive geometry and canonical form has played a central role in unravelling geometric structures of scattering amplitudes and beyond, including the amplituhedron, cosmological polytopes and the associahedron 
}

\affiliation[a]{School of Natural Sciences, Institute for Advanced Studies, Princeton, NJ, 08540, USA}
\affiliation[b]{Center of Mathematical Sciences and Applications, Harvard University, Cambridge, MA 02138, USA}
\affiliation[c]{CAS Key Laboratory of Theoretical Physics, Institute of Theoretical Physics, Chinese Academy of Sciences, Beijing, 100190, China}
\affiliation[d]{
School of Fundamental Physics and Mathematical Sciences, Hangzhou Institute for Advanced Study, UCAS, Hangzhou 310024, China\\
ICTP-AP
International Centre for Theoretical Physics Asia-Pacific, Beijing/Hangzhou, China}
\affiliation[e]{School of Physical Sciences, University of Chinese Academy of Sciences, No.19A Yuquan Road, Beijing 100049, China}
\affiliation[f]{Department of Mathematics, University of Michigan, 530 Church St, Ann Arbor, MI 48109, USA}
\affiliation[g]{Department of Mathematics, Massachusetts Institute of Technology, 77 Massachusetts Ave., Cambridge, MA 02139, USA}

\emailAdd{arkani@ias.edu}
\emailAdd{songhe@itp.ac.cn}
\emailAdd{tfylam@umich.edu}
\abstract{Canonical forms of positive geometries play an important role in revealing hidden structures of scattering amplitudes, from amplituhedra to associahedra. In this paper, we introduce  ``stringy canonical forms", which provide a natural definition and extension of canonical forms for general polytopes,  deformed by a parameter $\alpha'$. They are defined by real or complex integrals regulated with polynomials with exponents, and are meromorphic functions of the exponents, sharing various properties of string amplitudes. As $\alpha' \to 0$, they reduce to the usual canonical form of a polytope given by the Minkowski sum of the Newton polytopes of the regulating polynomials, or equivalently the volume of the dual of this polytope, naturally determined by tropical functions. At finite $\alpha'$, they have simple poles corresponding to the facets of the polytope, with the residue on the pole given by the stringy canonical form of the facet. There is the remarkable connection between the $\alpha' \to 0$ limit of tree-level string amplitudes, and scattering equations that appear when studying the $\alpha' \to \infty$ limit. We show that there is a simple conceptual understanding of this phenomenon for any stringy canonical form: the saddle-point equations provide a diffeomorphism from the integration domain to the interior of the polytope, and thus the canonical form can be obtained as a pushforward via summing over saddle points. When the stringy canonical form is applied to the ABHY associahedron in kinematic space, it produces the usual Koba-Nielsen string integral, giving a direct path from particle to string amplitudes without an a priori reference to the string worldsheet. We also discuss a number of other examples, including stringy canonical forms for finite-type cluster algebras (with type A corresponding to usual string amplitudes), and other natural integrals over the positive Grassmannian. 
}
\begin{document}

\maketitle
%\tableofcontents
\section{Introduction}\label{sec:intro}
Tree level $n$-point open superstring amplitudes are defined as integrals over a component $\M^+_{0,n}$ of the real points of the moduli space of $n$-points $z_1,z_2,\ldots,z_n$ on the Riemann sphere, associated with the Koba-Nielsen factor~\cite{Koba:1969kh}\footnote{Depending on the states that are scattering  there are additional factors that depend on external momenta and polarization, but in this paper we are focusing on the non-trivial structure of the worldsheet integral itself, and in the rest of the paper will refer to these as ``string integrals" or more loosely as ``string amplitudes".}:
\be \label{eq:In}
\bI_n (\{ s\})=(\alpha')^{n{-}3}~\int_{{\cal M}_{0,n}^+} \frac{d^{n{-}3} z}{z_{1,2} \cdots z_{n,1}}~\prod_{a<b} (z_{a,b})^{\alpha' s_{a,b}}\,,
%\bI_n (\{ s\})=(\alpha')^{n{-}3}~\int_{{\cal M}_{0,n}^+} \omega_n~\prod_{a<b} (a b)^{\alpha' s_{a,b}}\,,
\ee
where the $z$'s are ordered, so $z_{a, b}:=z_b-z_a>0$ for $a<b$.  The integral $\bI_n$ is a function of the Mandelstam invariants $s_{a,b}$. For $n = 4$, the open-string amplitude reduces  to the beta function (with $s:=\alpha' s_{1,2}$, $t:=\alpha' s_{2,3}$):
\be \label{eq:beta}
B(s,t)  = \int_0^1  \frac{d y}{y(1-y)}y^{s} (1-y)^{t} = \frac{\Gamma(s) \Gamma( t)}{\Gamma(s+t)}
\ee
whose properties were first studied by Euler and Legendre, and whose relevance to physics was discovered by Veneziano~\cite{veneziano1968construction}.  String amplitudes satisfy numerous remarkable properties that have been explored from many perspectives in the last fifty years ({\it c.f.}~\cite{GSW}).

In this work, we initiate the study of a vast generalization of string amplitudes, that we call {\bf stringy canonical forms}, or stringy integrals:
\begin{equation}\label{eq:stringy}
{\I}_{\{p\}} (\X, \{c\})=(\alpha')^d  \int_{\R_{>0}^d} \prod_{i=1}^d \frac{d x_i}{x_i}~x_i^{\alpha' X_i}~\prod_I p_I (\x)^{-\alpha' c_I}\,,
\end{equation}
where $p_I(\x)= p_I(x_1,\ldots,x_d)$ are Laurent polynomials with positive coefficients. We will show shortly that the string amplitude \eqref{eq:In} can be written in the form \eqref{eq:stringy}. The integral ${\cal I}_{\{p\}}$ analytically continues to a meromorphic function of ${\bf X}:=(X_1, \ldots, X_d)$ and the $c$'s. We call it a stringy canonical {\it form} because by putting an overall measure $d^d {\bf X}:=\wedge_{i=1}^d d X_i$ it becomes a (top-dimensional) differential form in ${\bf X}$ space. We find that our stringy canonical forms/integrals have properties analogous to those of open string amplitudes.

Such integrals have been studied since antiquity, and there is a large body of research for them both in mathematics and physics; more recently they have appeared in the literature, {\it e.g.} in relation to Euler-Mellin integrals and A-hypergeometric functions~\cite{Mellin, berkesch2014}, and in the study of the Hepp bound for Feynman integrals~\cite{Panzer:2019yxl}. We emphasize that in this paper, by the simple act of multiplying with $d^d {\bf X}$, we turn the integral into a {\it form} which has a number of important consequences. First it allows us to think about the singularity structures in a more invariant way, reflected in residues of the form on the poles. As we will see, this leads to connection with canonical forms of polytopes~\cite{Arkani-Hamed:2017tmz} as $\alpha' \to 0$, and a deformation of this important notion at finite $\alpha'$. Among other things, this perspective also makes the connection between critical points at $\alpha' \to \infty$  and the form as $\alpha' \to 0$ natural and geometric.  Before we explain these in detail, we first discuss the notion of {\it positive geometry} and {\it canonical form} that underlies both \eqref{eq:In} and \eqref{eq:stringy}.

A {\it positive geometry}~\cite{Arkani-Hamed:2017tmz} is a real, compact space (sitting inside a complex projective variety), that is a generalization of a convex polytope (sitting inside complex projective space). The defining property of a positive geometry is the existence of a unique, complex, top-dimensional differential form called the {\it canonical form}, defined to have simple poles only on the boundaries of the positive geometry, with the residue on each boundary in turn given by the canonical form for that boundary.  In recent years, positive geometries have been found to produce scattering amplitudes from a new geometric viewpoint. In this way, locality and unitarity are seen to emerge as derived concepts from the positive geometry, rather than taken as fundamental principles. This was first seen with the {\it amplituhedron}~\cite{Arkani-Hamed:2013jha} that produces all-loop scattering amplitudes for planar ${\cal N}=4$ SYM; originally defined via a map in terms of the positive Grassmannian~\cite{Postnikov:2006kva, ArkaniHamed:2012nw}, it has been reformulated directly~\cite{Arkani-Hamed:2017vfh} in momentum-twistor space~\cite{Hodges:2009hk}, and more recently in momentum space~\cite{He:2018okq, Damgaard:2019ztj}; another example is the {\it cosmological polytopes} that produce the wave function of the universe for a class of scalar theories in FRW cosmology~\cite{Arkani-Hamed:2017fdk}.

Of the most direct interest to us is the {\it associahedron}, which  is a convex polytope and thus a positive geometry. Recently a realization of the associahedron has been given, naturally defined in the kinematic space of Mandelstam invariants. The canonical form produces the tree-level S-matrix of bi-adjoint $\phi^3$ theory~\cite{Arkani-Hamed:2017mur}. While usual Feynman diagrams correspond to a particular way of computing the form, the geometry of such ABHY associahedra reveals hidden properties of amplitudes obscured by the Feynman diagrams.

It is also well known that compactifying the moduli space of the open-string worldsheet, ${\M}_{0,n}^+$, we obtain an $(n{-}3)$-dimensional (``curvy") associahedron, which is again a positive geometry. This positive geometry underpins three remarkable properties of the open-string amplitude, \eqref{eq:In}, that we highlight here. First,  {\it (a) as $\alpha' \to 0$, the field-theory limit of $\bI_n$ is the canonical function of the ABHY associahedron, which is the bi-adjoint $\phi^3$ amplitude; also (b) for finite $\alpha'$, $\bI_n$ factorizes as the product of lower-point amplitudes, on any massless pole which corresponds to a facet of the associahedron.} %There are natural extensions to generalized associahedra of finite type: {\it e.g.} for the polytope of type-${\cal D}$ realized in kinematic space, its canonical form computes one-loop amplitudes of bi-adjoint $\phi^3$ theory~\cite{Arkani-Hamed:2017mur}.

There is also a deep connection between ABHY associahedron and ${\M}_{0,n}^+$ which has been revealed in~\cite{Arkani-Hamed:2017mur}. The {\it scattering equations} of the Cachazo-He-Yuan (CHY) formulas~\cite{Cachazo:2013hca, Cachazo:2013iea} are the saddle-point equations of the Koba-Nielsen factor~\cite{Koba:1969kh} in $\bI_n$ in the $\alpha' \to \infty$, ``Gross-Mende" limit~\cite{Gross:1987kza}; it is fascinating that these equations underpin field-theory amplitudes in the opposite limit!  The third remarkable property of $\bI_n$ is a novel, geometric origin of the CHY formula for bi-adjoint $\phi^3$ amplitudes, conjectured in~\cite{Arkani-Hamed:2017mur}: {\it (c) scattering equations provide a diffeomorphism from ${\cal M}_{0,n}^+$ to the ABHY associahedron, and thus the canonical form of the latter is given by the pushforward of that of ${\M}_{0,n}^+$, by summing over the saddle points of $\bI_n$.}  

In this paper we will see that all these remarkable features (and some further ones as well) are properties of general stringy canonical forms. Indeed, understanding these features even for ordinary string amplitudes is most easily and conceptually done in this more general setting. To begin with, let's see how to write \eqref{eq:In} in the form of \eqref{eq:stringy}. Recall that one can remove the SL$(2, \R)$ redundancy of ${\cal M}_{0,n}^+$ by fixing three points, {\it e.g.} $(z_1, z_2, z_n)=(0, 1, \infty)$:
\be
{\cal M}_{0,n}^+=\{z_1< z_2< \cdots <z_n\}/{\SL}(2, \R)=\{0<1<z_3<\cdots<z_{n{-}1}<\infty\}\,,
\ee
The Koba-Nielsen factor $\prod_{i<j} (z_j-z_i)^{\alpha' s_{i,j}}$ is $\SL(2, \R)$-invariant due to the momentum conservation equations $\sum_{j\neq i} s_{i,j}=0$ for $i=1,2,\ldots, n$. The key for the rewriting is a {\it positive parametrization} of ${\cal M}_{0,n}^+$ by $\R_{>0}^{n{-}3}$, and a particularly simple way for doing this is the following:
\be
z_3=1+x_2\,, \quad z_4=1+x_2+x_3, \quad \ldots, \quad z_{n{-}1}=1+x_2+\cdots+x_{n{-}2}\,,
\ee
with $x_i>0$ for $i=2,3, \ldots, n{-}2$. With this change of variables, $\bI_n$ takes the form of \eqref{eq:stringy}:
\be\label{eq:string}
\bI_n(\{s\})=(\alpha')^{n{-}3}~\int_{\R_{>0}^{n{-}3}} \prod_{i=2}^{n{-}2} \frac{d x_i}{x_i}~x_i^{\alpha' s_{i,i{+}1}}~\prod_{i, j} p_{i,j}({\bf x})^{\alpha' s_{i,j}}\,.%\left(1+\sum_{a=2}^{i{-}1} y_a\right)^{\alpha' s_{1, i}}~\prod_{2\leq i<j\leq n{-}1} \left(\sum_{a=i}^{j{-}1} y_a\right)^{\alpha' s_{i,j}}\,. 
\ee
where $p_{i,j}:=\sum_{a=i}^{j{-}1} x_a$ for non-adjacent $i,j$ in the range $1\leq i<j\leq n{-}1$ (we define $x_1=1$); the Koba-Nielsen factor splits into two parts: the $n{-}3$ monomial factors $x_i^{\alpha' X_i}$ with $X_i=s_{i, i{+}1}$, for $i=2,\ldots, n{-}2$, and the remaining $\frac{(n{-}2)(n{-}3)}{2}$ polynomials ones $p_{i,j}^{-\alpha' c_{i,j}}$ with exponents $-c_{i,j}=s_{i,j}$, for non-adjacent pairs $i, j$. 

Now we are ready to summarize the main results of the paper. With the string amplitude in the new form \eqref{eq:string} as our motivating example, we show that its remarkable properties mentioned above generalize to all stringy canonical forms \eqref{eq:stringy}. 

\noindent
{\bf Convergence and Minkowski sums.}
In Sections \ref{sec:converge} and \ref{sec:multi}, we show that the integral \eqref{eq:stringy} converges absolutely when the point $\X%:=(X_1, \ldots, X_d)
$ lies in the polytope $\PP$ that is the Minkowski sum of the Newton polytopes of the $p_I(\x)$ (weighted by $c_I$). This was also shown in~\cite{Mellin, berkesch2014}. For example, for the beta function \eqref{eq:beta}, we recover the domain of convergence $s,t > 0$.  In the case of the string amplitude rewritten as \eqref{eq:string}, this Minkowski sum is exactly the kinematic associahedron of~\cite{Arkani-Hamed:2017mur}.

\noindent
{\bf Leading order (field-theory limit) and tropical function.} The leading order in $\alpha'$ of $\I_{\{p\}}$, namely, $\lim_{\alpha' \to 0} {\I}_{\{p\}} (\X, \{c\})$, is equal to the volume of the dual polytope to $\PP$, or equivalently, the canonical function of the positive geometry $\PP$.  In the same vein, the dual polytope of $\PP$ can be obtained as a halfspace cut out by the tropicalization of the integrand.  

\noindent
{\bf Residues, recurrence relations, stringy properties.} Stringy canonical forms are natural $\alpha'$ deformations of canonical form of a polytope $\PP$: at finite $\alpha'$, the residue on any pole corresponding to a facet of $\PP$ is given by a stringy canonical form for that facet. Thus the class of integrals/forms is closed under the operation of taking residues at such ``massless" poles. This elementary property becomes manifest after we present recurrence relations that the integrals satisfy at finite $\alpha'$.  Furthermore, stringy canonical forms exhibit analytic properties similar to open-string amplitudes.  Any stringy integral is exponentially suppressed in the limit that all exponents become large (``high energy limit"); moreover, it satisfies the analog of ``channel duality" and ``Regge behavior" as we sketch on in Section \ref{sec:finitealpha}.

\noindent 
{\bf Scattering equations and twisted (co-)homology} In Section \ref{sec:saddle} we show that for any stringy integral \eqref{eq:stringy}, the saddle-point equations for the $\alpha'\to \infty$ limit, provide a diffeomorphism from the integration domain to the interior of the polytope $\PP$.  Applying the results of~\cite{Arkani-Hamed:2017tmz}, the canonical form of $\PP$ is obtained via pushforward by summing over saddle points.  This reproduces scattering equations and CHY formulas when applied to ${\M}_{0,n}$. We also explain that the number of saddle points equals the dimension of $d$-dim twisted (co-)homology, or the number of independent integral functions. %It is therefore a completely general phenomenon that the $\alpha' \to 0$, ``field-theory" limit of any stringy canonical form can be computed by summing over saddle points, which appear in the $\alpha' \to \infty$!

\noindent 
{\bf Closed stringy integrals.} Closed string amplitudes are complex analogues of \eqref{eq:In} where the integral over $\M_{0,n}^+$ is replaced by an integral over $\M_{0,n}(\C)$.  We study a similar analogue for \eqref{eq:stringy}.  The integration domain $\R_{>0}^d$ is replaced by $\C^d$, and the new integrand is given by mod-squaring the integrand of \eqref{eq:string}, and more generally the exponents can be shifted by integers. We study the leading order {\it etc.} for such closed stringy integrals in Section \ref{sec:closed}.

\noindent
{\bf Dual $u$-variables.} The convergence of \eqref{eq:stringy} is usually a complicated condition on the exponent variables $\X$ and $c$.  In Section \ref{sec:big}, we show how to find dual {\it $u$-variables} $u_A$ and $U_A$ so that \eqref{eq:stringy} can be rewritten as $\int_{\R_{>0}^d} \prod \frac{dx_i}{x_i} \prod_A (u_A)^{\alpha^\prime U_A}$ and the convergence condition becomes simply $U_A > 0$.  In the case of the open-string amplitudes \eqref{eq:In}, the $u_A$ become the cross-ratios $u_{i\,j}$ of~\cite{Koba:1969kh, Brown:2009qja, Arkani-Hamed:2017mur}, given in \eqref{eq:uij}.

\noindent 
{\bf Tropical compactification.}
The moduli space $\M_{0,n}(\R)$ of $n$-points on the Riemann sphere has a well-known (Deligne-Knudsen-Mumford) compactification $\oM_{0,n}(\R)$~\cite{DM}.  Taking only the boundary strata of $\oM_{0,n}(\R) \setminus \M_{0,n}(\R)$ that ``touch" the positive component $\M_{0,n}^+$, we obtain an intermediate partial compactification $\M'_{0,n}(\R)$ sitting in between $\M_{0,n}(\R)$ and $\oM_{0,n}(\R)$.  The space $\M'_{0,n}(\R)$ (first studied in \cite{Brown:2009qja} where it is denoted ${\mathfrak M}_{0,n}^\delta$) is an affine variety with the same combinatorics as the associahedron (whereas $\oM_{0,n}(\R)$ has a more complicated stratification).  Starting from an integral \eqref{eq:stringy}, we synthetically construct spaces $U$ and $U^\circ$, called {\it tropical compactifications}, that are analogues of $\M'_{0,n}$ and $\M_{0,n}$, respectively.  Roughly speaking, the tropical compactification $U$ is determined by requiring that rational integrands for which \eqref{eq:stringy} almost converge are regular functions on $U$.

\medskip
Finally, we remark that stringy canonical forms provide a new intrinsic definition for the canonical form of any polytope\footnote{More precisely, for any polytope that can be realized with rational vertices.}, which comes naturally deformed with a ``string scale" $\alpha^\prime$. Given a polytope $\PP$, such stringy integrals are not unique: while their $\alpha' \to 0$ limit gives the same canonical function $\underline{\Omega}(\PP)$, they differ at finite $\alpha'$ as the integral of course depends on the coefficients of the polynomials in the integrand.  However, some polytopes are presented in a specific way that dictates the presentation of the stringy canonical form. For instance, the ABHY associahedron in kinematic space is naturally presented as a Minkowski sum of simple pieces. Remarkably, the stringy canonical form associated with this presentation precisely yields \eqref{eq:string} with Koba-Nielsen factor, giving a path from kinematic space to string amplitudes, making no reference either to bulk spacetime nor the string worldsheet as auxiliary constructs (see Section \ref{sec:5}).
Furthermore, for more general classes of polytopes, we know of choices for the integrand that produce stringy canonical forms with extra special properties.  In Section \ref{sec:examples}, we introduce the {\it cluster string integrals} and {\it Grassmannian string integrals}.  The cluster string integrals $\I_\Phi$, defined for any cluster algebra $\A(\Phi)$ of finite type, are stringy canonical forms for the generalized associahedron of the dual Dynkin diagram, $\PP(\Phi^\vee)$~\cite{fomin2003systems,chapoton2002polytopal}.~\footnote{The appearance of the dual Dynkin diagram is an important subtlety for non-simply laced cases, and we will discuss it in detail in~\cite{20201}.} The integrals $\I_\Phi$ have special factorization properties, particularly elegant dual $u$-variables and tropical compactifications, and will be explored further in~\cite{20193, 20201}.  In $\alpha'\to 0$ limit, the Grassmannian string integral $\I_{k,n}$ produces a polytope, which is combinatorially isomorphic to the tropical positive Grassmannian~\cite{SW}, and for $k =4$ has potential applications for non-perturbative geometries for ${\cal N}=4$ SYM amplitudes~\cite{ALS}.

\section{Stringy canonical forms and Newton polytopes}\label{sec:converge}
In this section, we define the {\bf stringy canonical form} of a (rationally realizable) polytope $\PP$.  As an integral, $\I(\X)$ depends on variables $\X$ and a new parameter $\alpha'>0$, whose $\alpha' \to 0$ limit recovers the canonical function $\underline{\Omega}({\PP}; {\X})$ of $\PP$. This provides a new way of computing the canonical form of the polytope $\PP$. We refer to Appendices~\ref{sec:canform} for background material on canonical forms, and to Section~\ref{sec:stringymulti} for a more leisurely introduction to normal fans.

\subsection{Newton polytopes and stringy integrals}
The coordinate simplex $\Delta_d \subset \PPP^d$ in projective space is the basic example of a positive geometry.  The (interior of the) simplex $\Delta_d$ can be parametrized as $\R^d_{>0}$, and the canonical form is simply $\Omega(\Delta_d)= \prod_{i=1}^d d\log x_i$. 
%To motivate stringy canonical forms, we start by considering integrals over the domain $\mathbb{R}^d_{>0}$. As defined in~\cite{}, this can be viewed as a {\it positive parametrization} of any $\Delta$-like positive geometry: $\Delta_d=\{0<x_i<\infty | i =1,2,\cdots, d\}$, and the canonical form is trivially $\Omega(\Delta_d)=\prod_{i=1}^d d\log x_i$.
The integral $\int_{\Delta_d} \Omega(\Delta_d) = \int_{\R^d_{>0}}  \prod_{i=1}^d d\log x_i$ does not converge: it has logarithmic divergences as $x_i \to 0$ and $x_i \to \infty$, and thus it needs to be regulated.  A natural way to regulate such divergences is to consider
\be
\I_p(\X, c)=(\alpha')^d \int_0^\infty  \prod_{i=1}^d\frac{d x_i}{x_i}~x_i^{\alpha' X_i}~p(\x)^{-\alpha' c}\,,\label{proto_int}
\ee
where we denote ${\x}:=(x_1, x_2, \ldots, x_d)\in \mathbb{R}_{>0}^d$ %the canonical form $\omega:=\prod_{i=1}^d d\log x_i$; 
and introduce into the integrand the ``regulator" $\left(\prod_{i=1}^d x_i^{\alpha' X_i}\right)~p(\x)^{-\alpha' c}$ with $c>0$: for $x_i\to 0$ the divergence is regulated by the factor $x_i^{\alpha' X_i}$ if $X_i>0$, and the divergence at infinity is regulated if the factor $p(\x)^{-\alpha' c}$ is chosen suitably.  For the regulator to not have any branch cut in the integration domain, let's assume that $p(\x)$ is a {\it subtraction-free} (Laurent) polynomial, {\it i.e.} 
\be \label{Newton}
p({\x})=\sum_\alpha a_\alpha  {\x}^{{\bf n}_\alpha}\,, \quad {\x}^{{\bf n}_\alpha}:=x_1^{n_{\alpha,1}} \cdots x_d^{n_{\alpha,d}}
\ee
where $a_\alpha>0$ and $\alpha$ labels terms in the polynomial: for each term we have a $d$-dim exponent vector ${\bf n}_\alpha \in \mathbb{Z}^d$ with the exponent of $x_i$ denoted as $n_{\alpha, i}$. The integral is a function of $\X:=(X_1, X_2, \ldots, X_d)$ and $c$, and the factor $\alpha'^d$ is for normalization. 

Both the convergence region of the integral, and its $\alpha'\to 0$ limit is controlled by the {\bf Newton polytope} $\New[p(\x)]$ of the polynomial $p(\x)$.
%
%In the next section, we will generalize the integral to include arbitrary numbers of factors of the form $p(\bf x)^{-\alpha' c}$. However, it turns out the generalization is essentially trivial and for now it suffices to focus on our prototype integral~\eqref{proto_int}. Just like string integrals, the integral $\I_p$ is a {\it meromorphic function} which share certain properties of string integrals at finite $\alpha'$. As we will show now, both the convergence region of the integral, and its $\alpha'\to 0$ limit is controlled by the so-called {\it Newton polytope} of the polynomial $p(\x)$.
The Newton polytope $\New[p(\x)]$ of a Laurent polynomial $p(\x)$ is defined to be the convex hull of the exponent vectors ${\bf n}_\alpha$ in \eqref{Newton}, 
\be
\New[p(\x)]:=\left \{\sum_\alpha \lambda_\alpha {\bf n}_\alpha | \lambda_\alpha \geq 0, \quad \sum_\alpha \lambda_\alpha=1\right \}\,.
\ee
Note that the definition does not depend on coefficients $a_\alpha$'s (which we assumed to be positive), and thus the Newton polytope remains the same if we set all coefficients of $p(\x)$ to be unity. For example, for $d=1$, we obtain intervals $\New [\sum_{i=-m}^n a_i x^i]=\New[ x^{-m}+ \cdots +x^n]=[-m,n]$.  Some more examples include: $\New[\frac 1 {x y}+x +y]$ is a triangle with vertices $(-1,-1), (1,0), (0,1)$ and $\New[1+3 x y^2+ x y^4+5 x^3y+2 x^3y^4+ x^4y^2
]$ is the following pentagon:
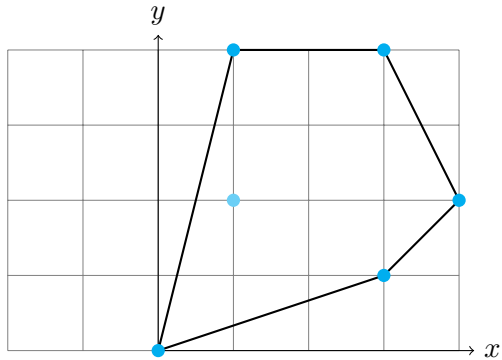
\begin{center}
\begin{tikzpicture}
\draw [help lines, step=1cm] (-2,0) grid (4,4);
 \draw[->] (0,0) -- (4.2,0) node[right] {$x$};
      \draw[->] (0,0) -- (0,4.2) node[above] {$y$};
\draw[thick] (0,0) -- (1,4) -- (3,4) -- (4,2) -- (3,1)-- cycle;
\node[fill=cyan!50,circle, inner sep=0pt,minimum size=5pt] at (1,2) {};
\node[fill=cyan,circle, inner sep=0pt,minimum size=5pt] at (3,1) {};
\node[fill=cyan,circle, inner sep=0pt,minimum size=5pt] at (4,2) {};
\node[fill=cyan,circle, inner sep=0pt,minimum size=5pt] at (3,4) {};
\node[fill=cyan,circle, inner sep=0pt,minimum size=5pt] at (1,4) {};
\node[fill=cyan,circle, inner sep=0pt,minimum size=5pt] at (0,0) {};
\end{tikzpicture}
\end{center}

The main result of this section is the following Claim for the integral $\I_p(\X, c)$.  Let us call the limit $\lim_{\alpha'\to 0} {\I}_p ({\X}, c)$ the {\bf leading order} of $\I_p(\X,c)$~\footnote{The condition for convergence of the integral was found in Theorem 1 of \cite{Mellin}.}.

%we assume that the $\N(\PP)$ is $d$ dimensional. 

\begin{claim} \label{claim:single} The integral \eqref{proto_int} converges if and only if the Newton polytope is top-dimensional (that is, $d$-dimensional) and ${\X}$ is in the interior\footnote{More generally, if we allow $\X$ and $c$ to be complex, the condition is that $\Re(\X)$ lies in the interior of $\Re(c)~\New[p(\x)]$.  For simplicity, we state our results assuming that $\X$ and $c$ are positive.} of the polytope $\PP = c~\New[p(\x)]$; the leading order of $\I_p(\X,c)$ is given by the canonical function of $c~\New[p(\x) ]$:
\be \label{eq:LS}
\lim_{\alpha'\to 0} {\I}_p ({\X}, c)=\underline{\Omega}(c~\New[p(\x)]; {\X})\,.
\ee
%{\color{red} General case without $x^X$ factors, and the volume of the dual polytope (then we say we restrict to special case in the following)...}
%\section{Formula for leading order}
Equivalently, 
%we can consider the product $\prod_{i=1}^d x_i^{\alpha' X_i} p(\x)^{-\alpha' c}$ to have Newton polytope $c\New(p(\x)) - \X$.  Now 
the condition for convergence is that the origin $0$ must be inside the polytope $c\New[p(\x)] - \X$, and the leading order is given by the volume $\Vol((\New[p(\x)] - \X)^\circ)$ of the dual polytope \eqref{eq:dualpoly}.  
\end{claim}
Throughout this work, $\Vol$ denotes the {\it normalized volume}: we have \be 
\Vol(\mbox{$d$-dimensional unit cube}) = d!.
\ee

Claim~\ref{claim:single} is established in the next section, after a review of the basic properties of dual polytopes.
%It turns out that this is a more general setting: instead of including explicitly $\prod_i x_i^{\alpha' X_i}$, we can consider general {\it Laurent} polynomial(s) which regulate divergences at $0$ and $\infty$.
%
% We shall show that for all $\alpha' > 0$ we have 
%\begin{equation}\label{eq:refsuggestion}
%\Vol(\PP^\circ) \leq {\I}_p ({\X}, c)\leq C \Vol(\PP^\circ) 
%\end{equation}
%for a positive constant $C$, and 
%\begin{equation}\label{eq:main}
%\Vol(\PP^\circ) = \lim_{\alpha' \to 0}  {\I}_p ({\X}, c)
%\end{equation}
%if $0$ belongs to the interior of $\PP$ (and then the integral is absolutely convergent for all $\alpha' > 0$).  
%Further the integral does not converge if $0 \notin \PP$.  
%Throughout the paper we will mostly focus on the special case with factors $\prod_i x_i^{\alpha' X_i}$, but for some purposes it is convenient to work this more general setting. Now we turn to the proof for \eqref{eq:main} with a single Laurent polynomial. 

\subsection{Volumes of dual polytopes as limits of stringy integrals}\label{ssec:decomp}
\begin{figure}
\begin{center}
\begin{tikzpicture}
\draw [help lines, step=1cm] (-2,-2) grid (2,2);
\draw[thick] (0,1) -- (1,1) -- (1,-1) -- (-1,-1) -- (-1,0)-- cycle;
\node[fill=black,circle, inner sep=0pt,minimum size=5pt] at (0,0) {};
\node[fill=cyan,circle, inner sep=0pt,minimum size=5pt] at (0,1) {1};
\node[fill=cyan,circle, inner sep=0pt,minimum size=5pt] at (1,1) {2};
\node[fill=cyan,circle, inner sep=0pt,minimum size=5pt] at (1,-1) {3};
\node[fill=cyan,circle, inner sep=0pt,minimum size=5pt] at (-1,-1) {4};
\node[fill=cyan,circle, inner sep=0pt,minimum size=5pt] at (-1,0) {5};
\begin{scope}[shift={(5,0)}]
\draw [help lines, step=1cm] (-2,-2) grid (2,2);
\node[fill=black,circle, inner sep=0pt,minimum size=5pt] at (0,0) {};
\draw[thick] (0,0) --(2,0);
\draw[thick] (0,0) --(-2,0);
\draw[thick] (0,0) --(0,-2);
\draw[thick] (0,0) --(0,2);
\draw[thick] (0,0) --(2,-2);
\node at (-1,-1) {$C_2$};
\node at (-1,1) {$C_3$};
\node at (1,1) {$C_4$};
\node at (1.3,-1/2) {$C_5$};
\node at (1/2,-1.3) {$C_1$};
\end{scope}
\begin{scope}[shift={(10,0)}]
\draw [help lines, step=1cm] (-2,-2) grid (2,2);
\node[fill=black,circle, inner sep=0pt,minimum size=5pt] at (0,0) {};
\draw[thick] (0,1) -- (1,0) -- (1,-1) -- (0,-1) -- (-1,0)-- cycle;
\node[fill=cyan,circle, inner sep=0pt,minimum size=5pt] at (0,1) {};
\node[fill=cyan,circle, inner sep=0pt,minimum size=5pt] at (1,0){};
\node[fill=cyan,circle, inner sep=0pt,minimum size=5pt] at (1,-1){};
\node[fill=cyan,circle, inner sep=0pt,minimum size=5pt] at (0,-1){};
\node[fill=cyan,circle, inner sep=0pt,minimum size=5pt] at (-1,0){};
\node at (0.7,0.7) {$F_4$};
\node at (1.2,-0.5) {$F_5$};
\node at (0.5,-1.2) {$F_1$};
\node at (-0.7,-0.7) {$F_2$};
\node at (-0.7,0.7) {$F_3$};
\end{scope}
\end{tikzpicture}
\caption{A pentagon $\PP$ containing the origin, its normal fan $\N(\PP)$ with maximal cones labeled by vertices of $\PP$, and the dual polytope $\PP^\circ$ with facets labeled by vertices of $\PP$.}
\label{fig:pentex}
\end{center}
\end{figure}
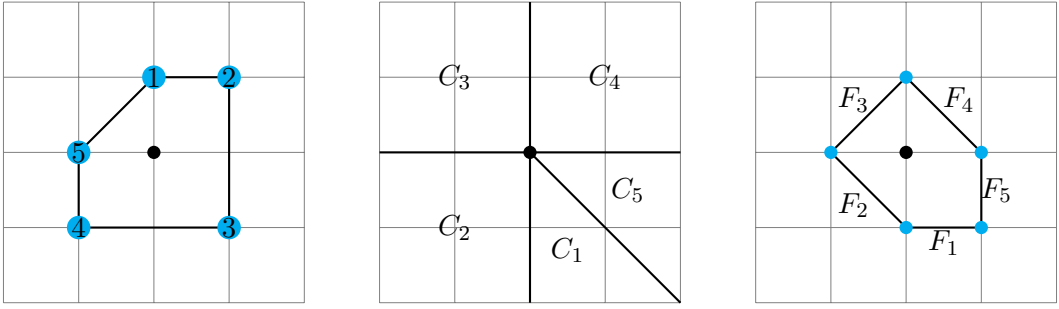

Let $\PP$ be a full-dimensional polytope in a vector space $\R^d$.  The {\it normal fan} $\N = \N(\PP)$ is a collection of cones $\{C_F \mid F \text{ a face of }\PP\}$ in $\R^d$ that completely tile space.  For a face $F$ of $\PP$, the cone $C_F$ consists of those $\bla \in \R^d$ such that the linear function $\uu \mapsto \uu \cdot \bla$ on $\R^d$ is minimized when $\uu \in F$.  The maximal dimensional cones of $\N(\PP)$ are the cones $\{C_v\subset \R^d \mid v \in \Vert(\PP)\}$ associated to vertices of $\PP$, given explicitly by
\begin{equation}\label{eq:vertexcone}
C_v := \{ \bla \in \R^d \mid \v \cdot \bla \leq \uu \cdot \bla \qquad \mbox{for all $\uu \in \PP$} \}.\end{equation}
Here, we write $\v$ when we consider the vertex $v$ as a vector in $\R^d$.  These cones $C_v$ have pairwise disjoint interiors, and tile $\R^d$.  All these cones are top-dimensional and pointed, that is, do not contain a line.   (This may no longer be true if $\PP$ is not full-dimensional.) The rays of $\N(\PP)$ are exactly the {\it inward-pointing} normals of $\PP$.  Also, define the dual cone 
\be C^\vee_v  := \{\y \in \R^d  \mid \y \cdot \bla \geq 0 \text{ for all } \bla \in C_v\} \subset \R^d.\ee %The cone $C_v$ is characterized by the following condition:
%\begin{equation}\label{eq:vertexcone}
%(v-u,\lambda) \leq 0 \qquad \mbox{for all $u \in P$ and $\lambda \in C_v$.}
%\end{equation}
%(The assumption $u \in P$ can be replaced by $u \in \Vert(P)-\{v\}$.)
%\subsection{Polar polytope}
Now assume that $\PP$ contains $0$ in its interior.   Let $\PP^\circ \subset \R^d$ denote the dual polytope:
\be \label{eq:dualpoly}
\PP^\circ := \{\bla \in \R^d \mid \uu\cdot \bla \geq -1 \text{ for all } \uu \in \PP\}.
\ee
The dual polytope 
$\PP^\circ$ is again full-dimensional and contains $0$ in its interior.  The normal fan $\N(\PP)$ is equal to the cone over $\PP^\circ$.  The facets $F_v$ of $\PP^\circ$ are in bijection with the vertices $v \in \Vert(\PP)$, and furthermore $C_v$ is the cone over $F_v$.  The convex hull $A_v :=\Conv(F_v \cup \{0\})$ is given by the intersection of $C_v$ with the half-space $\{\bla \in \R^d \mid \v \cdot \bla \geq -1\}$.    These constructions are illustrated in Figure \ref{fig:pentex}.
%Since $C_v$ is a pointed cone with nonempty interior, $C^\vee_v$ is also a pointed cone with nonempty interior, and we have $(C^\vee_v)^\vee = C_v$.  Note that \eqref{eq:vertexcone} is equivalent to $u-v \in C^\vee_v$.
%\begin{lemma}\label{lem:Cv}
The volume of $A_v$ is given by
\begin{equation}\label{eq:vollimit}
 \Vol(A_v) =  \int_{C_v} e^{\v \cdot \bla} d\bla, %= \lim_{\alpha' \to 0} (\alpha')^d \int_{C_v} e^{\alpha' \v \cdot \bla} d\bla , %= \frac{1}{\alpha^n} \Vol(A_v) + O(\frac{1}{\alpha^{n-1}})
 \end{equation}
 noting that $\v \cdot \lambda$ takes values in $[-1,0]$ in $C_v$.  The formula \eqref{eq:vollimit} is immediate if the cone $C_v$ is simplicial (that is, it is generated by $d$ rays).  In general, we obtain \eqref{eq:vollimit} by triangulating $C_v$ into simplicial cones. 
 
%\end{lemma}
%To prove this, we triangulate $C_v$ to reduce to the simplicial case.  

%\subsection{Proof of \eqref{eq:main}}
Now we turn to the proof of Claim~\ref{claim:single}.  Let $\PP=\New[p(\x)]$ denote the Newton polytope of $p(\x)$ in $\R^d$.  For integer valued $c$ and $\X$, we have $\New[p(\x)^c] = c \New[p(x)]$ and $\New[x^\X p(\x)] = \New[p(\x)]+ \X$.
By a continuity argument, to establish Claim~\ref{claim:single}, it suffices to consider the case $c = 1$ and $\X = 0$.  We set $\I_p:= \I_p({\bf 0},1) = (\alpha')^d \int_0^\infty  \prod_{i=1}^d\frac{d x_i}{x_i}~p(\x)^{-\alpha' }$, and begin by assuming that $\PP$ contains $0$ in its interior.

\begin{figure}
\begin{center}
\begin{tikzpicture}
\draw [help lines, step=2cm] (0,0) grid (6,6);
      \draw[->] (0,0) -- (6.2,0) node[right] {$x$};
      \draw[->] (0,0) -- (0,6.2) node[above] {$y$};
      \draw[domain=2:6,smooth,variable=\x,thick] plot ({\x},{4/\x});
      \draw[thick] (2,0)--(2,6);
      \draw[thick] (0,2)--(6,2);
      \node[fill=black,circle, inner sep=0pt,minimum size=5pt] at (0,0){};
      \node at (1,1) {$\exp(C_2)$};
      \node at (1,4) {$\exp(C_3)$};
      \node at (4,4) {$\exp(C_4)$};
      \node at (5,1.5) {$\exp(C_5)$};
      \node at (3,1/2) {$\exp(C_1)$};
    \end{tikzpicture}
    \caption{The regions $\exp(C_v)$ for the normal fan of Figure \ref{fig:pentex}.  The curve separating $\exp(C_1)$ from $\exp(C_5)$ is $xy=1$.}
        \label{fig:int}
    \end{center}

\end{figure}
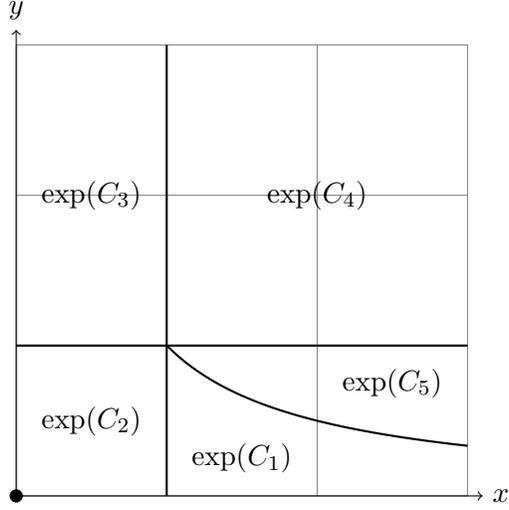

%Let $\PP^\circ \subset \R^d$ denote the {\it dual polytope} of $\PP$. 
Our integration domain is $\R_{>0}^d$, and we identify $\R^d$ with $\log \R_{>0}^d$.  The decomposition of $\R^d$ into the union $\bigcup_v C_v$ of cones $C_v$, gives the decomposition
$\R_{>0}^n = \bigcup_v \exp(-C_v)
$
where we ignore measure-zero overlaps.  Thus
\be
\int_{\R_{>0}^n} \mbox{(integrand)}= \sum_{v \in \Vert(\PP)} \int_{\exp(-C_v)} \mbox{(integrand)}.
\ee
Let $\y_1,\y_2,\ldots,\y_t$ be a minimal collection of generators of $C^\vee_v$.  Then 
\be
\x \in \exp C_v \; \iff \; \log \x \cdot \y_s \geq 0 \text{ for } s = 1,2,\ldots,t \;  \iff \; \x^{\y_s} \geq 1, \text{ for } s = 1,2,\ldots,t
\ee
In other words, each region $(-\exp C_v)$ is given by monomial conditions $\x^{\y_s} \leq 1$.  See Figure \ref{fig:int}.

Triangulate $C_v$ into pointed simplicial cones $C_1,C_2,\ldots,C_r$, and let $C$ be one of these cones.  Let $
\y_1,\y_2,\ldots,\y_d$ be a basis of $C^\vee$.  By allowing rational entries and reordering, we may assume that $\det(\y_i) = 1$.  Setting $w_i = \x^{\y_i}$, we have
\be
\Omega:= \prod_i \frac{dx_i}{x_i} = \prod_i \frac{dw_i}{w_i} \qquad \text{and thus} \qquad
\int_{\exp(-C)}\Omega  = \int_{[0,1]^n} \prod_i \frac{dw_i}{w_i}.
\ee
Since $\uu-\v \in C^\vee$ for $\uu \in \PP$, we have 
\be \label{eq:Phot}
p({\w}) = {\w}^{Y^{-1} \v}(a + \text{higher order terms}) 
\ee
where $Y$ is the matrix consisting of columns $\y_1,\y_2,\ldots,\y_d$, and $a>0$ is a constant, and the higher order terms have no constant term and belong to $\R[w_1,w_2,\ldots,w_d]$.  Letting $M = p(1,1,\ldots,1)$ be the sum of the coefficients of $p(\w)$, we have 
%$$
% {\w}^{Y^{-1} \v} \leq p(\w) \leq M  {\w}^{Y^{-1} \v}
%$$
%and thus
\begin{equation}\label{eq:twobounds}
 \int_{\exp(-C_v)} \Omega \,   {\w}^{-\alpha' Y^{-1} \v}
 \geq \int_{\exp(-C_v)} \Omega \, p(\x)^{-\alpha'} \geq M^{-\alpha'}  \int_{\exp(-C_v)} \Omega \,  {\w}^{-\alpha' Y^{-1} \v}
\end{equation}
Applying \eqref{eq:vollimit}, we get
\be
(\alpha')^d \, \int_{\exp(-C_v)} \Omega \,   {\w}^{-\alpha' Y^{-1} \v}
=(\alpha')^d \,  \int_{\exp(-C_v)} \Omega \, {\x}^{-\alpha' \v} = \Vol(A_v).
\ee
Substituting into \eqref{eq:twobounds}, and summing over $v$, we conclude that
\be
M^{-\alpha'} \leq \frac{ {\I}_p}{\Vol(\PP^\circ)} \leq 1.
\ee
This shows that $\I_p$ converges absolutely for all $\alpha' > 0$ and that 
\begin{equation}\label{eq:main}
\Vol(\PP^\circ) = \lim_{\alpha' \to 0}  {\I}_p = \lim_{\alpha' \to 0} (\alpha')^d \int_0^\infty  \prod_{i=1}^d\frac{d x_i}{x_i}~p(\x)^{-\alpha' }.
\end{equation}
Conversely, the same analysis shows that if $\PP$ does not contain $0$ in its interior then the integral over one of the domains $\exp(-C_v)$ will not converge.  This completes the proof of Claim~\ref{claim:single}.

\begin{remark}\label{rem:negcoeff}
Much of the analysis carries through even if $p(\x)$ has negative coefficients for lattice points in the interior of $\New[p(\x)]$.  The main difference is that the polynomial $p(\x)$ (and thus the factor $(c + \text{higher order terms})$ in \eqref{eq:Phot}) may have zeros on $\R_{>0}^d$, which may affect the convergence of the integral.  The form $\Omega$ has no poles at these extra zeros, so as long as $\alpha' >0$ is sufficiently small, the integral ${\I}_P ({\X}, c)$ will still converge.  The leading order remains the same and is given by \eqref{eq:LS}.
\end{remark}

\subsection{First examples}
We consider some simple examples. 

\begin{example}[Interval] Let's consider the simplest case, which is a $1$-dimensional integral
\be\label{1d}
\I_{\rm interval}=\alpha'~\int_0^\infty \frac {d x}{x} x^{\alpha' X} (1+ x)^{-\alpha' c},
\ee
where the Newton polytope is the interval $ c \New[1+x]=[0, c]$. The integral converges for $0<X<c$ and the leading order can be easily obtained as $\frac{c}{X (c{-}X)}$, which is $\underline{\Omega}([0,c]; X)$. Note that in this case the integral can be easily computed to give $\frac{\Gamma(\alpha' X)~\Gamma(\alpha' (c-X))}{\Gamma(\alpha' c)}$. Of course the integral is not unique: using instead $(1+ p x)^{-\alpha' c}$ or $(1+ p_1 x +\cdots + p_m x^m)^{-\alpha' \frac{c}{m}}$ does not change the Newton polytope, which gives identical leading order.

The beta function \eqref{eq:beta} can be recovered from this example.  With $y=x/(1+x)$ we have
\be
\int_0^\infty \frac {d x}{x} x^{\alpha' X} (1+ x)^{-\alpha' c} = \int_0^1 \frac{dy}{y(1-y)} y^{\alpha' X} (1-y)^{\alpha'(c-X)} = B(\alpha' X, \alpha'(c-X)).
\ee
Thus the condition $0 < X < c$ agrees with the known convergence of the beta function.  Also note that
\be
\int_0^\infty \frac {d x}{x} x^{\alpha' X} (1+ 2x+ x^2)^{-\alpha' c} = \int_0^\infty \frac {d x}{x} x^{\alpha' X} (1+x)^{-2\alpha' c} = B(\alpha' X, \alpha'(2c-X))
\ee
Replacing the polynomial $(1+2x+x^2)$ in the integrand by $(1+x+x^2)$ gives an integral with the same leading order as $\alpha' B(\alpha' X, \alpha'(2c-X))$.  However, the integral itself is some variant of the beta function.
\end{example}

\begin{example}[Triangle, quadrilateral and pentagon] We consider some examples of $d=2$ integrals. One possible integral for a triangle reads
\be
{\cal I}_{\rm triangle}=(\alpha')^2~\int_{\R_{>0}^2}\frac{d x d y}{x y}~x^{\alpha' X} y^{\alpha' Y} (1+x +y)^{-\alpha' c}\,,
\ee
where the Newton polytope $c \New[1+ x+  y]$ gives the triangle with vertices $(0,0), (0,c), (c,0)$. The integral converges if $(X,Y)$ is inside the triangle (or $X>0, Y>0, X+Y<c$), and the canonical function at $(X,Y)$ reads
\be
\lim_{\alpha'\to 0} {\I}_{\rm triangle}=\frac{1}{X Y}+\frac {1}{X (c{-}X{-}Y)}+\frac {1}{Y (c{-}X{-}Y)}\,,
\ee
which can also be directly read off from the explicit result of the integral 
\be
{\cal I}_{\rm triangle}=(\alpha')^2~\frac{\Gamma(\alpha' X)~\Gamma(\alpha' Y)~\Gamma(\alpha' (c-X-Y))}{\Gamma(\alpha' c)}\,.
\ee
Similarly we can consider integrals for a quadrilateral. For example, one choice is to use the polynomial $(1+x)(1+y)=1+x+y+xy$; a slightly more general choice is to use $1+ x^p+y^q+ x^s y^t$ where $(0,0), (p, 0), (s, t), (0, q)$ form a convex quadrilateral. The first choice is special in that the integral factorizes into two beta functions:
\begin{align}
\label{eq:quad}
{\I}_{\rm quadrilateral}&=(\alpha')^2~\int_{\R_{>0}^2} \frac{d x d y}{x y}~x^{\alpha' X} y^{\alpha' Y} (1+x+y+x y)^{-\alpha' c} \nonumber \\
&=(\alpha')^2~B(\alpha' X, \alpha' (c-X)) B(\alpha' Y, \alpha' (c-Y))\,,
\end{align}
while for more general choices the result is more complicated. The leading order of \eqref{eq:quad} is $
\underline{\Omega}([0,c]^2; (X,Y) )=\frac{c}{X (c-X)} \frac{c}{Y (c-Y)}$. 

Finally, let's consider a pentagon example. We choose the polynomial to be $1+x^2+y^2+ x^2 y+x y^2$, and the integral
\be
{\I}_{\rm pentagon}=(\alpha')^2~\int_{\R_{>0}^2} \frac{d x d y}{x y}~x^{\alpha' X} y^{\alpha' Y} (1+x^2+y^2+ x^2 y+x y^2)^{-\alpha' c}
\ee
converges if and only if $(X,Y)$ is inside the pentagon
\be
\PP = \Conv((0,0), (2 c ,0), (2 c, c), (c,2 c), (0,2 c)),
\ee which holds exactly when $X>0, Y>0$, $2c-X>0$, $3 c-X-Y>0$ and $2c-Y>0$. The LHS of these $5$ inequalities are the linear functions of the $5$ edges, and the canonical function is given by
\be
\lim_{\alpha'\to 0} {\cal I}_{\rm pentagon}=\frac{1}{X Y}+\frac{1}{Y (2 c{-}X)}+\frac{1}{(2c{-}X) (3c{-}X{-}Y)}+\frac{1}{(3c{-}X{-}Y) (2c{-}Y)}+\frac{1}{(2c{-}Y) X}\,.
\ee

\end{example}
\section{Stringy canonical forms at finite $\alpha'$\label{sec:finitealpha}%Factorizations, ``Regge" limit and more
}
Throughout the paper we will mostly focus on the $\alpha'\to 0$ limit (and consider saddle points for the opposite, $\alpha'\to \infty$ limit) of stringy canonical forms, but in this section we initiate some preliminary investigations about these integrals at finite $\alpha'$. More detailed discussions will be left for future works.

A remarkable feature of stringy canonical forms is that they exhibit various properties that are reminiscent of string amplitudes, not only as $\alpha'\to 0$, but also at finite $\alpha'$. To start with, when $\alpha'\to 0$, we recover the canonical function of the Newton polytope, $\underline{\Omega}(c \New[p(\x)]; {\X})$: it is a rational function where poles correspond to facets of the polytope, and the residue on a pole is given by the canonical function of the corresponding facet. This important property of the residues of canonical forms is crucial in its interpretation as a scattering amplitude, {\it e.g.} when the positive geometry is the amplituhedron (planar ${\cal N}=4$ SYM) or the associahedron (bi-adjoint $\phi^3$-theory).

\subsection{Recurrence relations and residues at massless poles}

Essentially the identical statement applies to the stringy canonical form $\I_p(\X,c)$ at finite $\alpha'$ when we consider the behavior near the poles of the leading order, which we will refer to as {\it massless poles}.  We show that for any massless pole, which corresponds to a facet $F$ of the Newton polytope $\PP = c\New[p(\x)]$, the residue of $\I_p ({\X}, c)$ is given by the stringy canonical form for that facet. As we will see shortly, the stringy canonical form $\I_p(\X,c)$ has an infinite number of poles~\footnote{The fact that any stringy integral has infinite number of poles, which can be obtained by translating the massless poles by integer shifts, was first shown in Theorem 2 of \cite{Mellin}; also see~\cite{Speer1975} for applications to Feynman integrals.}, and the phenomenon occurs on massless poles. This means the class of stringy canonical forms is closed under the operation of taking residues, which generalize similar property of the canonical form of positive geometries. 

A particularly nice way to study these poles and residues is by studying {\it recurrence relations} for stringy canonical forms.  These are linear relations satisfied by $\I_p(\X, c)$ with arguments shifted.  For every direction $X_i$, we define the translation operator $\hT_i$ so that ${\hat T}_i~\I_p(\ldots, X_i, \ldots, c) = \I_p(\ldots, X_i{+}1, \ldots,c)$.  Similarly, we define $\hC~\I_p(\ldots, c) =\I_p(\ldots, c{+}1)$. 

We have two types of recurrence relations connecting integrals at shifted arguments. The first type is essentially trivial; it says a shift induced by $\hC$ amounts to a polynomial of shifts by $\hT$'s:
\be\label{shift1}
\I =\hC~p({\bf T}) \I :=(\hC \sum_\alpha  a_\alpha~{\bf \hT}^{{\bf n}_\alpha} )~\I\,.
\ee
where we use the notation of \eqref{Newton}.
The second type is given by total derivative identities; for each direction $i$, the total derivative with respect to $x_i$ gives the following relation:
\be\label{shift2}
X_i~\I = c~\hC ( \sum_\alpha n_{\alpha, i}~a_\alpha {\bf \hT}^{{\bf n}_\alpha} )~\I\,.
\ee
The recurrence relations are important for many reasons. For now, a simple consequence of them is that it gives the location of {\it massless} poles and residues of the integrals at finite $\alpha'$.  These poles are given by facets of the Newton polytope $\PP = c\New[p(\x)]$.  

%It is generally difficult to determine the residues of the integrals at all possible poles, but it has a remarkably simple answer for those poles that already appear in the leading order, which we call massless poles. The poles are given by the facets of the polytope $\PP$, which include those at $X_i=0$ and the remaining ones. Let's see why these are the poles at finite $\alpha'$, and what are the residues, using the shift relations. 
%Note that any facet of the Newton polytope, or a massless pole of the integral (which is present at leading order), can be written projectively in $\mathbb{P}^d$ as ${\cal W} \cdot {\cal Y}=0$, where ${\cal Y}:=(1, \bf{X})$ and ${\cal W}$ denotes a facet. In our parametrization, there are two types of poles. The first type is for $X_i=0$ or equivalently with ${\cal W}=(0, {\bf e}_i)$ where ${\bf e_i}$ is the unit vector in the $i$-th direction, and such a pole arises from $x_i \to 0$. The second type of poles arise from some direction at infinity $|\x| \to \infty$, for each facet of the Newton polytope. It can be written as ${\bf w}\cdot \X-c=0$ (corresponding to ${\cal W}=(-c, {\bf w})$ for some vector ${\bf w}$): for all points inside the Newton polytope, we have ${\bf n} \cdot {\bf w} - 1 \geq 0$, with equality for the points on the facet ${\bf w}$. Now we show that the integral $\I_p$ indeed has these poles at finite $\alpha'$, and the residue is given by an $d{-}1$-dim integral for the corresponding facet. 

To begin, let us study the special case when $X_i = 0$ is a facet of $\PP$.  We may then take the limit $X_i \to 0$, staying within the interior of $\PP$ for which we know $\I$ converges.  Looking at \eqref{shift2} and taking $X_i \to 0$, we see that the LHS is 0 unless $\I$ has a pole. For $X_i=0$ to be a facet, we implicitly assumed that $p(x_i)$ is a polynomial of $x_i$ (not necessarily for other $x_j$), thus $n_{\alpha, i}>0$ and the RHS is the sum of shifted integrals (evaluated at a convergent point) with positive coefficients, so it must be positive. Thus the LHS must have a pole as $X_i \to 0$. The residue can be computed from the RHS in the limit and gives
\be
{\rm Res}_{X_i\to 0} \I=\lim_{X_i \to 0}~c~\hC ( \sum_\alpha n_{\alpha, i}~a_\alpha {\bf \hT}^{{\bf n}_\alpha} )~\I=\int_{\R_{>0}^{d-1}} \prod_{j\neq i} \frac{d x_j}{x_j}~x_j^{\alpha' X_j}~p(x_i=0)^{-\alpha' c}\,.
\ee
The Newton polytope of $p(x_i=0)$ is exactly the facet of $\PP$ corresponding to $X_i = 0$, so we have shown that the residue of $\I_p$ along $X_i = 0$ is the stringy canonical form for the corresponding facet.  (Note that $X_i = 0$ is not always a facet of $\PP$.  When it is not, the argument breaks because we cannot take the limit $X_i \to 0$ while staying inside the domain of convergence.)  A very similar argument applies for facets of $\PP$ of the form $ \X \cdot \bmu = 0$, where $\bmu$ is a normal vector for that facet.

Now let us consider a facet of $\PP$ given by $\X\cdot \bmu-c=0$, where $\bmu$ is an inward pointing normal vector for the facet.
We take a linear combination of the relations~\eqref{shift2}, which can be done by dotting it into the vector $\bmu \in \mathbb{R}^d$:
\be
(\X\cdot \bmu) \I = c~C (\sum_\alpha ({\bf n}_\alpha \cdot \bmu)~a_\alpha~{\bf T}^{{\bf n}_\alpha})~\I\,,
\ee
and by using \eqref{shift1}, we can rewrite this as
\be
(\X \cdot \bmu- c) \I = c~C \sum_\alpha ({\bf n}_\alpha \cdot \bmu- 1)~a_\alpha~{\bf T}^{{\bf n}_\alpha}~\I
\ee
Again, taking the limit $\X \cdot \bmu -c \to 0$ from inside $\PP$, the RHS is a sum of shifted integrals with positive coefficients (evaluated at a convergent point), so the LHS must have a pole when $\X \cdot \bmu - c \to 0$.  Let $p_\bmu(\x)$ be the sum of of all monomials $a_\alpha \x^{\n_\alpha}$ where $\n_\alpha \cdot \bmu = m$ takes minimum value $m$ (equal to 1).  Letting $Y_1,\ldots,Y_{d-1}$ be coordinates on the codimension one subspace $\X \cdot \bmu=c$, and $y_i = \x^{\n_\beta}$ be appropriate monomials satisfying $\n_\beta \cdot \bmu = m$ we have
\be
{\rm Res}_{\X \cdot \bmu -c \to 0} \I=\int_{\R_{>0}^{d-1}}  \prod_{j=1}^{d-1} \frac{d y_j}{y_j}~y_j^{\alpha' Y_j}~p_\bmu(\y)^{-\alpha' c}\,,
\ee
the stringy canonical form for the facet $\X \cdot \bmu -c = 0$. It is clear from our argument that for each massless pole (a facet of the Newton polytope), there is a family of infinite number of poles of the form $\X \cdot \bmu -c = N$ for integer $N$. However, this argument alone does not exclude the possibility of poles of other types, and as shown in~\cite{Mellin} all the poles of the integrals must be integer translation of the massless ones.

\subsection{Stringy properties}

Stringy canonical form integrals have a number of important qualitative features in common with usual string amplitudes. Here we content ourselves with briefly sketching some of these properties, leaving a fuller exposition to future work. As we have mentioned, stringy canonical forms are meromorphic functions of the ``kinematic" exponent variables~\cite{Mellin}, just as string amplitudes are. A standard saddle-point analysis also tells us that in the ``high-energy" limit where all the variables are large compared to $1/\alpha'$, or equivalently, in the $\alpha' \to \infty$ limit, the integral is exponentially suppressed, so are ``exponentially soft at high energies" just as string amplitudes are. In addition, we also have interesting analogs of ``Regge behavior" and``channel duality".  

Consider first the Regge behavior of usual open string amplitudes. We can consider the 4 particle scattering of colored scalar particles in type I string theory. The tree amplitude is $A^{open}(s,t)= (\alpha' s)^2 \Gamma(-\alpha^\prime s) \Gamma(- \alpha^\prime t)/\Gamma(1-\alpha^\prime s- \alpha^\prime t)$. Now, in the field theory limit when $\alpha^\prime s, \alpha^\prime t$ are small,  the amplitude from gluon exchange behaves as $s/t$. The residue on the $t = 0$ pole is simply $s$. Regge behavior tells us  that the residue of the full string amplitude as $t \to 0$ is exactly $s$, independent of $\alpha'$ , and in particular for {\it any} $s$ no matter how large compared to $1/\alpha'$. 

Note that $A^{open}(s,t)$ is not itself a canonical form, but is simply related to the Beta function $B(-\alpha' s,-\alpha' t)$ and hence the canonical form of an interval via $A^{open} (s,t) = -\alpha' s^2/(s+t) \times B(-\alpha' s,-\alpha' t)$. The property we just stated for the physical amplitude is obviously also true of the Beta function/canonical form. The residue of $B(-\alpha' s,-\alpha' t)$ at $\alpha' t \to 0$ is equal to 1, and is independent of $\alpha^\prime$. 

This is a simple general feature of all stringy canonical forms, which follows from the self-reduction properties of the residues we have highlighted, that are true even at finite $\alpha'$. In particular, if we take a residue of the stringy canonical form localizing to a vertex of the polytope, obtained by setting some set of kinematic variables to zero, the corresponding residue is fixed to unity, independent of $\alpha'$.

The full statement of Regge behavior extends this to small but finite values of the kinematic variables.  Again for four-particle scattering, it says that the behavior of the amplitude at fixed $t$ is polynomially bounded in $s$. Of course since the amplitude has poles on the real axis, this statement is understood to be strictly valid off the real axis.  For instance, the Stirling approximation shows that at fixed $t$, $B(-\alpha' s,-\alpha' t) \to \frac{ s^{\alpha^\prime t}}{t}$ and $A^{open} (s,t) \to \frac{ s^{ 1 + \alpha^\prime t}}{t}$ as $s \to \infty$. Note this behavior smoothly interpolates from the previous observation that residue strictly at $t \to 0$ is some power of $s$. Thus for small negative $t$, the large $s$ behavior is even further suppressed. 
This extends to general stringy canonical forms as well. If we move the exponent variables so as to be close to, but not exactly at, a vertex of the polytope, keeping away from the vertex but non-zero  only further suppresses the integral. 

Regge behavior is also important in establishing another intrinsically ``stringy" feature of string amplitudes. Consider again the familiar case of four-particle scattering. Regge behavior says that for fixed $t$, the amplitude is bounded by a power of $s$.  Cauchy's theorem then allows one to expand the amplitude as a sum over poles in the $s$-channel, but only when $t$ is small. Similarly, we can expand as a sum over poles in the $t$ channel when $s$ is small, but there is no regime where there is a double expansion in both channels. In the representation of the amplitude as a sum over poles in $s$, the $t$ poles are not manifest, but are encountered as we continue away from small $t$ (where the expansion is justified), arising from the infinite sum over $s$ poles.  This is ``channel duality", and contrasts with what we are familiar with in quantum field theory, where amplitudes are expressed as a (finite) sum over poles in all channels. 

The expansion as a sum over poles in $s$ or $t$, is also easy to see from the integral for the Beta function. For fixed (negative) $t$, the integral is convergent as $x \to 1$. We can then Taylor expand the integrand around $x = 0$, with the integrals $\int_0^1\frac{dx}{x} x^{-s} x^n = \frac{1}{n-s}$, giving the poles in $s$. This argument extends to general stringy canonical forms. Take any vertex of the polytope; as we have discussed we can choose positive co-ordinates for the integrand to explicitly see the poles associated with the facets of the polytope meeting at the vertex. Taylor expanding around this origin then gives us a sum over poles in these facet variables, keeping the other exponent variables fixed. Thus we {\it can} 
express any stringy canonical form as a sum over poles of variables associated with the facets meeting at any vertex of the polytope, but there is no "global" expression manifesting all the poles; instead the "missing" set of poles in any given representation arise from the infinite sum. 
We can illustrate this with a trivial example: the canonical form of a square, that is product of two intervals. The integral is just the product of two Beta functions $B(a,c) \times B(b,d)$, where $a,c$,$b,d$ are associated with opposing facets of the square. We can expand $B(a,c)$ as a sum over poles in $a$ or $c$, and similarly $B(b,d)$ as a sum over poles in $b$ or $d$. Thus, we have expansions for the product as a sum over poles in $(a,b),(a,d),(c,b),(c,d)$, corresponding the pairs of edges that meet at the four corners of the square. But there is no expansion in $(a,c)$ or $(b,d)$, which do not meet at a vertex of the square. 

We have highlighted some of the qualitative similarities between string amplitudes and stringy canonical forms for general polytopes. 
But of course string amplitudes are more magical than stringy canonical forms for generic polytopes. The chief difference is seen in the nature of their residues. As we have emphasized, all stringy canonical forms are ``self-reductive" under taking residues even at finite $\alpha^\prime$: the residues are in turn stringy canonical forms for facets of the polytope, where only the monomials living on that facet of the Newton polytope are kept. But the canonical form for a given combinatorial polytope is not fixed, it depends on the polynomial parameterization. Thus we can have the canonical form of a cube, and the residue on any square facet will be self-reductive, but while that residue will be the same for all squares as $\alpha^\prime \to 0$, at finite $\alpha^\prime$ ``a square isn't a square": the residue on two different squares will in general differ. 

This is not the case for string amplitudes. As we will discuss more fully below, $n$-particle string amplitudes can actually be discovered from a new point of view, as canonical forms for a special polytope--the associahedron ${\cal A}_n$--represented in a canonical way as a Minkowski sum of simplices, which is very natural from several points of view~\cite{postnikov, Arkani-Hamed:2017mur}. The associahedron has a remarkable boundary structure: facets of ${\cal A}_n$ are direct products, ${\cal A}_m \times {\cal A}_{n{-}m{-}1}$, of lower associahedra. As $\alpha^\prime \to 0$, this is reflected in factorization of the canonical form on massless poles, reflecting the locality and unitarity of the field theory limit of scattering amplitude. Now if we compute a stringy canonical form for a generic Newton polytope of an associahedron at finite $\alpha^\prime$, we will still have factorization, but what we mean for the forms of ${\cal A}_m, {\cal A}_{n{-}m{-}1}$ will be different for different facets; ``an associahedron isn't an associahedron". What is truly remarkable about string amplitude is that this factorization property holds at finite $\alpha^\prime$ and that the forms for the different ${\cal A}_m$ factors seen in the boundaries are all the same.  As we will also discuss, this magical property is not only restricted to string amplitudes; but are extended to stringy canonical forms associated with generalized associahedra, or cluster string integrals, as well.

\section{Minkowski sums and tropical functions}\label{sec:multi}
\subsection{Stringy integrals with many polynomials}\label{sec:stringymulti}
Now we proceed to more general stringy canonical forms, which are generalizations of \eqref{proto_int} to the case where in addition to the factor $\prod_i x_i^{\alpha' X_i}$, there are multiple polynomials. 
%This is an essentially trivial generalization since what we need is the Minkowski sum of all the Newton polytopes for these polynomials. 
We consider the integral
\begin{equation}
\I_{\{p\}} (\X, \{c\})=(\alpha')^d  \int_{\R_{>0}^d} \prod_{i=1}^d \frac{d x_i}{x_i}~x_i^{\alpha' X_i}~\prod_I p_I (\x)^{-\alpha' c_I}\,,\label{gen_int}
\end{equation}
where all the polynomials $p_I(\x)$ are subtraction-free Laurent polynomials.
%where again $X_i>0$ and we assume all polynomials $p_I(\x)$$ are subtraction free with $c_I>0$. %This can be stated as our first theorem, which is a direct consequence of the lemma. 

\begin{claim} \label{claim:multi}
Assuming that $X_i \geq 0$ and $c_I \geq 0$, the integral \eqref{gen_int} converges if and only if $\X$ is inside the {\it Minkowski sum} of the polytopes $c_I \New [ p_I(\x) ]$ defined as
\be \label{Minkowski}
\PP:=\sum_I c_I \New [p_I(\x)]=\left\{ \sum_I c_I \uu_I | \uu_I \in \New[p_I(\x)]\right\}\,,
\ee
and the leading order is given by its canonical function of $\PP$ at $\X$:
\be
\lim_{\alpha'\to 0} \I_{\{p\}} (\X, \{c\})=\underline{\Omega}(\PP; \X)\,.
\ee
\end{claim}

The condition for convergence of the integral was found in Theorem 2.3 of \cite{Mellin}. The Minkowski sum of any polytope $\PP$ with a point $\X$ is the translation of the polytope $\PP$ by the vector $\X$.  Note that the point $\X$ is itself the Newton polytope of the monomial $\prod_i x_i^{X_i}$, so Claim \ref{claim:multi} agrees with the sentence after \eqref{eq:LS}.  A few examples for Minkowski sums: for $(1+x)^a (1+y)^b$, the Minkowski sum of two intervals in the $x$ and $y$ directions, $a \New [1+x ]+ b \New [1+y ]$, gives a quadrilateral. Below are two more examples of Minkowski sums:
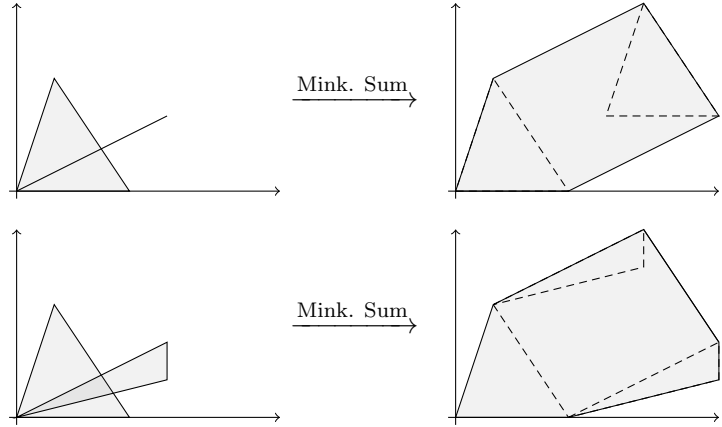
\begin{center}
\begin{tikzpicture}[baseline={([yshift=-.5ex]current bounding box.center)}]
\fill[gray!10] (3,-2) -- (2.5,-3.5) -- (4,-3.5) -- cycle;
\draw (3,-2) -- (2.5,-3.5) -- (4,-3.5) -- cycle;
\draw (2.5,-3.5) -- (4.5,-2.5);
\draw[->](2.5,-3.6) -- (2.5,-1); 
\draw[->](2.4,-3.5) -- (6,-3.5);
\end{tikzpicture}
$\xrightarrow{\text{Mink. Sum}}$\quad
\begin{tikzpicture}[baseline={([yshift=-.5ex]current bounding box.center)}]
\draw[fill=gray!10] (2.5,-3.5) -- (3,-2) -- (5,-1) -- (6,-2.5) -- (4,-3.5) -- cycle;
\draw[densely dashed] (3,-2) -- (2.5,-3.5) -- (4,-3.5) -- cycle;
\draw[densely dashed] (5,-1) -- (4.5,-2.5) -- (6,-2.5) -- cycle;
\draw[->](2.5,-3.6) -- (2.5,-1); 
\draw[->](2.4,-3.5) -- (6,-3.5);
\end{tikzpicture}
\end{center}

\begin{center}
\begin{tikzpicture}[baseline={([yshift=-.5ex]current bounding box.center)}]
\fill[gray!10] (3,-2) -- (2.5,-3.5) -- (4,-3.5) -- cycle;
\draw (3,-2) -- (2.5,-3.5) -- (4,-3.5) -- cycle;
\draw[fill=gray,fill opacity=0.1] (2.5,-3.5) -- (4.5,-2.5) -- (4.5,-3) -- cycle;
\draw[->](2.5,-3.6) -- (2.5,-1); 
\draw[->](2.4,-3.5) -- (6,-3.5);
\end{tikzpicture}
$\xrightarrow{\text{Mink. Sum}}$\quad
\begin{tikzpicture}[baseline={([yshift=-.5ex]current bounding box.center)}]
\draw[fill=gray!10] (2.5,-3.5) -- (3,-2) -- (5,-1) -- (6,-2.5) -- (6,-3) -- (4,-3.5) -- cycle;
\draw[densely dashed] (3,-2) -- (4,-3.5);
\draw[->](2.5,-3.6) -- (2.5,-1); 
\draw[->](2.4,-3.5) -- (6,-3.5);
\draw[densely dashed] (4,-3.5) -- (6,-2.5) -- (6,-3) -- cycle;
\draw[densely dashed] (3,-2) -- (5,-1) -- (5,-1.5) -- cycle;
\draw (5,-1) -- (6,-2.5);
\end{tikzpicture}
\end{center}

To prove Claim \ref{claim:multi} from Claim \ref{claim:single}, we first consider two polynomials $p_1$, $p_2$ with {\it rational} exponents, $c_1=r_1/s_1$, $c_2=r_2/s_2$.  We have 
\be
p_1^{c_1} p_2^{c_2}=(p_1^{r_1 s_2} p_2^{r_2 s_1})^{1/(s_1 s_2)}\,,
\ee
which corresponds to \eqref{proto_int} with a single (subtraction-free) polynomial $p:= p_1^{r_1 s_2} p_2^{r_2 s_1}$ and $c=1/(s_1 s_2)$. The convergence and the leading order is controlled by the Newton polytope $\frac 1 {s_1 s_2} \New [ p]$, and by definition this is exactly the Minkowski sum $\frac{r_1}{s_1} \New [p_1]+\frac{r_2}{s_2} \New [p_2]$. The result clearly generalizes to more polynomials with rational exponents, and by continuity it holds for any real exponents as well. We conclude that for any factor of the form $ p_1^{c_1} \cdots p_r^{c_r}$, we need the Minkowski sum $c_1 \New [ p_1]+ \cdots + c_r \New [ p_r]$.  Thus Claim \ref{claim:multi} follows from Claim \ref{claim:single}.
%
%This argument has actually been used when earlier when we absorb $\prod_i x_i^{\alpha' X_i}$ into the polynomial factor. Here again the equivalent statement is that we can consider the polytope from translating the Minkowski sum by $-\X$ (due to the factor $\prod_i x_i^{\alpha' X_i}$), then the integral converges if and only if $0$ is in its interior and the leading order is the volume of the dual polytope. 

Just as in Claim \ref{claim:single}, we can also consider the monomial factor $\prod_i x_i^{\alpha' X_i}$ as additional polynomials, instead of separating them.  Explicitly, we have the following statement.  Define
\be\label{general_Mink}
\I_{\{p\}}(S):= (\alpha')^d \int_{\R_{>0}^d} \prod_{i=1}^d \frac{d x_i}{x_i} \prod_{J=1}^r p_J(\x)^{-\alpha' S_J}.
\ee
where we assume that $S_J \geq 0$ for all $j$.  Then the integral $\I_{\{p\}}(S)$ converges if and only if $0$ is in the interior of the Minkowski sum $S_1\PP_1 + \cdots S_r\PP_r$ (where $\PP_J= \New [ p_J(\x) ]$ is the Newton polytope) and we have the formula
\be \label{eq:IS}
 \lim_{\alpha' \to 0} \I_{\{p\}}(S) = \Vol((S_1\PP_1 + \cdots S_r \PP_r)^\circ).
\ee
%\begin{example}[Pentagon] For a pentagon, there are again numerous ways to write down $2$-dimensional integrals, and a simple example reads
%\be\label{pentagon}
%\I_{\rm pentagon}=\alpha'^2~\int_0^\infty \frac{d x\, dy}{x\,y} x^{\hat{X}} y^{\hat{Y}} (1+x)^{-\hat{a}} (1+y)^{-\hat{b}}(1+x+ y)^{-\hat{c}}
%\ee
%where we use $\hat{X}:=\alpha' X$ {\it etc.} to suppress explicit dependence on $\alpha'$. By Claim \ref{claim:multi}, we need to consider the Minkowski sum of the intervals $0<X<a$, $0<Y<b$ and the triangle $0<Y<X<c$, and it is easy to see that the sum gives a pentagon $\PP_2$, which is also the region for convergence.  Explicitly, the pentagon is bounded by the $5$ facets $X_1:=X>0$, $X_2:=Y>0$,  $X_3:=a+c-X>0$, $X_4:=a+b+c-X-Y>0$, and $X_5:=b+c-Y>0$, and the canonical function which gives the leading order of the integral reads
%$%\lim_{\alpha' \to 0} \I_{\rm pentagon}=
%\underline{\Omega}(\PP_2)=\frac{1}{X_1  X_2}+ \frac{1}{X_2 X_3}+ \frac{1}{X_3 X_4}+ \frac{1}{X_4 X_5}+ \frac{1}{X_5 X_1}$. Again, there are many integrals that we can write down for this pentagon: we can consider {\it e.g.} $1+ p_1 x$, $1+p_2 y$, and $1+ p_3 x + p_4 y$ and the leading order remains unchanged. 
%\end{example}

Having seen that the leading $\alpha^\prime \to 0$ order behavior of the stringy canonical form is controlled by the Minkowski sum of the Newton polytopes of the polynomials $p_J(\x)$, let us illustrate how these Minkowski sums can be computed in practice, in some examples. Consider first the $n=5$ case of string amplitude \eqref{eq:string}:
\be\label{5ptstring}
{\bf I}_5=(\alpha')^2 \int_{\R^{n{-}3}_{>0}} \frac{d x_2}{x_2} \frac{d x_3}{x_3} x_2^{\alpha' X_{2 4}} x_3^{\alpha' X_{3 5}} p_{1,3}^{-\alpha' c_{13}} p_{1,4}^{-\alpha' c_{14}} p_{2,4}^{-\alpha' c_{24}}\,.
\ee
where, to be consistent with the notation of~\cite{Arkani-Hamed:2017mur}, we have used $X_{i,j}:=\sum_{i\leq a<b<j} s_{a,b}$ thus $X_{24}=s_{23}$, $X_{35}=s_{34}$ and $c_{i,j}=-s_{i,j}$ for non-adjacent $i, j$. 

This involves taking the Minkowski sum of the Newton polytopes  $\PP_{1,3}, \PP_{1,4}, \PP_{2,4}$ associated with the polynomials $p_{1,3} = 1 + x_2, p_{1,4}=1 + x_2 + x_3$, and $p_{2,4} =x_2 + x_3$. The Newton polytopes for $p_{1,3}$ and $p_{2,4}$ are line segments, while that of $p_{1,4}$ is a triangle. In this case, it is  easy to perform the Minkowski sum directly and obtain a pentagon, as in the figure
\begin{equation}\begin{tikzpicture}\small
\draw (-3.7,0.5) -- (-2.7,0.5);
\node[inner sep=1pt,circle,fill=black] at (-3.7,0.5) {};\node[below] at (-3.7,0.5) {$a$};
\node[inner sep=1pt,circle,fill=black] at (-2.7,0.5) {};\node[below] at (-2.7,0.5) {$b$};
\node at (-2.3,0.5) {$+$};
\draw (-1,1) -- (0,0);\draw (-2,1) -- (-2,0);\draw (-2,1) -- (-1,0);
\node[inner sep=1pt,circle,fill=black] at (-2,1) {};\node[above] at (-2,1) {$d$};
\node[inner sep=1pt,circle,fill=black] at (-2,0) {};\node[below] at (-2,0) {$c$};
\node[inner sep=1pt,circle,fill=black] at (-1,0) {};\node[below] at (-1,0) {$e$};
\node at (-1,0.5) {$+$};
\draw (-2,0) -- (-1,0);
\node[inner sep=1pt,circle,fill=black] at (-1,1) {};\node[above] at (-1,1) {$g$};
\node[inner sep=1pt,circle,fill=black] at (0,0) {};\node[below] at (0,0) {$h$};
\node at (0.2,0.5) {$=$};
\draw (0.6,1) -- (0.6,0) -- (2.6,0) -- (1.6,1) --cycle;
\node[below] at (0.6,0) {$a{+}c$};\node[inner sep=1pt,circle,fill=black] at (0.6,0) {};
\node[below] at (2.6,0) {$b{+}e$};\node[inner sep=1pt,circle,fill=black] at (2.6,0) {};
\node[above] at (0.6,1) {$a{+}d$};\node[inner sep=1pt,circle,fill=black] at (0.6,1) {};
\node[above] at (1.6,1) {$b{+}d$};\node[inner sep=1pt,circle,fill=black] at (1.6,1) {};
\node at (2.6,0.5) {$+$};
\draw (2.6,1) -- (3.6,0);
\node[above] at (2.6,1) {$g$};\node[inner sep=1pt,circle,fill=black] at (2.6,1) {};
\node[below] at (3.6,0) {$h$};\node[inner sep=1pt,circle,fill=black] at (3.6,0) {};
\node at (3.6,0.5) {$=$};
\draw (5.1,1.5) -- (4.1,1.5) -- (4.1,0.5) -- (5.1,-0.5) -- (7.1,-0.5) --cycle;
\node[below] at (5.1,-0.5) {$a{+}c{+}h$};\node[inner sep=1pt,circle,fill=black] at (5.1,-0.5) {};
\node[below] at (7.1,-0.5) {$b{+}e{+}h$};\node[inner sep=1pt,circle,fill=black] at (7.1,-0.5) {};
\node[right] at (4.1,0.5) {$a{+}c{+}g$};\node[inner sep=1pt,circle,fill=black] at (4.1,0.5) {};
\node at (5.3,1.8) {$b{+}d{+}g$};\node[inner sep=1pt,circle,fill=black] at (5.1,1.5) {};
\node at (3.9,1.8) {$a{+}d{+}g$};\node[inner sep=1pt,circle,fill=black] at (4.1,1.5) {};
\end{tikzpicture}
\end{equation} 
But let us illustrate some general facts about Minkowski sums already in this example, which will be useful in more general situations where we can't easily draw a picture. 

Suppose we have two polytopes $A$ and $B$, we would like to characterize the vertices and the facets of the Minkowski sum $(A+B)$ in terms of those of $A$ and $B$. For instance, we know that $(A+B)$ is the convex hull of all points of the form $v_A + v_B$ where $v_{A,B}$ range over the vertices of $A$, $B$. But not all pairs $v_A+ v_B$ will be vertices of $(A+B)$; some might be in the interior of $(A+B)$, so we'd like to know which pairs do end up as vertices of $(A+B)$.  To understand this it is useful to use the cones $C_v$ \eqref{eq:vertexcone} in the dual space to a polytope $\PP$.  Recall this is the space of all $\bla$ in $\R^d$, such for all $\uu$ in ${\PP}$, the function $\uu \cdot \bla $ attains its minimum value at $\uu = \v$, or equivalently, that $\v \cdot \bla \leq \v' \cdot \bla$ for all other vertices $\v'$ in ${\PP}$. Note that we might try to define this cone associated with any point $\y$ in the ${\PP}$, i.e. can look for all 
$\bla$ such that $\uu \cdot \bla$ is minimized exactly at $\uu=\y$. But this space is obviously empty unless $\y$ is one of the vertices of ${\PP}$.  Indeed, $\v$ is a vertex of $\PP$ precisely when the cone $C_v$ is non-empty and top-dimensional. 
The normal fan $\N(\PP)$ of ${\PP}$ is the collection of all the cones $C_v$, and these tile all of $\bla$ space. Every point $\bla$ belongs to one or more of the $C_v$. A generic point $\bla$ will belong to just one $C_v$, but on some co-dimension one surfaces it can belong to two cones, and so on for intersections on low-dimensional faces of the cones. Indeed the way in which the cones intersect on their lower-dimensional faces determines the face structure of the polytope: if some collection of cones $C_{v_1}, \ldots, C_{v_p}$ meet along a $q$-dimensional cone, the corresponding vertices ${v_1, \ldots, v_p}$ lie on a face of $\PP$ of co-dimension $q$. 

Now, let's turn to understanding whether $v_A + v_B$ is a vertex of $A+B$. For this to be the case, we must be able to find some $\bla$ for which $(\v_A+\v_B) \cdot \bla  \leq  (\uu_A + \uu_B) \cdot \bla$ for all $\uu_A \in A$ and $\uu_B \in B$.  But this means that $\bla$ must be in the cones of both $C_{v_A}$ and $C_{v_B}$, so we learn that $(\v_A + \v_B)$ is a vertex if and only if the cones have $C_{v_A}, C_{v_B}$ have top-dimensional intersection.  The general statement is that the normal fan of the Minkowski sum $(A+B)$ is the ``common refinement" of the normal fans $\N(A)$ and $\N(B)$, given by intersecting all the cones in $\N(A)$ with those in $\N(B)$. 

There is a closely related way of thinking about the cones and normal fan of a polytope.  Suppose $v$ is a vertex of a top-dimensional polytope. Then all the points in the cone $C_v$ can be written as a positive linear combination of some generators, $\bla = \sum_{i=1}^p w_i \bla_i$ where $p \geq d$. The vectors $\bla_i$ are nothing but the inward-pointing normal vectors to the $p \geq d$ facets of ${\PP}$, meeting at $v$. So we can think of the normal fan of ${\cal P}$ as being determined by giving all the normal vectors to the facets of ${\PP}$.  \footnote{The notion of a cone is however more natural and flexible when working with polytopes of different dimensionality, where the ``normal directions to facets" are ambiguous. Consider for instance a line segment between $(0,1)$ on the $X$-axis, but embedded in three-dimensional $(X, Y, Z)$ space. There is no canonical way to speak of the ``inward pointing normals to the faces" of this line segment in three dimensions. But we can perfectly well speak of the cones associated with the vertices. The cone associated with the point $(0,0,0)$  is just the half-space $X\leq 0$, while that associated to $(1,0,0)$ is the half-space $X \geq 0$. These cones are not associated with generators.  But now consider the Minkowski sum of three such intervals in the $X,Y,Z$ directions. The common refinement of the three fans just break up the space into octants, and each octant is associated with three generators, giving us a cube, with 8 vertices and 6 facets.}

Thus we see that the problem of finding the Minkowski sum $(A+B)$ is reduced to finding the common refinement of the normal fans of $A,B$, or what is the same, to intersecting the cones in these normal fans. Let's illustrate how this happens in our pentagon example. Let's put $\bla = (X, Y)$. As an example,  the cone associated with the vertex $(1,0)$ of $p_{1,4}$ is the region determined by $x\leq 0, x-y \leq 0$. We can determine all the rest of the cones in the same way, and find the common refinement just by drawing all the pictures on top of each other 
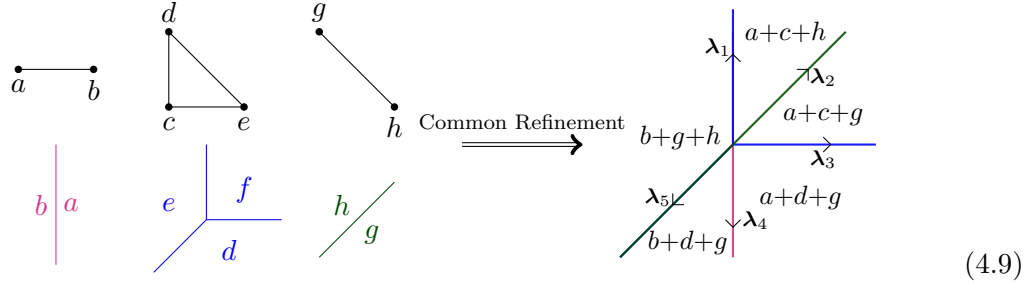
\begin{equation}
\begin{tikzpicture}
\draw (-4,-1) -- (-3,-1);
\node[inner sep=1pt,circle,fill=black] at (-4,-1) {};\node[below] at (-4,-1) {$a$};
\node[inner sep=1pt,circle,fill=black] at (-3,-1) {};\node[below] at (-3,-1) {$b$};
\draw (0,-0.5) -- (1,-1.5);\draw (-2,-0.5) -- (-2,-1.5);\draw (-2,-0.5) -- (-1,-1.5);
\node[inner sep=1pt,circle,fill=black] at (-2,-0.5) {};\node[above] at (-2,-0.5) {$d$};
\node[inner sep=1pt,circle,fill=black] at (-2,-1.5) {};\node[below] at (-2,-1.5) {$c$};
\node[inner sep=1pt,circle,fill=black] at (-1,-1.5) {};\node[below] at (-1,-1.5) {$e$};
\draw (-2,-1.5) -- (-1,-1.5);
\node[inner sep=1pt,circle,fill=black] at (0,-0.5) {};\node[above] at (0,-0.5) {$g$};
\node[inner sep=1pt,circle,fill=black] at (1,-1.5) {};\node[below] at (1,-1.5) {$h$};
\draw[magenta!90!black] (-3.5,-2) -- (-3.5,-3.6);
\node[magenta!90!black] at (-3.7,-2.8) {$b$};
\node[magenta!90!black] at (-3.3,-2.8) {$a$};
\draw[blue] (-1.5,-2) -- (-1.5,-3) -- (-2.2,-3.7);
\draw[blue] (-1.5,-3) -- (-0.5,-3);
\node[blue] at (-1.2,-3.4) {$d$};
\node[blue] at (-2,-2.8) {$e$};
\node[blue] at (-1,-2.6) {$f$};
\draw[green!30!black] (1,-2.5) -- (0,-3.5);
\node[green!30!black] at (0.3,-2.8) {$h$};
\node[green!30!black] at (0.7,-3.2) {$g$};
\draw[thick,magenta!90!black,opacity=0.9] (5.5,-0.5) -- (5.5,-3.5);
\draw[thick,blue,opacity=0.9] (5.5,-0.2) -- (5.5,-2) -- (4,-3.5);
\draw[thick,blue,opacity=0.9] (5.5,-2) -- (7.4,-2);
\draw[thick,green!30!black,opacity=0.9] (7,-0.5) -- (4,-3.5);
\draw[double distance = 1pt,->] (1.9,-2) -- (3.5,-2);
\node at (2.7,-1.7) {\scriptsize\text{Common Refinement}};
\node at (6.7,-1.6) {\small $a{+}c{+}g$};
\node at (6.4,-2.7) {\small $a{+}d{+}g$};
\node at (4.9,-3.3) {\small $b{+}d{+}g$};
\node at (4.8,-1.9) {\small $b{+}g{+}h$};
\node at (6.2,-0.5) {\small $a{+}c{+}h$};
\draw (5.4,-0.9) -- (5.5,-0.8) -- (5.6,-0.9);
\draw (6.35,-1) -- (6.5,-1) -- (6.5,-1.15);
\draw (6.7,-1.9) -- (6.8,-2) -- (6.7,-2.1);
\draw (5.4,-3) -- (5.5,-3.1) -- (5.6,-3);
\draw (4.7,-2.65) -- (4.7,-2.8) -- (4.85,-2.8);
\node at (5.3,-0.7) {\scriptsize$\bla_1$};
\node at (6.7,-1.1) {\scriptsize$\bla_2$};
\node at (6.7,-2.2) {\scriptsize$\bla_3$};
\node at (5.8,-3) {\scriptsize$\bla_4$};
\node at (4.5,-2.7) {\scriptsize$\bla_5$};
\end{tikzpicture}
\end{equation}
We have thus determined the vertices of the resulting pentagon,  and also determined the inward normal vectors to the facets, which are $\bla_1 = (0,1), \bla_2=(1,1), \bla_3=(1,0), \bla_4=(0,-1), \bla_5=(-1,1)$. 

Note that having determined the normals to the facets of $\PP = c_1 \PP_1 + \cdots + c_j \PP_j$, it is trivial to obtain the linear equations associated with a given normal  $\bla$, that collectively cut out $\PP$. Since every $\uu$ in $\PP$ is of the form $\uu = c_1 \uu_1 + \cdots + c_j \uu_j$ with $\uu_i$ in $\PP_i$, then $\uu \cdot \bla  = c_1 (\uu_1 \cdot  \lambda) + \cdots +c_j (\uu_j \cdot \bla)$.  But for each $j$, the function $(\uu_j \cdot \bla)$ has some minimum value $m_j$ as $\uu_j$ ranges inside the polytope $\PP_j$, which is just the minimum value of $\v \cdot \bla$ where $v$ ranges over all the vertices of $\PP_j$. Thus, we find that the equation for the facet with normal $\bla$ is $\uu \cdot \bla  \geq c_1 m_1 + \cdots+  c_j m_j$. Applying this simple algorithm to our example, we find that the five equations cutting out the pentagon $c_{13} {\PP}_{1,3} + c_{14} {\PP}_{1,4} + c_{24} {\PP}_{2,4}$, associated with $\bla_{1,\ldots,5}$, as (note the exponents for $x_2$ and $x_3$ are $X_{2,4}$ and $X_{3,5}$, respectively)
\begin{equation}
X_{2,4} \geq 0, X_{2,4}+ X_{3,5} \geq c_{24}, X_{3,5}  \geq 0, -X_{3,5} \geq - c_{14} - c_{24}, -X_{2,4} - X_{3,5} \geq - c_{13} - c_{14} - c_{24}\,. 
\end{equation}

Let's follow the same steps for the $n=6$ string amplitude
\be
{\bf I}_6=(\alpha')^3 \int_{\R^{3}_{>0}} \frac{d x_2}{x_2} \frac{d x_3}{x_3} \frac{d x_4}{x_4} x_2^{\alpha' X_{2, 4}} x_3^{\alpha' X_{3,5}}  x_4^{\alpha' X_{4,6}} p_{1,3}^{-\alpha' c_{13}} p_{1,4}^{-\alpha' c_{14}} p_{1,5}^{-\alpha' c_{14}} p_{2,4}^{-\alpha' c_{24}} p_{2,5}^{-\alpha' c_{25}} p_{3,5}^{-\alpha' c_{35}} 
\ee
whose leading oder, or the $n=6$ particle scattering amplitude is given in terms of the three-dimensional associahedron; the latter is given by $\sum c_{i,j} {\cal P}_{i,j}$, where ${\cal P}_{i,j}$ are the Newton polytopes for the polynomials $p_{1,3} = 1+x_2,  \;\;p_{1,4}=1 + x_2 + x_3, \;\;p_{1,5} = 1 + x_2 + x_3 + x_4, \;\; p_{2,4} = x_2 + x_3,  \;\;p_{2,5}=x_2 + x_3 + x_4, \;\; p_{3,5}=x_3 + x_4$. The sum of the three line segments ${\cal P}_{1,3}, {\cal P}_{2,4}, {\cal P}_{3,5}$ gives a parallelohedron, with 6 rays $\pm(0,0,1),\pm(1,1,1), \pm(0,1,1)$ in it's normal fan. ${\cal P}_{1,5}$ is just a simplex, with four vectors in its normal fan, $(-1,-1,-1),(0,0,1),(0,1,0),(1,0,0)$. Finally, the sum of the two triangles ${\cal P}_{1,4}$ and ${\cal P}_{2,5}$
 can easily be seen to have a normal fan generated by $ (1,0,0),(0,1,0),\pm(0,0,1), \pm(1,1,1)$. Interestingly, the intersection of these cones generate a single new ray $(1,1,0)$ in the common refinement, giving us a total of $9$ rays; 
$(1,0,0),(0,1,0),\pm (0,0,1),\pm(1,1,1),\pm(0,1,1), (1,1,0)$. From these, we derive the 9 facet equations cutting out the polytope:
\begin{align}
\begin{split}
&X_{2,4} \geq 0, \qquad X_{3,5} \geq 0, \qquad X_{4,6} \geq 0, \qquad X_{4,6}  \leq c_{15} + c_{25} + c_{35} \\
&-X_{2,4}-X_{3,5}-X_{4,6} \geq - c_{13} -  c_{14} - c_{15} - c_{24} - c_{25} - c_{35}, \\ &X_{2,4} + X_{3,5} + X_{4,6} \geq c_{24} + c_{25} + c_{35}, \qquad X_{3,5} + X_{4,6} \geq c_{35}, \\&-X_{3,5} - X_{4,6} \geq - c_{14} - c_{15} - c_{24} - c_{25} - c_{35},\qquad X_{2,4} + X_{3,5} \geq c_{24} 
\end{split}
\end{align}
%\begin{eqnarray}
%X_2 \geq 0 & X_3 \geq 0  \nonumber \\ 
% X_4 \geq 0 & -X_4  \geq c_{15} + c_{25} + c_{35} \nonumber \\
%-X_2 - X_3 - X_4 \geq - c_{13} -  c_{14} - c_{15} - c_{24} - c_{25} - c_{35} & X_2 + X_3 + X_4 \geq c_{24} + c_{25} + c_{35}\nonumber  \\ 
%X_3 + X_4 \geq c_{35} & -X_3 - X_4 \geq - c_{14} - c_{15} - c_{24} - c_{25} - c_{35}\nonumber  \\
%X_2 + X_3 \geq c_{24} & 
%\end{eqnarray}
We can do a simple consistency check on this result. If we set a single $c_{ij} \to 1$ and set all the rest to zero, this polytope must degenerate to the Minkowski summand ${\cal P}_{ij}$. Obviously for this to happen many of the equations must become redundant in the limit.  Let's see how this works when we set $c_{1,4} \to 1$ and the rest to 0. Note the fourth equation forces that $X_{4,6} \to 0$, and the equations for $(X_{2,4}+X_{3,5}+X_{4,6}), (X_{3,5}+X_{4,6}), (X_{2,4}+X_{3,5})$ become trivially satisfied. We are then left only with $0\leq X_{2,4} \leq 1$, $0 \leq X_{3,5} \leq 1$, $X_{2,4} + X_{3,5} \leq 1$, which gives exactly the triangle ${\cal P}_{1,4}$. 
 
\subsection{Tropical functions}\label{ssec:trop}
When we considered the integral \eqref{general_Mink}, we assumed that all $S_J\geq 0$. However, as we have already seen in some examples, it is also possible for $\I_{\{p\}}(S)$ to converge when some $S_J$-s are negative.  Minkowski subtraction is a somewhat subtle operation, so instead we formulate the general condition for convergence and the leading order in terms of tropical functions, for which the result is most elegant.  

Let $\PP = S_1 \PP_1 + \cdots  S_r \PP_r$ be the Minkowski sum of the Newton polytopes, and let $\N( \PP )$ be its normal fan, consisting of the maximal cones \eqref{eq:vertexcone}.  Any subtraction-free rational function $R(\x)$ gives rise to a piecewise-linear function $\Trop(R(\x))$ on $\R^d = (X_1,X_2,\ldots,X_d)$, obtained by formally substituting
\be \label{eq:tropsub}
x_i \mapsto X_i \qquad (+,\times,\div) \mapsto (\min,+,-).
\ee
The tropicalization procedure makes sense even if factors in $R(\x)$ have real exponents. 
For example, for real numbers $s$ and $t$ we have
\be \label{eq:tropex}
\Trop \left( \frac{x^s y^t}{1+x+y} \right) = sX + tY - \min(0,X,Y)
\ee
as a function on $(X,Y)$-space.  Note that for any $R(\x) = \prod_{J=1}^r p_J(\x)^{-S_J}$, the piecewise-linear function $\Trop(R(\x))$ has domains of linearity that is a coarsening of the normal fan $\N(\PP)$.  We call $\Trop(R(\x))$ {\it nonnegative} if it takes nonnegative values everywhere on $\R^d$.  We call $\Trop(R(\x))$ {\it positive} if $\Trop(R(\x)) > 0$ everywhere on $\R^d \setminus \{0\}$.  In \eqref{eq:tropex}, we see that $\Trop(R(\x))$ is nonnegative exactly when $(s,t)$ is inside the (closed) triangle with vertices $(0,0),(1,0),(0,1)$, and it is positive exactly when $(s,t)$ is in the interior of this triangle.  We call an integrand $R(\x)$ {\it convergent} if $\Trop(R(\x))$ is positive and we call $R(\x)$ {\it nearly convergent} if $\Trop(R(\x))$ is nonnegative.

\begin{claim} \label{claim:trop}
The integral $\I_{\{p\}}(S):= (\alpha')^d \int_{\R_{>0}^d} \prod_{i=1}^d \frac{d x_i}{x_i} \prod_{J=1}^r p_J(\x)^{-\alpha'S_J}$ is absolutely convergent exactly when $R(\x) =  \prod_{J=1}^r p_J(\x)^{-S_J}$ is convergent, and in that case the leading order is given by
\begin{equation}\label{eq:TropVol}
\lim_{\alpha' \to 0}  \I_{\{p\}}(S) = \Vol(\{\Trop(R(\x)) \leq 1\}).
\end{equation}
\end{claim}
The proof of Claim \ref{claim:trop} is essentially the same as that of \eqref{eq:main}.

\begin{example}
Let us consider the integrand $R(\x) = \frac{xy}{(1+x)(1+y)(1+x+xy)}$.  The reciprocal $1/R(\x)$ is a Laurent polynomial whose Newton polytope is the pentagon of Figure \ref{fig:pentex}.  The tropical function $\Trop(R(\x))$ is given by
\be
\Trop(R(\x)) = X+Y - \min(0,X) - \min(0,Y) - \min(0,X,X+Y).
\ee
The maximal domains of linearity of this piecewise linear function are the maximal cones of the normal fan in Figure \ref{fig:pentex}.  In Figure \ref{fig:trop}, we show which linear function $\Trop(R(\x))$ is equal to in each maximal cone.  One sees that $\Trop(R(\x))$ is positive, and the region $\{\Trop(R(\x)) \leq 1\}$ is the dual polytope $\PP^\circ$ in Figure~\ref{fig:pentex}.  In particular, the integral $\int_{\R_{>0}^2} \Omega R(\x)$ converges.
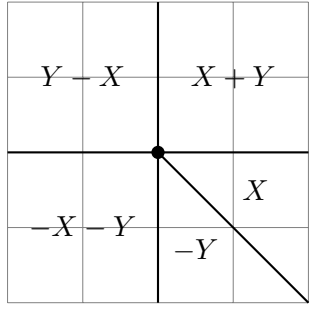
\begin{figure}
\begin{center}
\begin{tikzpicture}
\draw [help lines, step=1cm] (-2,-2) grid (2,2);
\node[fill=black,circle, inner sep=0pt,minimum size=5pt] at (0,0) {};
\draw[thick] (0,0) --(2,0);
\draw[thick] (0,0) --(-2,0);
\draw[thick] (0,0) --(0,-2);
\draw[thick] (0,0) --(0,2);
\draw[thick] (0,0) --(2,-2);
\node at (-1,-1) {$-X-Y$};
\node at (-1,1) {$Y-X$};
\node at (1,1) {$X+Y$};
\node at (1.3,-1/2) {$X$};
\node at (1/2,-1.3) {$-Y$};
%
%\begin{scope}[shift={(10,0)}]
%\draw [help lines, step=1cm] (-2,-2) grid (2,2);
%\node[fill=black,circle, inner sep=0pt,minimum size=5pt] at (0,0) {};
%\draw[thick] (0,1) -- (1,0) -- (1,-1) -- (0,-1) -- (-1,0)-- cycle;
%\node[fill=cyan,circle, inner sep=0pt,minimum size=5pt] at (0,1) {};
%\node[fill=cyan,circle, inner sep=0pt,minimum size=5pt] at (1,0){};
%\node[fill=cyan,circle, inner sep=0pt,minimum size=5pt] at (1,-1){};
%\node[fill=cyan,circle, inner sep=0pt,minimum size=5pt] at (0,-1){};
%\node[fill=cyan,circle, inner sep=0pt,minimum size=5pt] at (-1,0){};
%\node at (0.7,0.7) {$F_4$};
%\node at (1.2,-0.5) {$F_5$};
%\node at (0.5,-1.2) {$F_1$};
%\node at (-0.7,-0.7) {$F_2$};
%\node at (-0.7,0.7) {$F_3$};
%\end{scope}
\end{tikzpicture}
\caption{The domains of linearity of $ X+Y - \min(0,X) - \min(0,Y) - \min(0,X,X+Y)$.}
\label{fig:trop}
\end{center}
\end{figure}
\end{example}

\subsection{Applications to positive parametrizations}
A {\it positive reparametrization} or {\it positive coordinate-change} is an invertible coordinate change $\x = (x_1,\ldots,x_d) \to \x'=(x'_1,\ldots,x'_d)$ satisfying $\prod_{i=1}^d \frac{d x_i}{x_i} = \prod_{i=1}^d \frac{d x'_i}{x'_i}$, and such that each $x'_j$ (respectively, each $x_j$) is a subtraction-free rational expression in terms of the $x_i$ (respectively, the $x'_i$).  We call a positive reparametrization {\it Laurent} if in addition the rational expressions are all Laurent polynomials with positive coefficients.  The typical source of Laurent positive reparametrizations are coordinate changes between seeds of a cluster algebra, which are compositions of mutation transformations.  Note that while the Laurent-ness of cluster coordinate changes was known since the initial discovery of cluster algebras, the positivity of such coordinate changes is a much deeper result~\cite{LeeSchiffler}.

An immediate consequence of \eqref{eq:main} is the following statement: let $p(\x)$ be a subtraction-free Laurent polynomial and $\x \to \x'$ a positive Laurent coordinate-change, so that $p(\x')$ is also a Laurent polynomial.  Then $\New[p(\x)]$ contains the origin in its interior if and only if $\New[p(\x')]$ contains the origin in its interior\footnote{Instead of integrating over $\R_{>0}^d$, we could instead integrate over a compact (complex) torus around the origin.  We deduce that the constant term of $p(\x)$ is equal to the constant term of $p(\x')$, see~\cite[Corollary 2.9]{LamSpeyer}.}.  More generally, let $p(\x)$ be a subtraction-free rational function and $\x \to \x'$ be a positive coordinate change.  Then $\Trop(p(\x))$ is positive if and only if $\Trop(p(\x'))$ is positive.

Let us now consider a collection $\{p_1(\x),p_2(\x),\ldots,p_r(\x)\}$ of subtraction-free Laurent polynomials.  We call the sequence {\it standard}, if each $x_i$ is equal to a monomial $\prod_J p_J(\x)^{-S_J}$ for some choice of $S_J$-s.  Thus the collection $\{p_1(\x),p_2(\x),\ldots,p_r(\x)\}$ is standard if \eqref{general_Mink} can be put in the form \eqref{gen_int}.  Now let $\x \to \x'$ be a positive Laurent coordinate-change, and assume that $\{p_1(\x'),p_2(\x'),\ldots,p_r(\x')\}$ is also standard.  Then we have the following statement concerning the two Minkowski sums of Newton polytopes
\be \label{eq:comb}
\PP=\sum_{J=1}^r \New[p_J(\x)] \text{ and } \PP'= \sum_{J=1}^r \New[p_J(\x')] \text{ are combinatorially isomorphic.}
\ee
To see \eqref{eq:comb}, we first note that $\I_{\{p(\x)\}}(S) = \I_{\{p(\x')\}}(S)$ and by Claim \ref{claim:multi}, the two rational functions $\underline{\Omega}(\PP)$ and $\underline{\Omega}(\PP')$ are the same.  (Here, we consider these as rational functions in all $S$-s, instead of fixing the $c_I$-s and varying the $\X$.)  The combinatorial isomorphism type of $\PP$ is completely reflected in the poles and repeated residue structure of the rational function $\underline{\Omega}(\PP)$, and thus we obtain \eqref{eq:comb}.

\section{From kinematic associahedra to open-string integrals}\label{sec:5}

We have motivated stringy canonical forms as a way to produce the canonical form for a polytope, as well as generalizing the notion of a canonical form endowed with a parameter $\alpha^\prime$ giving it stringy properties. From the discussion in the previous section it is evident that stringy canonical forms should be especially interesting for polytopes that are naturally built as Minkowski sum of a number of simpler pieces. In this section, we will see that this observation applies perfectly to the ABHY construction of the associahedron in kinematic space~\cite{Arkani-Hamed:2017mur}. Recall that the ABHY associahedron was motivated by finding ``the amplituhedron" for the bi-adjoint $\phi^3$ theory, directly in kinematic space.  Quite remarkably, this construction naturally presents the associahedron as a Minkowski sum (see also~\cite{postnikov}), and when these summands are used in the construction of the stringy canonical form, the result is precisely the open-string Koba-Nielsen integral.  Thus, beginning with the desire for a picture of {\it particle scattering} associated with a geometry in kinematic space making no reference to unitary evolution in the bulk of spacetime as an auxiliary construct, we are directly led to it's generalization to {\it string amplitudes}, without referring to the worldsheet as an auxiliary construct! 

Recall that the ABHY construction describes the associahedron ${\cal A}$ by intersecting the positive region in kinematic space $X_{ij} \geq 0$, with a particular subspace 
\begin{equation}
X_{ij} + X_{i+1 j+1} - X_{i j+1} - X_{i+1 j} = c_{ij}
\end{equation}
for $1 \leq i < j-1 <n-1$, with $c_{ij} > 0$ (recall that $X_{i, i{+}1}=0$ for $i=1,2,\cdots, n$). 

Now, this form is already suggestive that the polytope might be expressed as a Minkowski sum $\sum c_{i j} {\cal A}_{i j}$, where each ${\A}_{ij}$ is the degeneration of the polytope where $c_{ij} \to 1$ and all the other $c$'s are set to zero. Indeed consider a much more general problem, where we have variables $Z_I$, and we construct some polytope ${\cal Q}$ by intersecting the positive region $Z_I \geq 0$ with the subspaces $L_i(Z) = c_i$ where each $L_i$ is linear in the $Z_I$, with $c_i >0$.  Let ${\cal Q}_I$ be the degeneration of the polytope where $c_i \to 1$ with all other $c$'s set to zero. Clearly, the Minkowski sum $\tilde{{\cal Q}} \equiv \sum c_i {\cal}Q_i$ lies inside $\Q$, since $Z_I$ are positive in $\tilde {\cal Q}$, and for all $Z$ in $\tilde{{\cal Q}}$, $L_i(Z) = c_i$. But in general, $\tilde{\cal Q}$ will not cover all of ${\cal Q}$. 

This is associated with another observation. In general, the shape of the polytope ${\cal Q}$ -- it's facet structure -- will not be independent of the specific choice of the positive constants $c_i$. Consider a simple example where we cut out a two-dimensional space by $X,Y,Z,W\geq 0$, with the equations $X + Y + Z = c_1$ and $2X + Y + W = c_2$. In the $(X,Y)$ plane, the region is cut out by $X,Y>0$ and $X+Y < c_1, 2X + Y < c_2$. But the shape of this region is not the same for all positive $c_1,c_2$: if $c_2 > 2 c_1$, the second inequality is immediately implied by the first and we get a triangle, while for $c_2<2c_1$ we get a quadrilateral. 

On the other hand, the shape of a Minkowski sum of polytopes $\sum c_i {\cal P}_i$ for positive $c_i$ is manifestly independent of the $c_i$. Indeed in this example we can see that if we set either $c_1 \to 0$ or $c_2 \to 0$, the resulting polytopes are just the point at the origin, and so adding these summands just gives us a point and not the full polygon.

But the ABHY associahedron ${\cal A}_{n{-}3}$ is much more special. Its shape is completely independent of the $c_{ij}$, so long as they are positive, and it {\it is} given as a Minkowski sum
\begin{equation}
{\cal A}_{n{-}3}= \sum_{1 \leq i < j-1 <n-1} c_{i\,j}~{\cal A}_{ij}
\end{equation}
Both facts can be deduced inductively, as following from the remarkable, defining feature of the associahedron, that on its boundaries it factors into the product of lower associahedra. 
\begin{figure}
\begin{center}
\begin{tikzpicture}[baseline={([yshift=0ex]current bounding box.center)}]
%\begin{tikzpicture}[scale=0.75,baseline={([yshift=-3.5ex]current bounding box.center)}]
%\draw (0,0) -- (3,0) -- (3,4) -- (2,4) -- (1,3) -- (0,1) -- cycle ;
%\node at (-0.2,0.5) {\scriptsize $c_2$};
%\node at (1.5,-0.3) {\scriptsize $c_1+c_3+c_4$};
%\node[rotate=-90] at (3.29,2) {\scriptsize $c_2+c_3+2 c_4$};
%\node at (2.5,4.2) {\scriptsize $c_1$};
%% \draw[densely dotted] (0,1) -- (1,1) -- (1,3);
%% \draw[densely dotted] (1,3) -- (2,3) -- (2,4);
%%\node at (1.3,1.9) {\scriptsize $2d$};
%% \node at (2.2,3.5) {\scriptsize $c$};
%% \node at (1.5,2.8) {\scriptsize $c$};
%% \node at (0.5,0.8) {\scriptsize $d$};
%
%\draw[->] (-0.1,0) -- (3.5,0);
%\draw[->] (0,-0.1) -- (0,4.5);
%\node at (3.6,-0.3) {\scriptsize $X$};
%\node at (-0.3,4.5) {\scriptsize $Y$};
%
%\draw[gray,->] plot[smooth, tension=.7] coordinates {(0.5,2.1) (-0.1,2.4) (-0.4,3)};
%\node at (-0.3,3.2) {\scriptsize $Y{-}c_2=2X$};
%\draw[gray,->] plot[smooth, tension=.7] coordinates {(1.5,3.6) (1.4,4.2) (1,4.9)};
%\node at (1.2,5.1) {\scriptsize $Y{-}(c_2{+} c_4)=X$};
%\end{tikzpicture}
%\end{center}
\draw [thick](-3.5,1) -- (-1.5,1);
\draw (-2.5,1) node[below]{${\cal A}_{13}$};
\draw [thick](-4,3.5) -- (-4,1.5);
\draw (-2,2.5) node[below]{${\cal A}_{14}$};
\draw [thick](-3.5,1.5) -- (-1.5,3.5) -- (-1.5,1.5) -- cycle;
\draw (-4.5,2.5) node[below]{${\cal A}_{24}$};
\end{tikzpicture}
$\quad \xrightarrow{\text{ Minkowski~sum}%c_{13}P_{13} + c_{24}P_{24}+c_{14} P_{14}}
}$
\begin{tikzpicture}[baseline={([yshift=-4ex]current bounding box.center)}]
\draw [thick](0.5,0.5) -- (1.5,1.5) -- (3,1.5) -- (3,-1) -- (0.5,-1) -- cycle;
\draw [->](0.3,-1) -- (3.5,-1) node[right]{$X_{25}$};
\draw [->](0.5,-1.2) -- (0.5,2) node[above]{$X_{35}$};
\node at (0.2,-0.3) {\scriptsize $c_{2,4}$};
\node at (1.7,-1.2) {\scriptsize $c_{1,3}+c_{1,4}$};
\node[rotate=-90] at (3.2,0.3) {\scriptsize $c_{1,4}+c_{2,4}$};
\node at (2.2, 1.65) {\scriptsize $c_{1,3}$};
\end{tikzpicture}
\end{center}
\caption{The ABHY associahedron ${\cal A}_2$ (pentagon) as a Minkowski sum.}
\label{fig:pentagon}
\end{figure}
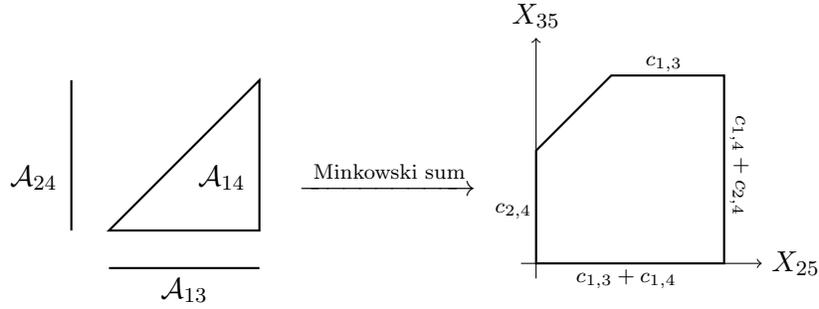

Let's determine what the summands ${\cal A}_{ij}$ look like for the case of $n=5$, and proceed from there to determining the stringy canonical form in this case. We work with the coordinates $X_{25},X_{35}$ and solve for the remaining $X$'s in terms of these. The pentagon we obtain in this way is cut out by $X_{25} \geq 0, X_{35} \geq 0$, together with $X_{14} = c_{14} + c_{24} - X_{35} \geq 0, X_{13} = c_{13} + c_{14} - X_{25} \geq 0, X_{24} = c_{24} + X_{25} - X_{35} \geq 0$. Let's determine
 the summand ${\cal A}_{13}$ when $c_{13} \to 1$ with the rest set to zero. The $X_{14}$ inequality forces $X_{35} \to 0$, and we are left just with an interval, $0 \leq X_{25} \leq 1$. Similarly, ${\cal A}_{24}$ is the interval $0 \leq X_{35} \leq 1$. Finally, 
${\A}_{14}$ is the triangle bounded by $0 \leq X_{35} \leq X_{25}<1$. The pentagon and the three summands are shown in Figure \ref{fig:pentagon}.  The special polynomials associated with these Newton polytopes are 
\begin{equation}
p_{13} = 1 + y_2, \qquad p_{24} = 1 + y_3, \qquad p_{14} = 1 + y_2 + y_2 y_3\,.
\end{equation}
%where we introduce $(x_2,x_3)$ to cover $(X_{25},X_{35})$ space. 

We can now immediately write the stringy canonical form integral associated with this polytope as 
\begin{equation}
{\bf I}_5 = \int \frac{dy_2}{y_2} \frac{dy_3}{y_3} y_2^{\alpha^\prime X_{25}} y_3^{\alpha^\prime X_{35}} (1 + y_2)^{-\alpha^\prime c_{13}} (1 + y_2 + y_2 y_3)^{-\alpha^\prime c_{14}} (1 + y_3)^{-\alpha^\prime c_{24}}
 \end{equation}
On the hand other hand, the Koba-Nielsen integral \eqref{eq:string} for $n=5$ is given by \eqref{5ptstring}, % (using $s_{ii+1} = X_{ii+2}$, and $s_{ij} = - c_{i j}$)
%\begin{equation} \int \frac{dx_2}{x_2} \frac{dx_3}{x_3} x_2^{\alpha^\prime X_{24}} x_3^{\alpha^\prime X_{35}} (1 + x_2)^{-\alpha^\prime c_{13}} (1 + x_2 + x_3)^{-\alpha^\prime c_{14}} (x_2 + x_3)^{-\alpha^\prime c_{24}} \end{equation}
and we can easily see the equality between these forms upon making the simple change of variable $x_2 = y_2, x_3 = y_2 y_3$! Note that beginning from the Koba-Nielsen form, after the change of variable to $(y_2,y_3)$ the power of $y_2$ becomes $X_{24} + X_{35} - c_{24} = X_{25}$, matching what we get from the $y$ integral. 

This result generalizes to all $n$ in the obvious way. The Minkowski summands ${\cal A}_{ij}$ are just simplices, and, defining the variables $(y_2, \ldots, y_{n-2})$ to cover $(X_{2,n},\ldots, X_{n-2,n})$ space, the Newton polynomials for ${\cal A}_{i j}$ are just~\footnote{We remark that the fact that the ABHY associahedron is given by the Minkowski sum of these simplices is a special case of the construction of generalized permutohedra in~\cite{postnikov2005}.} 
\begin{equation}
p_{ij} = 1 + y_{i+1} + y_{i+1} y_{i+2} + \cdots  + y_{i+1} \cdots y_{j-1}
\end{equation}
The stringy canonical form integral associated with this Minkowski sum 
\begin{equation}\label{eq:KNy}
{\bf I}_n ={\cal I}_{{\cal A}_{n{-}3}}= (\alpha')^{n-3}~\int \prod_{i=2}^{n-2} \frac{dy_i}{y_i} y_i^{\alpha^\prime X_{i n}} \prod_{ij} p_{ij}(\y)^{-\alpha^\prime c_{ij}}
\end{equation}
exactly matches the Koba-Nielsen $n$-point integral, as manifested by the variable change $x_2 = y_2, x_3 = y_2 y_3, \ldots, x_j = y_2 \cdots y_j$. 

A remarkable property of the integral $\bf{I}_n$ is that it factorizes nicely at massless poles, or the $n(n{-}3)/2$ planar poles, $X_{ij}=0$,  in a much more special way than for a general stringy canonical form. On any massless pole, the residue is the product of two open-string integrals for the left and right associahedra (which correspond to two polygons divided by $(i,j)$):
\be
{\rm Res}_{X_{i j}=0} \I_{{\cal A}_{n{-}3}} =\I_{{\cal A}_L} \times \I_{{\cal A}_R}\,.
\ee
Note that if we had chosen some other polynomials that give ${\A}_{n{-}3}$, {\it e.g.} polynomials different from $p_{ij}$ but whose Newton polytopes still give ${\A}_{ij}$, although it produces the correct leading order, we would not have these perfect factorizations at massless poles. We emphasize that the $n$-point open-string integral, $\I_{{\cal A}_{n{-}3}} $, is a very special stringy canonical form for the $(n{-}3)$-dim ABHY associahedron. Any stringy canonical form has the correct $\alpha' \to 0$ limit, which is the canonical form of ABHY associahedron (or bi-adjoint $\phi^3$ amplitude), and the residue at any massless pole is still given by the stringy canonical form of that facet. However, unlike in the special case of $\I_{{\cal A}_{n{-}3}} $, in general the residues do not factorize into two lower-point integrals of the same form, and such ``perfect" factorization is a special feature of string integrals. 
This will become more transparent in Section \ref{sec:big} when we talk about dual variables for stringy canonical forms: the dual variables of string integrals satisfy beautiful equations which gives what we call a ``binary geometry", and the dual variables make the this ``perfect" factorization property of string integrals completely manifest.

\section{Further examples}\label{sec:examples}

We have seen that there are at least two interpretations of the open-string integrals \eqref{eq:string}: (a) stringy canonical forms for the ABHY associahedron, and (b) natural regulated integrals over $\G_+(2,n)/T$. In this section, we generalize open-string integrals along these two directions. We define and study (a) cluster string integrals which are stringy canonical forms of generalized associahedra, and (b) and Grassmannian string integrals which are natural regulated integrals over $\G_+(k, n)/T$. 
%which are stringy canonical forms for associated polytopes. 

\subsection{Cluster string integrals}\label{sec:cluster}

We define a special class of stringy canonical forms for the generalized associahedron of a finite type cluster algebra, called {\bf cluster string integrals}. %We define it by choosing any positive parametrization such that all the cluster ${\cal A}$ coordinates are (subtraction-free) polynomials, which regulate the integral. In~\cite{} it has been proposed that the Minkowski sum of Newton polytopes for these polynomials gives the corresponding generalized associahedron in an ABHY realization. 
Let $\Phi$ be a crystallographic root system of rank $d$, and thus belonging to one of the infinite families $A_d, B_d,  C_d, D_d$ or equal to one of the exceptional types $E_6,E_7,E_8,F_4,G_2$. %{\cal E}_7,{\cal E}_8,{\cal F}_4,{\cal G}_2$.  
Let $\A(\Phi)$ denote a cluster algebra of type $\Phi$, with a full rank choice of coefficients\footnote{We work with skew-symmetrizable cluster algebras with geometric coefficients.  The coefficients are chosen so that the {\it extended exchange matrix} has full rank.  For example, $\A(\Phi)$ can be chosen to have {\it principal} coefficients.}.
%In many cases, such as $\Phi=A_{2n}$, this holds with no coefficients.}.  
Let $x_\gamma$, $\gamma \in\Gamma$ denote the cluster variables of $\A(\Phi)$ and let $\A(\Phi)_+$ denote the positive part of the cluster variety.  The cluster string integral is defined to be
\be
\I_\Phi =(\alpha')^d~\int_{\A(\Phi)_+/T} (\omega/T)~\prod_{\gamma \in \Gamma}x_\gamma^{\alpha' c_\gamma},
\ee
where $T$ is the (positive part of the) torus of cluster automorphisms of $\A(\Phi)$, and $\omega$ is the natural top-form on a cluster algebra, defined to be $\omega = \prod_i dx_i/x_i$ in any cluster $(x_1,\ldots,x_r)$.  Equivalently, $\omega/T$ is a dlog-form for a positive parametrization of $\A(\Phi)_+/T$.

Let $\PP(\Phi)$ denote the corresponding {\it generalized associahedron}, which is a $d$-dimensional {\it simple} polytope~\cite{fomin2003systems, chapoton2002polytopal}.  The facets $F_\gamma$ of the generalized associahedron $\PP(\Phi)$ are in bijection with cluster variables $x_\gamma$ of $\A(\Phi)$.  Each vertex of $\PP(\Phi)$ corresponds to a seed with $d$ cluster variables (the $d$ facets adjacent to it).  We let $N(\Phi) = |\Gamma|$ denote the number of facets.  The formula for $N(\Phi)$ for all finite types reads:
\ba
&N(A_d)=d(d{+}3)/2\,,\quad N(B_d)=N(C_d)=d(d{+}1)\,,\quad N(D_d)=d^2\,,\nonumber\\
&N(E_6)=42\,,\quad N(E_7)=70\,,\quad N(E_8)=128\,,\quad N(F_4)=28\,, \quad N(G_2)=8\,.
\ea
%We set $m:= N-d$.
%\ba
%&N({\cal A}_d)=d(d{+}3)/2\,,\quad N({\cal B}_d)=N({\cal C}_d)=d(d{+}1)\,,\quad N({\cal D}_d)=d^2\,,\nonumber\\
%&N({\cal E}_6)=42\,,\quad N({\cal E}_7)=70\,,\quad N({\cal E}_8)=128\,,\quad N({\cal F}_4)=28\,, \quad N({\cal G}_2)=8\,.
%\ea
%For all these polytopes, we have obtained ABHY realizations~\cite{} and in particular the canonical function for type ${\cal B}/{\cal C}$ and that for type ${\cal D}$ can be interpreted as sum of tadpole diagrams and one-loop planar $\phi^3$ amplitude, respectively. 
As will be explained in \cite{20201}, there are many positive parametrizations of the cluster string integral.  For concreteness, let us formulate the integral by choosing $\A(\Phi)$ to be the principal coefficient cluster algebra, with principal coefficients $\y = (y_1, \ldots, y_d)$ giving a positive parametrization of $\A(\Phi)_+/T$.  After picking an initial seed, for each cluster variable $\gamma$, we have an {\it $F$-polynomial} $\F_\gamma(\y)$, which is a polynomial in $\y$ with positive integer coefficients.  The $F$-polynomials for the initial cluster variables are trivial, and the remaining ones we denote $\F_I(\y)$, for $I = d+1,\ldots,N = N(\Phi)$.  For the open-string integrals, {\it i.e.} the $\Phi = A_{n-3}$ case, the $F$-polynomials for a suitable initial cluster are nothing but the $p_{ij}(\y)$'s in \eqref{eq:KNy}. 
%
%After choosing such a parametrization, one can express the $N$ cluster variables as subtraction-free polynomials known as {\it $F$-polynomials}; $d$ of them are trivial, ${\cal F}_i=1$ for $i=1,2, \cdots, d$, and the remaining $N-d$ are non-trivial polynomials (always start with $1$), ${\cal F}_I (\{x\})$ for $I=d{+}1, \cdots, N$. 
The cluster string integral can thus be written as
\be\label{cluster}
\I_\Phi(\X, \{c\})=(\alpha')^d~\int_{\mathbb{R}_{>0}^d} \prod_{i=1}^d \frac{dy_i}{y_i}~y_i^{\alpha' X_i}~\prod_{I=d{+}1}^N \F_I(\y)^{-\alpha' c_I}\,,
\ee
where we have denoted exponents of $y$'s as $X$'s and those for $F$-polynomials as $c$'s.  As we will show in \cite{20201}, the Minkowski sum $ \sum_I c_I \New[\F_I(\y)]$ is the generalized associahedron $\PP(\Phi^\vee,c)$ of the dual Dynkin diagram $\Phi^\vee$.  For example, if $\Phi = B_n$ then $\Phi^\vee= C_n$.  Thus the leading order of \eqref{cluster} is the canonical function of the generalized associahedron:
\be
\lim_{\alpha' \to 0} \I_{\Phi} (\X, \{c\})=\underline{\Omega}( \PP(\Phi^\vee,c); \X).
\ee

Note that similar to ABHY associahedron (which is for type $A$), here we obtain the very special realizations of generalized associahedra, which was first proposed in~\cite{bazier2018abhy} and will be discussed more in~\cite{20191} where the relevance for scattering amplitudes {\it etc.} is discovered.  For type $B$ and $C$, the canonical function gives the sum of all one-loop tadpole diagrams, while for type $D$, it gives the integrand of one-loop bi-adjoint $\phi^3$ amplitude~\cite{20191}. Similar to the open-string integral of Section \ref{sec:5}, the cluster string integral satisfies a number of remarkable properties not satisfied by a general stringy canonical form for the same general associahedron $\PP(\Phi)$.  Some of these properties depend crucially on the exact coefficients of $\F_\gamma(\y)$, and not just their Newton polytopes.  %We will discuss cluster string integrals further in Section \ref{sec:cluster2}.
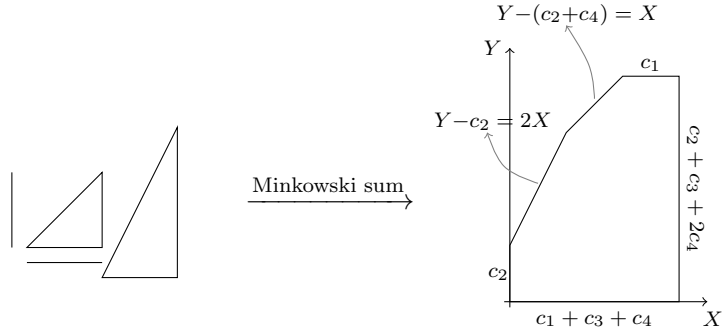
\begin{figure}
\begin{center}
\begin{tikzpicture}[baseline={([yshift=-.5ex]current bounding box.center)}]
\draw (-1.8,0) -- (-1.8,1);
\draw (-1.6,-0.2) -- (-0.6,-0.2);
\draw (-1.6,0) -- (-0.6,0) -- (-0.6,1) -- cycle;
\draw (-0.6,-0.4) -- (0.4,-0.4) -- (0.4,1.6) -- cycle;
\end{tikzpicture}
\quad \quad $\xrightarrow{\text{Minkowski sum}}$
\begin{tikzpicture}[scale=0.75,baseline={([yshift=-3.5ex]current bounding box.center)}]
\draw (0,0) -- (3,0) -- (3,4) -- (2,4) -- (1,3) -- (0,1) -- cycle ;
\node at (-0.2,0.5) {\scriptsize $c_2$};
\node at (1.5,-0.3) {\scriptsize $c_1+c_3+c_4$};
\node[rotate=-90] at (3.29,2) {\scriptsize $c_2+c_3+2 c_4$};
\node at (2.5,4.2) {\scriptsize $c_1$};
% \draw[densely dotted] (0,1) -- (1,1) -- (1,3);
% \draw[densely dotted] (1,3) -- (2,3) -- (2,4);
%\node at (1.3,1.9) {\scriptsize $2d$};
% \node at (2.2,3.5) {\scriptsize $c$};
% \node at (1.5,2.8) {\scriptsize $c$};
% \node at (0.5,0.8) {\scriptsize $d$};

\draw[->] (-0.1,0) -- (3.5,0);
\draw[->] (0,-0.1) -- (0,4.5);
\node at (3.6,-0.3) {\scriptsize $X$};
\node at (-0.3,4.5) {\scriptsize $Y$};

\draw[gray,->] plot[smooth, tension=.7] coordinates {(0.5,2.1) (-0.1,2.4) (-0.4,3)};
\node at (-0.3,3.2) {\scriptsize $Y{-}c_2=2X$};
\draw[gray,->] plot[smooth, tension=.7] coordinates {(1.5,3.6) (1.4,4.2) (1,4.9)};
\node at (1.2,5.1) {\scriptsize $Y{-}(c_2{+} c_4)=X$};
\end{tikzpicture}
\caption{The generalized associahedron (or cyclohedron) for $\A(B_2)$ as a Minkowski sum.}
\label{fig:hexagon}
\end{center}
\end{figure}

Let's present the simplest example beyond type $A$, namely the stringy canonical form for the hexagon $\PP(B_2) = \PP(C_2)$.  We have $N=6$ and thus $m=4$ non-trivial ${\F}_I$ polynomials, and denoting the principal coefficients by $x,y$, the cluster string integral $\I_{B_2=C_2}$ is given by
\be\label{eq:IB2}
%\I_{B_2=C_2} = 
(\alpha')^2 \int_{\mathbb{R}_{>0}^2} \frac{d x~d y}{x~y} x^{\alpha' X} y^{\alpha' Y} (1{+}x)^{-\alpha' c_1} (1{+}y)^{-\alpha' c_2} (1{+}x{+}x y)^{-\alpha' c_3} (1{+}x{+}2 x y{+}x y^2)^{-\alpha' c_4}\,.
\ee
As shown in Figure~\ref{fig:hexagon}, the generalized associahedron $\PP(B_2)$ is thus given by the Minkowski sum of two line intervals and two triangles, $\sum_I c_I \New[ {\cal F}_I]$.  Remarkably, just as for the type $A$ case, we see that for finite $\alpha'$, the residue at any edge of the hexagon gives the $A_1$ integral, {\it i.e.} the beta function, which would not be true if we change the coefficients of these polynomials. It is remarkable that exactly as string amplitudes, the cluster string integral factorizes perfectly at ``massless" poles corresponding to facets of the generalized associahedron. As will be elaborated in~\cite{20201}, these integrals can be viewed as generalized open-string amplitudes whose factorizations correspond to removing a node of the corresponding Dynkin diagram. For example, for finite $\alpha'$ and $\Phi=A_n$ the residue of $\I_{A_n}$ at a massless pole is equal to $\I_{A_r} \times \I_{A_{n-r-1}}$ for some $r$.  Similarly, for finite $\alpha'$ and $\Phi=B_n$ or $\Phi = C_n$, the residue of the cluster string integral at any facet of the {\it cyclohedron} (the generalized associahedron $\PP(B_n)$) is given by (corresponding to removing a node of type $B_n$ or $C_n$ Dynkin diagram)
\be
\I_{{ B}_n} \to \I_{{\cal A}_m} \times \I_{{B}_{n{-}m{-}1}}\,, \quad \I_{{C}_n} \to \I_{{A}_m} \times \I_{{C}_{n{-}m{-}1}}\,;
\ee
%\be
%\I_{{\cal B}_n} \to \I_{{\cal A}_m} \times \I_{{\cal B}_{n{-}m{-}1}}\,, \quad \I_{{\cal C}_n} \to \I_{{\cal A}_m} \times \I_{{\cal C}_{n{-}m{-}1}}\,,
%\ee
the residues for a type $D_n$ integral on a facet of $\PP(D_n)$ can be of the following types:
\be
\I_{{D}_n} \to \I_{{A}_m} \times \I_{{ D}_{n{-}m{-}1}}\,, \quad \I_{{A}_{n{-}3}} \times \I_{{A}_1} \times \I_{{A}_1}\,, \quad \I_{{A}_{n{-}1}}\,,
\ee 
which correspond to removing a node of the type $D_n$ Dynkin diagram. 

Finally, let's list the $f$-vectors of generalized associahedra for the three cases $\Phi=D_4,E_6,E_8$ that are closely related to the Grassmannian string integrals that we study shortly. 
%These are computed from taking the Minkowski sum of Newton polytopes of $m$ $F$-polynomials, which give ABHY realizations of generalized associahedra. 
We have $m=12$ for ${D}_4$ and the $f$-vector reads $(1, 50, 100, 66, 16, 1)$; for ${E}_6$ we have $m=36$ and the $f$-vector is $(1, 833, 2499, 2856, 1547, 399, 42, 1)$; finally for $E_8$, we have $m=120$ and the $f$-vector reads $(1, 25080, 100320, 163856, 1440488, 67488, 17936, 2408, 128,1)$. 

%For the type ${\cal D}_4$ integral, there are $16-4=12$ non-trivial ${\cal F}_I$ polynomials of  and we list them here for completeness. $x_1+1,x_2+1,x_1 \left(x_2+1\right)+1,x_3+1,\left(x_2+1\right) x_3+1,x_3+x_1 \left(\left(x_2+1\right) x_3+1\right)+1,x_4+1,\left(x_2+1\right) x_4+1,x_4+x_1 \left(\left(x_2+1\right) x_4+1\right)+1,x_4+x_3 \left(\left(x_2+1\right) x_4+1\right)+1,\left(x_3+1\right) \left(x_4+1\right)+x_1 \left(x_4+x_3 \left(\left(x_2+1\right) x_4+1\right)+1\right),x_4+x_3 \left(\left(x_2+1\right) x_4+1\right)+x_1 \left(\left(x_2+1\right) x_3+1\right) \left(\left(x_2+1\right) x_4+1\right)+1$
%{\color{blue} Examples for ${\cal B}_2={\cal C}_2$, ${\cal D}_4$, {\it etc.}? factorization at finite $\alpha'$ and $u$ variables?}

\subsection{Grassmannian string integrals} 
The positive component $\M_{0,n}^+$ of the moduli space of $n$-points on a Riemann sphere is also isomorphic to the quotient space ${\rm G}_+(2,n)/T$ of the positive Grassmannian $\G_+(2,n)$ by the positive torus $T$.
More generally, it is natural to define integrals over ${\rm G}_+(k,n)/T$, which has dimension $d:=(k{-}1)(n{-}k{-}1)$. These integrals are studied in more detail in \cite{ALS}.  They are closely related to cluster string integrals and for $k=4$, to possible non-perturbative geometries for ${\cal N}=4$ SYM amplitudes. The canonical form of ${\rm G}_+(k,n)/T$ is given by~\cite{ArkaniHamed:2009dn, ArkaniHamed:2012nw}
\be
\omega_{k,n}=\Omega({\rm G}_+(k,n)/T)= \frac{d^{k \times n} C}{{\rm vol} SL(k) \times GL(1)^n} \frac{1}{(12 \cdots k) \cdots (n 1 \cdots k{-}1)}\,.
\ee
The open-string integral corresponds to $k=2$, which has been regulated by all the $n \choose 2$ minors.  For the {\bf Grassmannian string integral}, we regulate the integral with all $n \choose k$ minors, $(a_1, a_2, \ldots, a_k)$ for $a_1 < a_2 < \cdots< a_k$, with exponents denoted as $s_{a_1, a_2, \ldots, a_k}$ 
\be
R_{k,n}%({\rm G}_+(k,n)/T)
:=\prod_{a_1, a_2, \ldots, a_k} (a_1, a_2, \ldots, a_k)^{\alpha' s_{a_1, a_2, \ldots, a_k}}\,.
\ee 
Under the torus action, there are $n$ linear constraints on the exponents similar to the ``momentum conservation" equations (in our notation, $s_{a_1,\ldots, a_k}$ is symmetric with respect to the $k$ indices):
\be
\sum_{a_2, \ldots, a_k \neq a_1} s_{a_1, a_2, \ldots, a_k}=0\,,\quad {\rm for}~a_1=1, \ldots, n\,,
\ee
and thus only $D:={n \choose k}-n$ of the exponents are independent. 
%$U$-space integral for a finite-type ${\cal T}$ of rank $d$, with $0<x_i<\infty$ a {\it positive parametrization} of $U^+({\cal T})$ and $F_I$ the $F$-polynomials for $I=1,2, \cdots, N-d$ ($N$ is the number of clusters):
%\be
%\I_{\cal T} (\X, \{c\})=\int_{U^+({\cal T})}~\Omega(U^+({\cal T}))~\prod_\alpha u_a^{\alpha' Y_a}
%=\int_{U^+({\cal T})}~\prod_{i=1}^d \frac{d x_i}{x_i}~x_i^{\alpha' X_i}~\prod_{I=1}^{N{-}d} F_I (\x)^{-\alpha' c_I}\,,
%\ee
%{\color{blue} Natural $u$ equations, leading order $\to$ ABHY,  ``perfect" factorizations at finite $\alpha'$. $\cdots$}
Explicitly by choosing a positive parametrization of ${\rm G}_+(k,n)/T$ (see~\cite{ALS}), we have
\be
\I_{k,n} (\X, \{c\}):=(\alpha')^d \int_{{\rm G}_+(k,n)/T} \omega_{k,n} R_{k,n}=\int_{\mathbb{R}_{>0}^d} \prod_{i=1}^d \frac{dx_i}{x_i}~x_i^{\alpha' X_i}~\prod_I^{D-d} \F_I(\x)^{-\alpha' c_I}%\Omega({\rm G}_+(k,n)/T)~M({\rm G}_+(k,n)/T)=
\ee
where we have used the fact that in $R_{k,n}$ we have $m=D{-}d$ minors which contain non-trivial polynomials ${\cal F}_I$, and we denote their exponent as $-c_I=s_I$, while the remaining $d$ minors are monomials of $x$'s, and we collect the product as $\prod_{i=1}^d x_i^{\alpha' X_i}$ with $X$ being the sum of certain $s$ variables. For example, in the parametrization of ${\rm G}_+(2,n)/T$ coming from \eqref{eq:string}, the non-trivial minors are $(a b)$ for $2<a<n<n{-}1$, and we have $n{-}3$ planar variables $X_i$. 

In the $\alpha'\to 0$ limit, the leading order of $\I_{k,n}$ gives the canonical function of a polytope, which we denote as $\PP(k,n):=\sum_I c_I \New[{\F}_I]$. Beyond the $k=2$ case $\PP(k,n)$ is not a simple polytope ({\it i.e.} there are vertices that do not belong to exactly $d$ facets).  Let's denote the facets as $F_a$ for $a=1,2,\ldots, N$ and for $k>2$ we have $N>d{+}m=D$. In addition to the facets with $F_i=X_i$ for $i=1, 2, \ldots, d$, the remaining $F$'s are linear combinations of $X$'s and $c$'s (equivalently each of them is a sum of certain $s$ variables). The computation of the Minkowski sum is straightforward but tedious as $k$ and $n$ increases.  The normal fan $\N(\PP(k,n))$ is combinatorially isomorphic to the tropical positive Grassmannian, which was originally studied in~\cite{SW} and revisited more recently in~\cite{ALS, Drummond:2019qjk}, see also~\cite{Cachazo:2019ngv}. 

It is particularly interesting to compute $\PP(k,n)$ for $k=3$, $n=6,7,8$ since they are known to be related to the generalized associahedra $\PP(D_4)$, $\PP(E_6)$, and $\PP(E_8)$ respectively.  In fact, both the polytopes and the Grassmannian string integrals are degenerations of those of the cluster case. The cluster variables for $D_4$, $E_6$ and $E_8$ can be written in terms of the minors of G${}_+(3,6)$, G${}_+(3,7)$ and G${}_+(3,8)$. In particular, for $D_4$ the $16$ mutable cluster variables can be expressed in terms of $D=14$ (non-cyclic) minors, together with $2$ additional ``cross-products" of the form $(1\times 2, 3\times 4, 5\times 6)$, and $(2\times 3, 4\times 5, 6\times 1)$\footnote{Recall the definition $( 1\times 2, 3\times 4, 5\times 6 )=(134)(256)-(234)(156)$.}. There is a very simple motivation for adding these extra factors. The positive Grassmannian $\G_+(k,n)$ has a remarkable symmetry under the ``twist map" related to the parity duality of amplitudes~\cite{GalashinLam}: given a positive matrix $[C_1, \ldots, C_n]$, we get a new positive matrix $[\tilde C_1, \ldots, \tilde C_n]$ where $\tilde{C}_i$ is the dual vector $\tilde C_i = (C_{i+1} \cdots C_{i+k})$ made from wedging the $k$ columns to the right of $C_i$.  It is natural to define a stringy integral to preserve this symmetry.  For $\G_+(3,6)$, the twist of most of the minors simply give other minors (up to cyclic minors). There are just two exceptions: the minors $(135)$ and $(246)$ are mapped to  $( 1 \times 2, 3 \times 4, 5 \times 6 )$ and $(2 \times 3, 4 \times 5, 6 \times 1)$ respectively. Thus it is natural to add these factors to the stringy canonical form integral; this turns out to yield the cluster string integral for $D_4$. 

If we denote the exponents of the two cross-products as $c$ and $c'$, then to go from the $D_4$ integral to that for $\G_+(3,6)/T$, we simply set $c=c'=0$, 
\be
{\I}_{3,6}={\I}_{D_4} (c=c'=0)\,.
\ee  
We will defer a more detailed discussion of the polytopes for $D_4$ and  $\G_+(3,6)/T$ to Section \ref{sec:big}, where we can use a more powerful and insightful language to describe it.

Similarly, the $42$ factors for $E_6$ include $D=28$ independent (unfrozen) minors and $14$ additional ones: $7$ of them are the cyclic rotations of $( 1\times 2, 3\times 4, 5\times 6)$ and the other $7$ are cyclic rotations of $(2\times 3, 4\times 5, 6\times 1)$.  Let's denote their (negative) exponents to be $c_i$ and $c_i'$ for $i=1,\ldots, 7$, respectively. The relation between the $E_6$ and $\G_+(3,7)/T$ integrals is simply 
\be
{\I}_{3,7}={\I}_{E_6} (c_1=\cdots=c_7=c_1'=\cdots=c_7'=0)\,.
\ee 
The $128$ factors for $E_8$ can be written in terms of $D=56$ independent minors and $72$ additional ones which we do not write explicitly here. Again we have that the Grassmannian case is given by the cluster case by setting $72$ of the exponents to zero.

Therefore, the Minkowski sum for the Grassmannian string integral is given by that of the cluster string integral with Minkowski summands removed, and thus $\PP(3,6)$, $\PP(3,7)$, and $\PP(3,8)$ are degenerations of the generalized associahedron for $D_4$, $ E_6$ and $E_8$, respectively.  Further degenerations are studied in~\cite{ALS}.  These polytopes are no longer simple, and the facet structure becomes very different.  Let's record the $f$-vectors for these three cases. The $f$-vector for $\PP(3,6)$ is $(1, 48, 98, 66, 16, 1)$, that for $\PP(3,7)$ is $(1, 693, 2163, 2583, 1463, 392, 42, 1)$, and for $\PP(3,8)$ it is $(1, 13612, 57768, 100852, 93104, 45844, 14088, 2072, 120, 1)$. 

\section{Scattering equations and pushforward }\label{sec:saddle}

In this section we move to the study of saddle-point equations for stringy canonical forms. The saddle-point equations for the string amplitudes $\bI_n$ \eqref{eq:In}, namely the vanishing of $d \log$ of the Koba-Nielsen factor, give the CHY scattering equations 
\be\label{CHYSE}
\frac{\partial}{\partial z_a}\log \left(\prod_{a<b} z_{a,b}^{\alpha' s_{a,b}}\right)=\alpha' \sum_{b\neq a} \frac{s_{a,b}}{z_a-z_b}=0\,,\quad \text{ for all } a\,. 
\ee
Equivalent to CHY formulas for bi-adjoint $\phi^3$ amplitudes, it has been shown in~\cite{Arkani-Hamed:2017mur} that by the sum of the canonical form of $\M_{0,n}^+$, over $(n{-}3)!$ solutions, gives canonical form of the ABHY associahedron. In this section, we show that it is a completely general phenomenon (for which the string amplitude is a special example) that the limit as $\alpha'\to 0$ of any stringy canonical form can be obtained as a pushforward using saddle point equations. 

The idea that the canonical form of a polytope can be obtained as a pushforward was proposed in~\cite{Arkani-Hamed:2017tmz}, where it was called the Newton-polytope map (see Appendix~\ref{sec:canform}).  It is based on the following general claim (sec. 7.3.3 of~\cite{Arkani-Hamed:2017tmz}): suppose one is given a map $\phi$ which restricts to a diffeomorphism from the interior of a positive geometry ${\cal A}(X)$ to another positive geometry ${\cal B}(Y)$, then $\Omega({\cal B}; Y)$ is given by the {\it pushforward} of $\Omega({\cal A}; X)$ by summing over all pre-images $X=\phi^{-1}(Y)$~\cite{Arkani-Hamed:2017tmz}. For open-string amplitudes \eqref{eq:string}, the scattering equations can be rewritten to give a diffeomorphism from $\M_{0,n}^+$ to the interior of ABHY associahedron, thus the canonical form of the latter is obtained as a pushforward by summing over saddle points of the integral. 

The saddle points are of course important for many reasons. We will argue that the number of saddle points equals the dimension of the space of integral functions as the contour varies, or the dimension of a $d$-dim twisted (co-)homology group. An interesting observation is that for any stringy canonical forms for the associahedron, the string amplitude $\bI_n$ seems to have the smallest number of saddle points, which is the well-known $(n{-}3)!$ (the number of independent open-string integrals~\cite{BjerrumBohr:2009rd, Stieberger:2009hq, Mafra:2011nw}). We conjecture that it is true for generalized associahedra as well, and we briefly comment on the number of saddle points for these cases. 

\subsection{Saddle point equations as a diffeomorphism}

The {\it scattering equations} are the saddle-point equations of the integral \eqref{gen_int} in the limit $\alpha'\to \infty$. Let us denote the regulator of the integral as $R(\x)=\prod_{i=1}^d x_i^{X_i}~\prod_I p_I(\x)^{-c_I}$ (with $\alpha'$ suppressed) and the equations read
\be \label{SE}
d\log R=0\,;\quad {\rm or}\quad \frac{X_i}{x_i}=\sum \frac{c_I}{p_I(\x)}~\frac{\partial p_I(\x)}{\partial x_i} \,, \quad {\rm for}~i=1,2,\ldots, d\,.
\ee
These are $d$ equations for $d$ variables $x_1, x_2, \ldots, x_d$ and it is natural to rewrite \eqref{SE} as a map from $\x$-space to $\X$-space.

\begin{claim} 
Denote by $\Delta_d:=\{0<x_i<\infty| i=1,\ldots, d\}$ the interior of a (projective) simplex, and by $\Int(\PP)$ the interior of the polytope $\PP$ from \eqref{Minkowski}.  The {\it scattering-equation} map $\Phi: \Delta_d \to \Int(\PP)$ defined as
\be \label{SEmap}
X_i=\sum_I c_I \frac{\partial \log p_I(\x)}{\partial \log x_i}=\sum_I \frac{c_I}{p_I(\x)} \frac{x_i \partial p_I(\x)}{\partial x_i}\,,
\ee
is a diffeomorphism, and we have the pushforward formula for the canonical form of $\PP$:
\be
\Omega(\PP)=\Phi_*\left(%\bigwedge
\prod_{i=1}^d \frac{d x_i}{x_i}\right)\,,
\ee
where $\Phi_*$ involves a sum over solutions of \eqref{SE}. This is equivalent to the following formula for the canonical function:
\be
\underline{\Omega}(\PP; X)=\int_{\C^d} \prod_{i=1}^d \frac{d x_i}{x_i}~\delta\left(X_i-x_i \sum_I \frac{c_I}{p_I} \frac{\partial p_I}{\partial x_i} \right)\,.
\ee
\end{claim}
Similar to the proof of Claim \ref{claim:multi}, it suffices to prove this for the case with single polynomial $p$, \eqref{proto_int}, where the map is a diffeomorphism from $\Delta_d$ to the interior of the Newton polytope $c \New[p(\x)]$. At this point we remark that in fact such a map has been studied in~\cite[Section 7.3.3]{Arkani-Hamed:2017tmz} as the {\it Newton polytope map}, reviewed in Appendix \ref{sec:canform}.  Let's now connect our result for a single polynomial to the Newton polytope map.  Consider the map $\Phi: \Delta_d \to {\bf Int}(\PP)$:
\be\label{eq:ourSE}
X_i=\frac{c}{p(\x)}\frac{x_i \partial p(\x)}{\partial x_i}\,,\quad {\rm for}~i=1,2,\ldots,d
\ee
Recall that the polynomial reads $p(\x)=\sum_\alpha a_\alpha \x^{{\bf n}_\alpha}$, thus the map can be written as $X_i=\frac{c}{p(\x)} \sum_\alpha a_\alpha n_{\alpha, i} \x^{{\bf n}_\alpha}$. By combining these two equations projectively, it is natural to define $Y^I=(p, p~\X)$ ($Y^0=p$ and $Y^i/Y^0=X_i$) and we have
\be
Y^I=\sum_\alpha a_\alpha (1, c~{\bf n}_{\alpha}) \x^{{\bf n}_\alpha}=\sum_\alpha y^I_{\alpha} \x^{{\bf n}_\alpha}\,,\quad y^I_\alpha:=a_\alpha (1, c~{\bf n}_\alpha)\,.
\ee
Thus we have seen that the projectivized map gives a projective polytope $Y^I \in \mathbb{P}^d$, which is the convex hull of all points $y^I_\alpha$, with coefficients given by $\x^{{\bf n}_\alpha}$. This is nothing but eq. (7.95) and (7.96) of~\cite{Arkani-Hamed:2017tmz} (see also \eqref{eq:Newton}): $y^I_\alpha$ are the vertices of a projective polytope which has the same oriented matriod as the integer matrix given by $n^I_\alpha$ with $n^I:=(1, {\bf n})$. It has been proven in~\cite{Arkani-Hamed:2017tmz} that it provides a diffeomorphism from $\Delta_d$ to $\Int (\PP)$,  and thus our scattering-equation map \eqref{eq:ourSE} is just a rewriting of the Newton polytope map.  The generalization to multiple polynomials proceeds as before, and this completes the proof of the Claim.

\begin{example}[Interval] We consider the $1$-dim integral, \eqref{1d}.  The scattering-equation map $X=c \frac{x}{1+x}$ indeed provides a diffeomorphism from $0<x<\infty$ to $0<X<c$, and the pushforward, which gives the canonical form, is given by:
\be
\Phi_*\left(\frac{d x}{x}\right)=d\log x |_{x=\frac{X}{c-X}}=d\log (\frac{X}{c-X})=\frac{c~d X}{X~(c-X)}\,,%d X (\frac {1}{X}-\frac{1}{c{-}X})\,,
\ee
and the canonical function is given by $\underline{\Omega}=\int \frac{d x}{x} \delta(X-c \frac{x}{1+x})=\frac {1}{X}+\frac{1}{c-X}$.
\end{example}

\begin{example}[Pentagon] We can consider a pentagon integral similar to the $n=5$ string amplitude \eqref{5ptstring}:
\be\label{pentagon}
\I_{\rm pentagon}=\alpha'^2~\int_0^\infty \frac{d x\, dy}{x\,y} x^{\alpha' X} y^{\alpha' Y} (1+x)^{-\alpha' a} (1+y)^{-\alpha' b}(1+x+ y)^{-\alpha' c}\,.
\ee
%where we use $\hat{X}:=\alpha' X$ {\it etc.} to suppress explicit dependence on $\alpha'$. 
By Claim \ref{claim:multi}, we need the Minkowski sum of the intervals $0<X<a$, $0<Y<b$ and the triangle bounded by $X>0, Y>0, X+Y<c$, which gives a pentagon $\PP_2$ bounded the $5$ facets $X_1:=X>0$, $X_2:=Y>0$,  $X_3:=a+c-X>0$, $X_4:=a+b+c-X-Y>0$, and $X_5:=b+c-Y>0$. %$%\lim_{\alpha' \to 0} \I_{\rm pentagon}=
%\underline{\Omega}(\PP_2)=\frac{1}{X_1  X_2}+ \frac{1}{X_2 X_3}+ \frac{1}{X_3 X_4}+ \frac{1}{X_4 X_5}+ \frac{1}{X_5 X_1}$. Again, there are many integrals that we can write down for this pentagon: we can consider {\it e.g.} $1+ p_1 x$, $1+p_2 y$, and $1+ p_3 x + p_4 y$ and the leading order remains unchanged. 
The scattering-equation map reads
\be
X=x~\left(\frac{a}{1+x}+\frac{c}{1+x+y}\right)\,,\quad Y=y\left(\frac{b}{1+y}+ \frac{c}{1+x+y}\right)\,,
\ee
which, as one can check, is a diffeomorphism from $\Delta_2$ to the pentagon $\PP_2$, and the canonical form $\Omega(\PP_2)$ is given by summing over two solutions
\be
\Phi_*\left(\frac{d x d y}{x y}\right)=d X d Y \left(\frac{1}{X_1  X_2}+ \frac{1}{X_2 X_3}+ \frac{1}{X_3 X_4}+ \frac{1}{X_4 X_5}+ \frac{1}{X_5 X_1}\right)\,.
\ee%{\color{blue} change to another example!}
\end{example}
In general, the scattering-equation map, \eqref{SEmap} (related to $\alpha'\to \infty$), always provides a pushforward formula for the canonical form (related to $\alpha' \to 0$)! In the special case of ABHY associahedron, they become literally the scattering equations and pushforward/CHY formula. Let's take a look at the CHY scattering equations, \eqref{SE}, which give saddle points of (the Koba-Nielsen factor in) open-string amplitude $\bI_n$. By writing a basis of planar variables, {\it e.g.} $X_{i,n}$ as a function of $x$'s of the positive parametrization (and the $c$'s), indeed we find the CHY scattering-equation map from ${\cal M}_{0,n}^+$ to ${\cal A}_{n{-}3}$:
\be\label{CHYSEmap}
X_{i,n}=x_i \sum_{a,b} \frac{c_{a,b}}{p_{a,b}} \frac{\partial p_{a,b}}{\partial x_i}\,.
\ee 
By our claim, this is a diffeomorphism when restricted to the interior, which proves the conjecture made in~\cite{Arkani-Hamed:2017mur}. It is well known (but quite non-trivial) that there are $(n{-}3)!$ solutions; by summing over these saddle points, the pushforward of $\omega^{(n{-}3)}:=\wedge_i d\log x_i$ gives $\Omega({\A}_{n{-}3})$ %(which is equivalent to CHY formulas for any $m(\alpha|\beta)$):
\be
\sum_{\rm sol.} \omega^{(n{-}3)}=\Omega({\A}_{n{-}3})=\prod_{i=2}^{n{-}2} d\log X_{i,n}~m_n\,
\ee
where $m_n:=\underline{\Omega}({\cal A}_{n{-}3})$ is the planar $\phi^3$ amplitude. For example for $n=5$, we see that the pentagon ${\A}_2$ (see Figure \ref{fig:pentagon}), which we have obtained as a Minkowski sum in the $\alpha'\to 0$ limit, is also given by the image of the map 
\be
X_{2,5}=x \left(\frac{c_{1,3}}{1+x}+\frac{c_{1,4}(1+y)}{1+x +x y}\right)\,, \qquad X_{3,5}=y \left(\frac{c_{2,4}}{1+y}+\frac{c_{1,4} x}{1+x+xy}\right)\,,
\ee
thus pushforward by summing over two solutions give $\sum_{\rm sol.} \frac {d x d y}{x y}=d^2 X~m_5$ as expected.

Before proceeding, let us comment on the number of solutions for ``scattering equations" for general stringy canonical forms, and those for string amplitudes. It is known \cite{Ber,Kou} that for a single Laurent polynomial $p(\x)$, the number of solutions in $(\C^*)^d$ for the collection of (Laurent) polynomial equations, $p(\x) X_i=c x_i \partial p(\x)/\partial x_i$ is given by the normalized volume of $\New[\PP]$, assuming certain non-degeneracy conditions on the coefficients is satisfied.  For example, for the $1$-d integral, \eqref{1d}, we have one solution, which is the length of the interval $[0,1]$; if we generalize to the degree-$m$ polynomial, we have $m$ solutions for the interval $[0, m]$ of length $m$.  %Here \eqref{1d} gives the minimal number of solutions. 

%On the other hand, for multiple polynomials, the number of solutions for the polynomial form of scattering equations is given by the mixed volume of Newton polytopes~\cite{Ber}. 
On the other hand, when the coefficients of the polynomial are special, the number of solutions can be different.
Let's see how this works for our pentagon integral \eqref{pentagon}. We can consider first the simplified case with $a=b=c$, in which case the factors reduces to a single polynomial $p(\x)=(1+x)(1+y)(1+x+y)=1+ 2 x + x^2+ 2 y+y^2+ x^2 y+x y^2+3 x y$. The Newton polytope has (normalized) volume $7$, and we expect $7$ solutions to the polynomial equations. However, one computes that $5$ of the $7$ solutions are spurious since they are also solutions of $p(\x)=0$\footnote{These spurious solutions should be counted with multiplicity.}, thus there are $2$ solutions to \eqref{SE}, as mentioned above. The interesting observation is that we do have more solutions for general polynomials with the same Newton polytopes.  For example, we can consider $p(\x)'=1+ 2 a_{1,0} x + a_{2,0} x^2+ 2 a_{0,1} y+ a_{0,2} y^2+ a_{2,1} x^2 y+ a_{1,2} x y^2+3 a_{1,1} x y$, or $p(\x)''=(1+ a_1 x)(1+a_2 y)(1+ a_3 x + a_4 y)$. For  $p'$ with generic coefficients $a>0$, the number of solutions to \eqref{SE} is indeed $7$ which agrees with the volume of Newton polytope, while for $p(\x)''$ with generic coefficients, we see that $3$ solutions are spurious thus we have $4$ solutions to \eqref{SE}.  We notice that \eqref{pentagon} gives the minimal number of solutions, just $2$, among these cases. 

The same observation applies to the scattering equations of string amplitudes, and we conjecture that ${\bf I}_n$ has the smallest number of saddle points, $(n{-}3)!$, among all stringy canonical forms for the ABHY associahedron. Just as in the above $n=5$ example, the number of solutions for generic choice of polynomials is (much) bigger than $(n{-}3)!$ for higher $n$.  If we consider more general positive coefficients of the polynomials $p_{a,b}$ (but still in a factorized form like $p(\x)''$), or even put a single polynomial with arbitrary positive coefficients like $p(\x)'$, we see that the number of solutions undergoes huge reductions from the single-polynomial $p(\x)'$ case to the factorized $p(\x)''$ case, and further reductions to the $(n{-}3)!$ for \eqref{CHYSEmap}. It is remarkable that ${\cal M}_{0,n}^+$ gives the ``most efficient" stringy canonical form and pushforward formula in that it has the smallest number of solutions. 

It is of great interest to study in more detail the ``scattering equations" and pushforward for cluster string integrals, and Grassmannian string integrals. Here let's briefly comment on the number of solutions, which is related to the topology of the underlying spaces. We leave the more extensive discussions, which involves counting the number of points over a finite field, to~\cite{20193, 20201}, and just list the counting of solutions for some cases. Beyond type $A_{n{-}3}$ or equivalently $\G_+(2,n)/T$, the simplest cases are type ${B}_n$ and ${C}_n$. We find that the number of saddle points is $n^n$ and $(2n)!!/2$, respectively. The former can be obtained (just as that for type ${A}$) from a hyperplane arrangement for ${B}_n$, and the latter has been studied also in~\cite{Li:2018mnq} with a worldsheet picture. Beyond that, we have computed the point count for certain cluster cases, and {\it e.g.} we find the number of saddle points is $13$ for ${G}_2$, $55$ for ${D}_4$ and $674$ for ${D}_5$. We have studied scattering equations for certain $\G_+(k,n)/T$ cases (which was also considered recently in~\cite{Cachazo:2019ngv}). For example, for $k=3$, and $n=6,7,8$, we find the number of solutions to be $26$, $1272$ and $188112$, respectively. The counting for G$_+(3,n)/T$ with $n=6,7,8$ have been verified very recently from studying solutions under soft limits~\cite{Cachazo:2019ble}. 

\subsection{Dimension of the space of integral functions}

In this subsection we explain that the number of solutions of scattering equations (or saddle points) equals the dimension of the space of integral functions as the contour varies.  Consider the integral
\be
\I_p(\X,c)= \int \prod_{i=1}^d \frac{d x_i}{x_i} x_i^{X_i} \, p(\x)^{-c} =  \int  \prod_{i=1}^d \frac{d x_i}{x_i} \exp({F(\x)}) 
\ee
where $F(\x) = \sum_{i=1}^d X_i \log(x_i) -c \log p(\x)$.  We would like to understand the dimension of the vector space of such integrals as the integration cycle (or contour) varies.  By using duality between homology (cycles) and cohomology (forms), we can instead consider all integrals of the form
\be \label{eq:integrals}
\int \prod_{i=1}^d \frac{d x_i}{x_i} Q(\x) p(\x)^{-c} =  \int  \prod_{i=1}^d \frac{d x_i}{x_i} Q(\x) \exp({F(\x)}) 
\ee
where $Q(\x)$ are various rational functions.  Let $W:=(\C^\times)^d \setminus \{p =0\}$.  The operator $d + dF \wedge$ sends $r$-forms on $W$ to $(r+1)$-forms on $W$.  If $\omega$ is a holomorphic $(d-1)$-form on $W$, then by Stokes' theorem the integral
\be
\int d(  \exp({F(\x)})  \omega) = \int (dF \wedge \omega + d\omega)  \exp({F(\x)}) 
\ee
vanishes for all appropriate integration cycles.  This leads us to consider the following {\it twisted algebraic deRham complex}.  Let $\Omega^r$ denote the space of algebraic holomorphic $r$-forms on $W$.  Then we have the twisted algebraic deRham complex

\be \label{eq:twistedDR}
\Omega^0 \xrightarrow{d+dF\wedge} \Omega^1 \xrightarrow{d+dF\wedge} \cdots  \xrightarrow{d+dF\wedge} \Omega^d.
\ee
The space of integrals \eqref{eq:integrals} is equal to the twisted cohomology group $H^d(W, d+dF\wedge)$ where
\be
H^r(W, d+dF\wedge) := \frac{\ker(d + dF \wedge \text{ on } \Omega^r) }{(d + dF \wedge) \Omega^{r-1}}.
\ee
When $F$ depends on transcendental parameters such as $c$, or if we assume $c$ is a generic complex number, then we expect that the $H^r(W, d+dF\wedge)$ is $0$ unless $r = d$ ({\it c.f.}~\cite{aomoto1974,kita_1994}), and furthermore $\dim H^r(W, d+dF\wedge)$ equals the Euler characteristic of $W$.

We give a heuristic explanation for the relation between critical points of $F(\x)$ (assuming they are isolated and multiplicity-free) and $\dim H^d(W, d+dF\wedge)$.
Consider the operator $\hbar d + dF\wedge$ (where $\hbar = 1/\alpha'$), which corresponds to replacing $\exp({F(\x)})$ by $\exp({F(\x)}/\hbar)$ or $p(\x)^{-c} \prod_i x_i^{X_i} $ by $p(x)^{-\alpha'c} \prod_i x_i^{\alpha' X_i}$.  Naively, assuming that $c, \X$ are generic, we expect that for a small $\hbar$, we have
\be
\dim H^d(W, \hbar d+dF\wedge) = \dim  H^d(W, dF\wedge).
\ee
Suppose that $F(\x)$ has $m$ isolated multiplicity-free critical points $q_1,\ldots,q_m$ in $W$.  We now argue that $ \dim  H^d(W, dF\wedge) = m$.  Writing 
\be
\Omega^{d-1} = \left\{ \sum_{j=1}^d Q_j(\x) \prod_{i\neq j}^d \frac{d x_i}{x_i}  \right\},
\ee
where $Q_j(\x)$ are regular functions on $W$, we see that
\be
H^d(W, dF\wedge) =\frac{ \text{regular functions $Q(\x)$ on $W$}}{(\frac{\partial F}{\partial x_1}, \frac{\partial F}{\partial x_2},\ldots,\frac{\partial F}{\partial x_d})}
\ee
where the $(\frac{\partial F}{\partial x_1}, \frac{\partial F}{\partial x_2},\ldots,\frac{\partial F}{\partial x_d})$ is the subspace consisting of all functions of the form
\be
\sum_{j=1}^d Q_j(\x) \frac{\partial F}{\partial x_j}.
\ee
To see that $\dim  H^d(W, dF\wedge) = m$, we consider the linear map $H^d(W, dF\wedge)  \to \C$ given by 
\be \label{eq:imageQ}
Q(\x) \longmapsto (Q(q_1),\ldots,Q(q_m)) \in \C^m.
\ee 
This is well-defined since $\frac{\partial F}{\partial x_j}$ is 0 at $q_i$ for any $i,j$ and under the isolated and multiplicity-free assumptions it gives an isomorphism between $H^d(W, dF\wedge)$ and $\C^m$.  Thus we expect that the dimension of the space of integrals is equal to the number of critical points.

%{\color{red} Explain how the following example is related to recurrence relations from earlier?}

\begin{example}
\def\im{\operatorname{im}}
Consider $d = 1$ and $F(x) = X \log(x) - c\log(1+x)$, that is, the case $p(x) = 1+x$.  We have $W = \C^\times \setminus \{-1\}$, which has Euler characteristic $1$.  For generic $X, c$, we have $\dim H^1(W,d + dF \wedge) = 1 $ and $\dim H^0(W,d + dF \wedge) = 0$.   Let us consider the complex $(\Omega^\bullet(W), dF \wedge)$, which can be identified with the map
\be 
\varphi: 
\C[x,\frac{1}{x}, \frac{1}{1+x}] \to \C[x,\frac{1}{x}, \frac{1}{1+x}], \qquad f(x) \mapsto \left(\frac{X}{x} - \frac{c}{1+x} \right)f(x).
\ee
The kernel of $\varphi$ is trivial, giving $\dim H^0 = 0$, and the image $\im(\varphi)$ consists of those elements divisible by $cx - X(1+x)$ (inside $\C[x,\frac{1}{x}, \frac{1}{1+x}]$).  We claim that the element $1$ spans $H^1$.  To see this, writing $(cx - X(1+x))x^n = \alpha x^n + \beta x^{n+1}$, we deduce inductively that $1 + \im(\varphi)$ contains all powers of $x$, and thus all Laurent polynomials.  Writing $(cx-X(1+x))/(1+x)^n = \gamma/(1+x)^n + \delta/(1+x)^{n-1}$ we deduce that $1 + \im(\varphi)$ contains all elements of the form $1/(1+x)^n$.  It is then easy to see that $1+ \im(\varphi) = \C[x,\frac{1}{x}, \frac{1}{1+x}]$.  So $\dim H^1 = 1$.  \footnote{For generic $X,c$, a similar argument gives the same dimensions ($\dim H^0 = 0$ and $\dim H^1=1$) for the operator $d+ dF \wedge$.}

Indeed, for generic $X,c$, we have that $F(x)$ has a single critical point $q:= X/(c-X)$.  So our calculation is consistent with \eqref{eq:imageQ}: any function $f(x) \in  \C[x,\frac{1}{x}, \frac{1}{1+x}]$ such that $f(q) \neq 0$ would span $H^1(W, dF \wedge)$.

Note that this computation fails if $X,c$ are not generic, for example if $X = 0$ and $c = 1$.  Then $\varphi$ is given by multiplication by $-1/(1+x)$, and we see that $\dim H^1(W, dF \wedge) = 0$.  In this case, $F(x)$ has no critical points in $W$.

\end{example}
\begin{example}
We give a simple two-dimensional example of a slightly different form where the relations $(\frac{\partial F}{\partial x_1}, \frac{\partial F}{\partial x_2},\ldots,\frac{\partial F}{\partial x_d})$ can be seen explicitly.
Suppose that $F(x,y) = x + y + 1/xy$.  (In this case, we can take $W = (\C^\times)^2$ since the only denominators in $dF$ are powers of $x,y$).  Then $F$ has three critical points, $(1,1), (\omega,\omega), (\omega^2,\omega^2)$ where $\omega$ is a primitive cube root of unity.  Let us check directly that the quotient $Q = \C[x^{\pm 1}, y^{\pm 1}]/(xy^2-1, x^2y-1)$ has dimension three, where the ideal is obtained from the components of the gradient of $F$.  A basis for $\C[x^{\pm 1},y^{\pm 1}]$ is given by $\{x^i y^j \mid (i,j) \in \Z^2\}$.  The relations $xy^2=1$ and $x^2y=1$ say that the basis element $(i,j)$ is equivalent to both $(i+2,j+1)$ and $(i+1,j+2)$ inside $Q$.  Thus, $(i,j)$ is equivalent to $(0,j-2i)$.  Also, $(0,j)$ is equivalent to $(2,j+1)$ which is equivalent to $(0,j-3)$.  Thus every $(i,j) \in \Z^2$ is equivalent to one of the three basis elements $(0,0),(0,1),(0,2)$.  Since the equivalence relations preserve the quantity $i+j \mod 3$, these three elements are linearly independent, and form a basis for $Q$.  Thus $\dim(Q) = 3$.
\end{example}

\section{Complex closed-stringy integrals}\label{sec:closed}
The beta function \eqref{eq:beta}
%$$
%B(s,t) = \int_0^1  \frac{dx}{x(1-x)} x^s (1-t)^t = \frac{\Gamma(s) \Gamma(t)}{\Gamma(s+t)}
%$$
has a complex analogue
\be \label{eq:complexbeta}
B_\C(s,t) = \int_{\C}\frac{dz d \bar z}{|z|^2|1-z|^2}   |z|^{2s} |1-z|^{2t} = (-2\pi i)  \frac{\Gamma(s) \Gamma(t) \Gamma(1-s-t)}{\Gamma(s+t)\Gamma(1-s)\Gamma(1-t)}. 
\ee
We now consider various complex analogues of our stringy integral.

\subsection{Mod-squared stringy canonical forms}\label{ssec:modsquare}
We first consider the complex integral obtained by taking the mod-squared of the integrand of \eqref{gen_int}:
\be
\I^{|\cdot|^2}_{\{p\}}(\X, \{c\})= (\alpha')^d \int_{\C^d} \prod_{i=1}^d \frac{dz_i d\z_i}{|z_i|^2} |z_i|^{2\alpha' X_i}  \prod_I |p_I(\zz)|^{-2\alpha' c_I}
\ee
and let $\PP$ be given by \eqref{Minkowski}.
As in Section \ref{ssec:decomp}, we decompose the integration domain $\C^d$ into regions  ($\log |z|:=(\log |z_1|, \cdots, \log |z_d|)$)
\be \label{eq:Rv}
R_v := \{ z \in \C^d \mid \log|z| \in -C_v\}
\ee
for each vertex $v$ of the polytope $\PP$.  Now, even if $p_I(\zz)$ has positive coefficients, it will still have zeroes in $\C^d$, but as long as $\alpha'$ is sufficiently small, these zeros will not affect the absolute convergence or leading order of the integral (see Remark \ref{rem:negcoeff}).  The analysis in Section \ref{ssec:decomp} can be repeated, where we note that 
\be
\alpha' \int_{|z| \leq 1}  \frac{dz d\z}{|z|^2}    |z|^{2 \alpha' c}  = -2i \alpha' \int_0^1 \int_0^{2\pi}   dr d\theta \; r^{2 \alpha' c - 1} %=  \alpha' 4 \pi  \int_0^1 r^{2\alpha'c -1} 
= -2\pi i \frac{1 }{c} = -2\pi i \left(\alpha' \int_{0}^1 \frac{dx }{x}    x^{\alpha' c} \right).
\ee
Thus 
\be
\lim_{\alpha' \to 0} \I^{|\cdot|^2}_{\{p\}}(\X, \{c\}) = (-2 \pi i)^d \lim_{\alpha' \to 0} \I_{\{p\}}(\X, \{c\}) = (-2\pi i)^d \Vol((\PP-\X)^\circ)
\ee
with absolute convergence when $\X \in \Int({\cal P})$ and $\alpha' >0$ is sufficiently close to 0.

\subsection{Volumes of duals of unbounded polyhedra}\label{ssec:warmup}

As a warmup, we first consider a simple generalization of stringy canonical forms, by ``shifting" some factors in the regulator:
\be \label{eq:openshift}
\I_{\{p\}}(\{c, n\})= (\alpha')^d \int_{\mathbb{R}^d_{>0}} \Omega \; \prod_J p_J(\x) ^{-\alpha' c_J} \prod_I p_I(\x)^{-(\alpha' c_I+ n_I)} 
\ee
where some of the exponents $c_I$ have been shifted by integers $n_I > 0$ (without the prefactor $\alpha'$). 
Applying \eqref{eq:IS} and Claim \ref{claim:trop}, we have
\begin{align}\label{eq:loshift}
\lim_{\alpha' \to 0} {\I}_{\{p\}}(\{c, n\}) &= \lim_{\hbar \to \infty} \Vol((\sum_J c_J \PP_J + \sum_I (c_I +\hbar n_I) \PP_I)^\circ) \\
&= \lim_{\hbar \to \infty} \Vol(\{ -\sum_J c_J \Trop(p_J) -  \sum_I (c_I +\hbar n_I) \Trop(p_I) \leq 1\})
\end{align}
where $\hbar = 1/\alpha'$.  Note that it is important in the formula above to consider the limit 
\be
\lim_{\hbar \to \infty} \sum_I (c_I +\hbar n_I) \Trop(p_I) = \sum_I c_I \Trop(p_I) + \lim_{\hbar \to \infty}  \hbar \sum_I n_I \Trop(p_I)
\ee
Instead of the simpler limit $\lim_{c_I \to \infty} \sum_I c_I \Trop(p_I)$.  This is because in certain directions the sum $ \sum_I n_I \Trop(p_I)$ could be 0, and then the term $\sum_I c_I \Trop(p_I)$ (and thus the whole function) would take a finite value.

The polytope $\sum_J c_J \PP_J +  \sum_I (c_I +\hbar n_I) \PP_I$ is typically unbounded as $\hbar \to \infty$, but the dual polytope $(\sum_J c_J \PP_J + \sum_I (c_I +\hbar n_I) \PP_I)^\circ$ can still have a finite non-zero volume.  For example, let us consider the Minkowski sum of intervals inside $\R$:
\be
\lim_{b \to \infty} c[-1,0] + b[0,1] =c[-1,0] + \infty [0,1] = [-c, \infty).
\ee
The dual of this polyhedron (unbounded polytope) is $[-1/c,0]$ which has volume $1/c$.

In general, the function $f = \lim_{\hbar \to \infty}  (-\sum_J c_J \Trop(p_J) - \sum_I (c_I +\hbar n_I) \Trop(p_I)) $ is a piecewise-linear function on $\R^d$ that takes values in $\R \cup \{\infty\}$, and for convergence we require that it takes positive (or infinite) values on $\R^d \setminus \{0\}$.  The function $f$ is determined by its values on the rays of the normal fan $\N(\PP)$.  If $f$ takes value $\infty$ on any of the bounding rays of a cone $C_v$ of $\N(\PP)$, then the cone $C_v$ contributes 0 to the volume $\Vol(\{f \leq 1\})$.  But if the value of $F$ is finite on all the bounding rays of a cone $C_v$, then that cone will contribute the volume of a corresponding cone (obtained by intersecting $C_v$ with $\{f \leq 1\}$.

\begin{example}

Consider $\PP_1 = \Conv((0,0),(1,0),(0,1))$, $\PP_2 = \Conv((0,0),(-1,0))$ and $\PP_3 = \Conv((0,0),(0,-1))$, and the Minkowski sum $\PP = \infty \PP_1 + a \PP_2 + b\PP_3$ is the unbounded region pictured in Figure \ref{fig:unbounded}.  It is bounded by %the line segment joining $(-a,0)$ and $(0,-b)$, and 
two rays $R_1 = \{(-a,t) \mid t \geq 0\}$ and $R_2 = \{(t,-b) \mid t \geq 0\}$.  The dual $\PP^\circ$ is the triangle $\Conv((0,0),(1/a,0),(0,1/b))$ which has volume $1/ab$.  Thus the integral
\be \label{eq:openshift1}
\I = \alpha'^2 \int_{\mathbb{R}^d_{>0}} \Omega \;  \left(1+\frac{1}{x}\right)^{-\alpha' a} \left(1+\frac{1}{y}\right)^{-\alpha' b}  (1+x+y)^{-(\alpha'c+1)}
\ee
converges when $a,b >0$ (with $c>0$) and has leading order equal to $1/ab$.

\begin{figure}
\begin{center}
\begin{tikzpicture}
%\draw [help lines, step=1cm] (-2,-2) grid (2,2);
\fill[lightgray] (-2/3,2) -- (-2/3,0) -- (0,-1) -- (2,-1) --(2,2)--cycle;
\draw [help lines, step=1cm] (-2,-2) grid (2,2);
\draw[thick] (-2/3,2) -- (-2/3,0) -- (0,-1) -- (2,-1);
\node[fill=black,circle, inner sep=0pt,minimum size=5pt] at (0,0) {};
\node[fill=cyan,circle, inner sep=0pt,minimum size=5pt] at (-2/3,0) {};
\node at (-1.2,-1/4) {$(-a,0)$};
\node[fill=cyan,circle, inner sep=0pt,minimum size=5pt] at (0,-1) {};
\node at (-.2,-1.4) {$(0,-b)$};
\begin{scope}[shift={(5,0)}]
\filldraw[fill=gray,thick] (0,0)--(3/2,0)--(0,1)--cycle ;
\draw [help lines, step=1cm] (-2,-2) grid (2,2);
\node[fill=black,circle, inner sep=0pt,minimum size=5pt] at (0,0) {};
\node[fill=cyan,circle, inner sep=0pt,minimum size=5pt] at (3/2,0) {};
\node[fill=cyan,circle, inner sep=0pt,minimum size=5pt] at (0,1) {};
\node at (2.1,-0.3) {$(1/a,0)$};
\node at (0,1.4) {$(0,1/b)$};
\end{scope}
\begin{scope}[shift={(10,0)}]
%\filldraw[fill=gray,thick] (0,0)--(3/2,0)--(0,1)--cycle ;
\draw [help lines, step=1cm] (-2,-2) grid (2,2);
\node[fill=black,circle, inner sep=0pt,minimum size=5pt] at (0,0) {};
\draw[thick] (0,0)--(2,0);
\draw[thick] (0,0) --(0,2);
\node at (-1,-1) {$\infty$};
\node at (1,1) {$aX+bY$};
\end{scope}

%\begin{scope}[shift={(5,0)}]
%\draw [help lines, step=1cm] (-2,-2) grid (2,2);
%\node[fill=black,circle, inner sep=0pt,minimum size=5pt] at (0,0) {};
%\draw[thick] (0,0) --(2,0);
%\draw[thick] (0,0) --(-2,0);
%\draw[thick] (0,0) --(0,-2);
%\draw[thick] (0,0) --(0,2);
%\draw[thick] (0,0) --(2,-2);
%\node at (-1,-1) {$C_2$};
%\node at (-1,1) {$C_3$};
%\node at (1,1) {$C_4$};
%\node at (1.3,-1/2) {$C_5$};
%\node at (1/2,-1.3) {$C_1$};
%\end{scope}
%\begin{scope}[shift={(10,0)}]
%\draw [help lines, step=1cm] (-2,-2) grid (2,2);
%\node[fill=black,circle, inner sep=0pt,minimum size=5pt] at (0,0) {};
%\draw[thick] (0,1) -- (1,0) -- (1,-1) -- (0,-1) -- (-1,0)-- cycle;
%\node[fill=cyan,circle, inner sep=0pt,minimum size=5pt] at (0,1) {};
%\node[fill=cyan,circle, inner sep=0pt,minimum size=5pt] at (1,0){};
%\node[fill=cyan,circle, inner sep=0pt,minimum size=5pt] at (1,-1){};
%\node[fill=cyan,circle, inner sep=0pt,minimum size=5pt] at (0,-1){};
%\node[fill=cyan,circle, inner sep=0pt,minimum size=5pt] at (-1,0){};
%\node at (0.7,0.7) {$F_4$};
%\node at (1.2,-0.5) {$F_5$};
%\node at (0.5,-1.2) {$F_1$};
%\node at (-0.7,-0.7) {$F_2$};
%\node at (-0.7,0.7) {$F_3$};
%\end{scope}
\end{tikzpicture}
\caption{The polyhedron $\PP = \infty \PP_1 + a \PP_2 + b\PP_3$, its dual $\PP^\circ$, and the piecewise linear function $-\infty \min(0,X,Y) - a \min(0,-X) + b \min(0,-Y)$.}
\label{fig:unbounded}
\end{center}
\end{figure}
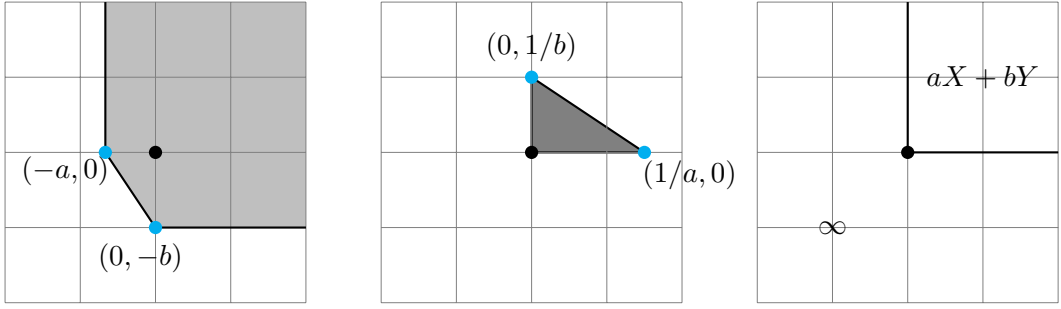

We repeat the computation using piecewise linear functions.  Let 
\begin{align}
\begin{split}
f(X,Y) &=\lim_{\hbar \to \infty} \Trop( (1+x+y)^{-\hbar} (1+1/x)^{-a} (1+1/y)^{-b}) \\
& = -\infty \min(0,X,Y) - a \min(0,-X) + b \min(0,-Y).
\end{split}
\end{align}
The tropical function $f$ is positive and takes value $\infty$ if one of $X$ or $Y$ is negative.  In the positive quadrant, it takes value $aX + bY$.  The region $\{f \leq 1\}$ is exactly the dual polytope $\PP^\circ$, see Figure \ref{fig:unbounded}.
\end{example}

\subsection{Complex stringy integrals}
We now consider the complex stringy integral
\be \label{eq:closed_int}
\I^{\C}_{\{p\}}(\X,\{c, n\})=(\alpha')^d \int_{\C^d} \prod_{i=1}^d \frac{dz_i d\z_i}{|z_i|^2} \prod_{i=1}^d z_i^{\alpha' X_i} \z_i^{\alpha'X_i + n_i}
%\bar z_i^{\alpha' X_i+n_i} 
\prod_I p_I(\zz)^{-\alpha' c_I} p_I(\bar \zz)^{-\alpha' c_I-n_I}
\ee
where the exponents have been shifted by parameters $n_i, n_I$.  For simplicity, we assume that $n_i, n_I$ are nonnegative integers.  (The integrand is multi-valued unless the shifts are integers.)%For the integral to converge in the neighborhood of a zero of $p_I(z)$, we suppose that the integers $n_I$ are all nonnegative.

The integral \eqref{eq:closed_int} has a number of convergence issues, and usually does not absolutely converge.  One issue is that the polynomials $p_I(\zz)$ typically have zeros in $\C^d$.  We shall assume that these are renormalized to make no contribution to the leading order when we take the limit $\lim_{\alpha' \to 0}$.
Next, let us decompose the integration domain as in \eqref{eq:Rv}.  Instead of the integral  $\int_{|z| \leq 1}  \frac{dz d\z}{|z|^2}    |z|^{2 \alpha' c}$ of Section~\ref{ssec:modsquare}, we are faced with higher-dimensional versions of integrals of the form
\be
\int_{|z| \leq 1}  \frac{dz d\z}{|z|^2}  |z|^{2 \alpha' c} z^{n} (1+ O(z))
\ee
where the integer $n$ is equal to a linear combination of the $n_i$ and $n_I$, and can be of either
sign. These integrals do not converge absolutely.  Nevertheless, let us compute
\be
\int_{|z| \leq 1}  \frac{dz d\z}{|z|^2}  |z|^{2 \alpha' c} z^{n}
 = (-2i) \int_0^1 \int_0^{2\pi}   dr d\theta \; e^{n i \theta} r^{2 \alpha' c +n - 1} = (-2i) \int_0^1 dr\; r^{2 \alpha' c +n - 1} \int_0^{2\pi} d\theta \; e^{n i \theta}  %=  \alpha' 4 \pi  \int_0^1 r^{2\alpha'c -1} 
 \ee
 and note that $\int_0^{2\pi} d\theta \; e^{n i \theta} = 0$ if $n \neq 0$.  We define the integral $\I^{\C}_{\{p\}}(\X,\{c, n\})$ so that this phase cancellation occurs in every integration region $R_v$ where the integral has an integer shift ({\it i.e.} $n \neq 0$), thus contributing 0.  In the remaining integration regions, the analysis is the same as for the mod-squared integral $\I^{|\cdot|^2}_{\{p\}}(\X, \{c\})$.  With this regularization, we conclude that
\be\label{eq:locomplex}
\lim_{\alpha' \to 0} \I^{\C}_{\{p\}}(\X,\{c, n\}) = (-2\pi i)^d \lim_{\alpha' \to 0} (\alpha')^d \int_{\R_+^d}\Omega \prod_i x_i^{\alpha' X_i+n_i} \prod_I p_I(x)^{-(\alpha' c_I+n_I)}
\ee
which is in the form of the integral studied in Section \ref{ssec:warmup}.  The leading order of the complex stringy integral $ \I^{\C}_{\{p\}}(\X,\{c, n\})$ is thus the canonical form of a (possibly unbounded) polyhedron.
   
\subsection{Closed-string amplitudes}

Last but not least, let's go back to ${\cal M}_{0,n}$ and consider closed-string amplitudes with two Parke-Taylor forms of different orderings and similarly open-string ones where the Parke-Taylor form and integration domain have different orderings. The real moduli space ${\cal M}_{0,n}(\mathbb{R})$ is known to have $(n{-}1)!/2$ connected components corresponding to different orderings (up to the dihedral symmetry), and we denote each such component with ordering $\alpha$ by ${\cal M}_{0,n}^+ (\alpha)$.  We denote the canonical form for it to be $\omega(\alpha)$, and it is natural to consider integrating
$\omega(\beta)$ over ${\cal M}_{0,n}^+ (\alpha)$. Similarly we consider $\omega(\alpha)$, $\omega^*(\beta)$ (complex conjugate) in the integral over ${\cal M}_{0,n}(\mathbb{C})$ (with Koba-Nielsen factor mod squared):
\ba
&&\bI_n(\alpha|\beta):=(\alpha')^{n{-}3}~\int_{{\cal M}_{0,n}^+(\alpha)} \omega(\beta) \prod_{a b}~|(a b)|^{\alpha' s_{a b}}\,,\\
&&\bI_n^{\rm closed}(\alpha|\beta):=(-\frac{\alpha'}{2\pi i})^{n{-}3}~\int_{{\cal M}_{0,n}(\mathbb{C})} \omega(\alpha)\,\omega^*(\beta)~\prod_{a,b} |(a b)|^{2 \alpha' s_{a b}}\,,
\ea
Let's apply our general discussion to such ``off-diagonal" open-string integrals and closed-string integrals\footnote{The open-string integals have appeared in the context of $Z$ theory~\cite{Carrasco:2016ygv, Mafra:2011nw, Broedel:2013tta}, and both have appeared in the study of single-valued projection~\cite{Schlotterer:2018zce, Brown:2018omk}.}.
We have seen that for the same shifts the leading order for closed-string integrals is identical to the open case, so we will focus on the closed-string case.

We fix $\alpha=(12\cdots n)$, and consider all possible $\beta$. This corresponds to the formula above with $d=n{-}3$, and we choose $z_i$'s to be positive parametrization for ordering $\alpha$, introduced in Section~\ref{sec:5}; the shifts in $\bar{z}_i$'s and polynomials $P_{a,b}(\bar{\bf z})$, which we denote as $n_i$ and $n_{a,b}$, are determined by the ordering $\beta$:
\be\label{eq:Iclosed}
{\cal I}(\alpha|\beta)=(-\frac{\alpha'}{2\pi i})^{n{-}3} \int_{\C^{n{-}3}} \prod_{i=2}^{n{-}2} \frac{dz_i d\z_i}{|z_i|^2} \prod_{i=2}^{n{-}2} z_i^{\alpha' X_{i, n}} \z_i^{\alpha'X_{i,n} + n_i} \prod_{a,b} p_{a,b} (\zz)^{-\alpha' c_{a,b}} p_{a,b}(\bar \zz)^{-\alpha' c_{a,b}-n_{a,b}}
\ee
where we recall that $P_{a,b}({\bf z})=1+\sum_{j=a{+}1}^{b{-}1} \prod_{i=a{+}1}^j z_i$. If we have $\beta=\alpha$, {\it i.e.} the mod-squared integral considered above, then all shifts are zero, $n_i=n_{a,b}=0$. For $\beta \neq \alpha$, the 
shifts can be easily read off from the ratio of Parke-Taylor forms, $\omega(\beta)/\omega(\alpha)$, which is a SL(2)-invariant function of $z_i$'s. It is straightforward to see that $n_i$ and $n_{a, b}$ and take values only in $\{0,1\}$.  Let's spell out some explicit examples for $n=4,5$.

\begin{example}
For $n=4$ there is essentially only one non-trivial example different from the mod-squared integral.  However, we can get slightly different presentations as an integral of the form \eqref{eq:closed_int}.  Let us take $\alpha =(1234)$ where $\M_{0,4}$ is parametrized as $(z_1,z_2,z_3,z_4) = (0,1,z,\infty)$ as in Section~\ref{sec:intro}, with Parke-Taylor form $\omega(\alpha) = dz/z$.  For $\beta=(1324)$, we have $\omega(\beta) = \frac{dz}{z(1+z)}= \frac{1}{1+z}\omega(\alpha)$, and
\be
{\cal I}(1234|1324) = 
(-\frac{\alpha'}{2\pi i}) \int_{\C} \frac {dz d\bar z}{|z |^2}~z^{\alpha' X} \bar{z}^{\alpha' X} (1+z)^{-\alpha' c} (1+\bar{z})^{-\alpha' c-1}.
\ee  
By \eqref{eq:loshift} and \eqref{eq:locomplex} the leading order of the integral is given by the canonical function of a ray $X \in [0, \infty)$; equivalently, the Minkowski sum in question is $\infty[0,1] - X = [-X,\infty)$, and the dual is $[-\frac 1 X, 0]$ which has volume $1/X$. 

For $\beta=(1342)$, we have $\omega(\beta) = -\frac{dz}{1+z}= -\frac{z}{1+z}\omega(\alpha)$, and
\be \label{eq:1342}
{\cal I}(1234|1342) = 
(\frac{\alpha'}{2\pi i}) \int_{\C} \frac {dz d\bar{z}}{|z |^2}~z^{\alpha' X} \bar{z}^{\alpha' X+1} (1+z)^{-\alpha' c} (1+\bar{z})^{-\alpha' c-1}.
\ee
The Minkowski sum we should consider is $\lim_{\hbar \to \infty} {-}(X{+}\hbar){+}(c{+}\hbar)[0,1]$ which is somewhat subtle because of the presence of both $\hbar$ and $-\hbar$.  It is simpler to consider the piecewise linear function on $Z$-space,
\be
f(Z) = \lim_{\hbar \to \infty} \left((X+\hbar)Z - (c +\hbar) \min(0,Z)\right)
\ee
which takes the value $(X-c)Z$ when $Z< 0$ and the value $\infty$ when $Z > 0$.  Thus, with our regularization, the integral \eqref{eq:1342} converges when $(X-c) <0$ and the leading order is equal to $1/(X-c)$.
%
%By \eqref{eq:loshift} and \eqref{eq:locomplex} the leading order of the integral is given by the canonical function of a ray $X \in [0, \infty)$; equivalently, the Minkowski sum in question is $\infty[0,1] - X = [-X,\infty)$, and the dual is $[-\frac 1 X, 0]$ which has volume $\frac{1}{X}$. 
\end{example}

\begin{example}

For $n=5$, things get even more interesting.  To be explicit, let's write \eqref{eq:Iclosed} again for $n=5$ (omitting the factor of $(\frac{\alpha'}{2\pi i})^2$):
\begin{align}
%(-\frac{\alpha'}{2\pi i})^2 
\int_{\C^2}  &\frac{d^2 z_2 d^2 z_3}{|z_2|^2 |z_3|^2}~z_2^{\alpha' X_{2,5}}  z_3^{\alpha' X_{3,5}} (1+z_2)^{-\alpha' c_{1,3}} (1+z_3)^{-\alpha' c_{2,4}} (1+z_2+ z_2 z_3)^{-\alpha' c_{1,4}}\\
& \times \bar{z}_2^{\alpha' X_{2,5}+n_2} \bar{z}_3^{\alpha' X_{3,5}+n_3} (1+\bar{z}_2)^{-\alpha' c_{1,3}-n_{1,3}} (1+\bar{z}_3)^{-\alpha' c_{2,4}-n_{2,4}} (1+\bar{z}_2+ \bar{z}_2 \bar{z}_3)^{-\alpha' c_{1,4}-n_{1,4}}\,,\nonumber
\end{align}
where the shifts $n_2, n_3, n_{1,3}, n_{2,4}, n_{1,4}$ depend on $\beta$. 

We first consider $\beta=(12435)$.  Then $\omega(\beta) = \frac{1}{1+z_3}\omega(\alpha)$, so the only non-zero shift is $n_{2,4}=1$.  In the Minkowski sum (see Figure~\ref{fig:unbound}) we see the interval in $X_{3,5}$ direction becomes a ray, while the other two pieces remain finite, thus the resulting shape is an unbounded quadrilateral with vertices $(0,0)$ and $(0, c_{1,3}+c_{1,4})$ (the other two vertices are sent to infinity). The canonical function at $(X_{2,5}, X_{3,5})$ is given by
\be
\lim_{\alpha'\to 0} \I(12345| 12435)=-\frac 1 {X_{3,5}} \left(\frac 1 {X_{2,5}}+ \frac 1 {c_{1,3}+c_{1,4}-X_{2,5}}\right)\,.
\ee
Equivalently, the dual polytope after translating by $-(X_{2,5}, X_{3,5})$, is given by the product of $[-1/X_{3,5}, 0]$ and $[-1/X_{2,5}, 1/(c_{1,3}+c_{1,4}-X_{2,5})]$. Its area gives the same result. 

For some other orderings, such as $\beta=(13245)$ or $(12354)$ {\it etc.}, we also end up with an unbounded quadrilateral which has two vertices at infinity, though the shifts look different. 

A distinct type of ordering is given by {\it e.g.} $\beta=(13425)$, for which we have $n_{1,3}=n_{2,4}=1$ and others vanish. The Minkowski sum involves the finite triangle with two rays in the $X_{2,5}$ and $X_{3,5}$ directions, and the result is the first quadrant in $X_{2,5},X_{3,5}$ space, see Figure~\ref{fig:unbound}. The canonical function at $(X_{2,5}, X_{3,5})$ is 
\be
\lim_{\alpha'\to 0} \I(12345| 13425)=\frac 1 {X_{2,5} X_{3,5}}.
\ee
Equivalently we can translate it by $-X_{2,5}$ and $-X_{3,5}$, and the dual of the cone is given by the product of $[0, 1/X_{2,5}]$ and $[0, 1/X_{3,5}]$, whose area gives the same result. 

Similarly if we consider $\beta=(12534)$, the shifts are different but the Minkowski sum of all factors again gives a cone, and the dual is given by $[-1/(c_{1,3}{+}c_{1,4}{-}X_{2,5}), 0] \times [-1/X_{3,5}, 0]$. The area gives the correct leading order, $\frac{1}{X_{3,5} (c_{1,3}+c_{1,4}-X_{2,5})}$. 

Finally, let's consider the ordering $\beta=(13524)$, which has all shifts one, $n_2=n_3=n_{1,3}=n_{2,4}=n_{1,4}=1$. It is clear that the Minkowski sum of all factors give the entire space, thus the volume of the dual vanishes. This is the only ordering for $n=5$ (out of $12$ in total), for which the leading order of the integral vanishes. 

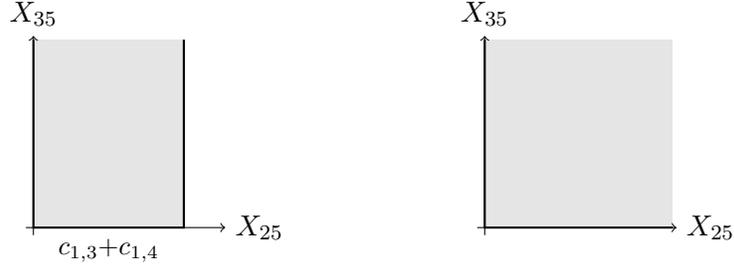
\begin{figure}
\begin{center}
\begin{tikzpicture}
\draw[->] (0,-0.1) -- (0,2.55) node[above]{$X_{35}$};
\draw[->] (-0.1,0) -- (2.55,0) node[right]{$X_{25}$};
\fill[gray,opacity=0.2] (2,2.5) -- (0,2.5) -- (0,0) -- (2,0);
\draw[thick] (0,2.5) -- (0,0) -- (2,0) -- (2,2.5);
\node at (1,-0.3) {\small $c_{1,3}{+}c_{1,4}$};
\draw[->] (6,-0.1) -- (6,2.55) node[above]{$X_{35}$};
\draw[->] (5.9,0) -- (8.55,0) node[right]{$X_{25}$};
\fill[gray,opacity=0.2] (8.5,2.5) -- (6,2.5) -- (6,0) -- (8.5,0);
\draw[thick]  (6,2.5) -- (6,0) -- (8.5,0);
\end{tikzpicture}
\end{center}
\caption{Unbounded polyhedra appearing for $\beta=(12435)$ and $\beta=(13425)$.}
\label{fig:unbound}
\end{figure}

\end{example}

We remark that both the leading order of $\bI_n(\alpha|\beta)$ and that of $\bI^{\rm closed}_n (\alpha|\beta)$ is given by doubly-partial amplitudes of bi-adjoint $\phi^3$ theory, or $m(\alpha|\beta)$~\cite{Cachazo:2013iea} (we have focused on $m(\alpha|\alpha)$ in earlier sections). More importantly, these (unbounded) polyhedra are identical to the ABHY realization for $\alpha\neq \beta$ which has facets at infinity. They have been discussed in Section 3 of~\cite{Arkani-Hamed:2017mur}, and their dual volumes has also been studied in~\cite{Frost:2018djd}. Here, we see that these unbounded polytopes and their dual volumes naturally appear as limits of Minkowski sums and their duals.
%We end with two more comments about stringy canonical forms. As we show $\I_n^{U^+}$ is special in that it factorizes in a canonical way, and the same is no longer true for stringy canonical form for a general polytope. We will study such integrals and weaker version of factorizations {\it etc.} in~\cite{20192}. Besides, one can again generalize such integrals to complex case and the leading order is given by ``intersection" of two Minkowski sums. In the special case of $\I^c_n (\alpha|\beta)$, it is well known that the leading order, $m(\alpha|\beta)$, is given by the CHY formula which can be interpreted as intersection numbers~\cite{Mizera:2017rqa}.

\section{Big polyhedron and dual $u$-variables}\label{sec:big}

Let us return to our general integral of the form 
\begin{equation}
\I(\X,\{c\}) = (\alpha')^d \, \int_{\R_{>0}^d} \prod_{i=1}^d \frac{dx_i}{x_i} x_i^{\alpha^\prime X_i} \prod_{I=1}^m p_I(\x)^{-\alpha^\prime c_I}
\end{equation}
We would like to re-express this integral in a way which makes the convergence properties completely manifest. We will first describe this in a completely pedestrian way, recognizing the role of an important ``big polyhedron" in the large $(d+m)$-dimensional space of all the variables $(\X, \{c\})$, before giving a much more elegant treatment in terms of tropical functions.

\subsection{Big polyhedron and $u$-variables}
 We know that for $c_I$ positive, the integral $\I(\X,\{c\})$ is convergent when ${\bf X}$ is contained inside the Minkowski sum ${\PP} = \sum c_I {\PP}_I$, where ${\PP}_I = \New[p_I(\x)]$ are the Newton polytopes of the polynomials $p_I(\x)$. This enforces $N$ linear inequalities on the $(X_i, c_I)$, where $N$ is the number of facets of ${\PP}$. Let's define $S_J$ to be the $r = (d+m)$-dimensional vector $S_J=(-X_1,\ldots,-X_d, c_1, \ldots, c_m)$.  The polytope ${\PP}$ is cut out by equations of the form $W^J_{a} S_J \geq 0$ for $a=1,\ldots, N$ ranging over all facets of ${\PP}$. For ease of notation we will write the above integral as 
\begin{equation}
\I(S) =(\alpha')^d \int_{\R_{>0}^d} \prod_{i=1}^d \frac{dx_i}{x_i} \prod_J p_J(\x)^{-\alpha^\prime S_J}
\end{equation} 
and for the integral to be convergent, we must have $W_a^J S_J \geq 0$ for all $a$. 

Now, so far our polytopal discussions have focused on the $d$-dimensional polytope associated with Minkowski sums {\it etc. } But we now see  there is {\it another} natural polytope, in a larger space. This ``big polytope" $\B$ or more properly, {\bf big polyhedron}, since it is unbounded, lives in $r=d+m$ dimensions, and we have just described how to cut it out with inequalities $W_a^J S_J \geq 0$. As always, it is also natural to think of the dual, vertex-based description of ${\B}$, as the cone generated by a collection of $v \geq r$ vectors $V^A_ J$, for $A=1, \cdots v$. Any ${\bf S}$ inside ${\B}$ can be written as $S_J = U_A V^A_J$ with $U_A \geq 0$. Thus, we have a way of parameterizing our integral as
\begin{equation}
(\alpha')^d\int_{\R_{>0}^d} \prod \frac{dx_i}{x_i} \prod_A (u_A)^{\alpha^\prime U_A} \, \quad {\rm where} \, u_A = \prod_J p_J^{-V^A_J} , \, {\rm and} \,\, U_A > 0
\end{equation}
In this way, we have traded the variables $S_J$, which satisfy complicated conditions for convergence of the integral,  for a new set of variables $U_A$, which must merely be positive for convergence. These are also associated with the new ``u-variables" $u_A$ defined above.   

In general, the big polyhedron ${\B}$ is not a simplex, which means that there is no unique way of writing $S_J$ as a positive linear combination of the vertices $V^A_J$. In this case the $u_A$ variables are not independent, and satisfy multiplicative (monomial) relations. But there do exist special situations where ${\B}$ {\it is} a simplex (or more precisely, a simplicial cone). Note this is quite non-trivial: in the language of our usual $d$-dimensional polytope ${\PP}$ written as a Minkowski sum of $m$ summands, we must have that the number of summands $m$ is equal to $N - d$, where $N$ is the number of facets of the Minkowski sum ${\PP}$. This turns out to happen for the ABHY realizations of all generalized associahedra associated with all finite-type cluster algebras. In this case, the variables $u_A$ are all independent, and furthermore, every facet of ${\PP}$ can be associated with a single $u_A$ going to zero, giving a ``binary" geometry associated with these cluster algebras we will describe in \cite{20193, 20201}. 

\subsection{The $u$-variables for the open-string amplitude}\label{sec:u}

Let us illustrate the construction of $u$-variables and the big polyhedron for the $n=5$ ABHY associahedron described in Sections \ref{sec:stringymulti} and \ref{sec:5}. To begin with the integral is parametrized by the five variables $(X_{25},X_{35},c_{13},c_{14},c_{24})$. The five facets of the polygon are  cut out by $X_{25}>0,X_{35}>0$ as well as  $ X_{13} := c_{13} + c_{14} - X_{25} \geq 0, X_{14} := c_{14} + c_{24} - X_{35} \geq 0, X_{24} := c_{24} + X_{25} - X_{35} \geq 0$. We have the special situation where the number of variables in the exponents of the integral (5) exactly matches the number of facets of the Minkowski sum polytope (also 5). The $U$-variables in this case are nothing but all the $X_{ij}$, and to determine the associated $u_{ij}$, we must simply invert and solve for $(X_{25},X_{35},c_{13},c_{14},c_{24})$ in terms of $(X_{25},X_{35},X_{13},X_{14},X_{24})$. Of course this inversion is directly what is given by the ABHY subspace, that tells us $c_{ij} = X_{ij} + X_{i+1 j+1} - X_{i j+1} - X_{i+1 j}$! Thus we find 
\begin{equation}\label{eq:I5again}
{\bf I}_{n=5} = (\alpha')^2 \int_{\R_{>0}^2} \frac{dy_2}{y_2} \frac{dy_3}{y_3} \prod u_{ij}^{\alpha^\prime X_{ij}}
\end{equation}
with
\begin{eqnarray}
u_{13} = \frac{1}{p_{13}} = \frac{1}{1 + y_2}, \, u_{14} = \frac{p_{13}}{p_{14}} = \frac{1 + y_2}{1 + y_2 + y_2 y_3}\,,u_{24}=\frac{p_{14}}{p_{13} p_{24}} = \frac{1 + y_2 + y_2 y_3}{(1+y_2)(1+y_3)},  \nonumber \\
u_{25} = \frac{y_2 p_{24}}{p_{14}}=\frac{y_2(1+y_3)}{1 + y_2 + y_2 y_3}\,,u_{35} = \frac{y_3}{p_{24}} = \frac{y_3}{1+y_3}
\end{eqnarray}
Note that \eqref{eq:I5again} is identical to \eqref{5ptstring} under the transformation $(x_2, x_3) \to (y_2,y_2y_3)$.

Quite beautifully, each of the massless poles of $\bI_5$ where some $X_{ij} \to 0$, is associated with a region in integration space where the corresponding $u_{ij} \to 0$. Indeed, the $u_{ij}$ have a striking property of providing a {\it perfect binary} 
representation of the geometry of the pentagon, which can be verified from the relations 
\begin{equation}
u_{13} + u_{24} u_{25}=1,  \, + \text{ cyclic}
\end{equation}
Note  all the $u_{ij}$ are positive functions on $\R_{>0}^2$ and by the above they are also smaller than one. But then when e.g. $u_{13} \to 0$, the $u's$ associated with the incompatible facets $X_{24},X_{25}$ of the pentagon, are forced to go to $1$. 

The same analysis holds for all $n$, where we begin with the integral 
\begin{equation}
{\bf I}_n = (\alpha')^{n-3}\int_{\R_{>0}^{n-3}} \frac{dy_2}{y_2} \cdots \frac{d y_{n-2}}{y_{n-2}} \prod_i y_i^{\alpha^\prime X_{i n}} \prod p_{ij}(\y)^{-\alpha^\prime c_{ij}}
\end{equation}
Writing $c_{ij} = X_{ij} + X_{i+1 j+1} - X_{i j+1} - X_{i+1 j}$, and recognizing $\omega = \frac{dy_2}{y_2} \cdots \frac{dy_{n-2}}{y_{n-2}}$ as the usual Parke-Taylor canonical form, we can write this more invariantly~\cite{Arkani-Hamed:2017mur} (see also \cite{Brown:2018omk}) as
\begin{equation} 
\bI_n = (\alpha')^{n-3}\int_{\R_{>0}^{n-3}}  \omega~\prod_{1\leq i<j{-}1<n} %\prod_{{\rm all \, non-adjacent\,  i,j}} 
u_{i\,j}^{\alpha^\prime X_{i\,j}}
\end{equation} 
where the $u_{ij}$ can be written gauge-invariantly as the cross-ratios
\begin{equation}\label{eq:uij}
u_{i\,j} = \frac{(i{-}1\,j)(i\,j{-}1)}{(i{-}1\,j{-}1)(i\,j)}
\end{equation} 
The $u_{ij}$ satisfy the remarkable equation (known since the early days of dual resonance models, see~\cite{Koba:1969kh}, and introduced again in~\cite{Brown:2009qja}):
\begin{equation} \label{eq:u+u}
u_{i\,j} + \prod_{(k\,l) \, \text{crossing} \, (i\,j)} u_{k\,l} = 1
\end{equation}
where the product is over all chords $(k\,l)$ crossing $(i\,j)$. Once again, given $u_{i\,j} > 0$ these equations also tell us that all the $u$'s are bounded between $0$ and $1$. But if a single $u_{i\,j} \to 0$, we learn that the $u$ variables associated with the incompatible chords of the polygon that cross $(i\,j)$ must be sent to 1! 
As already mentioned, this phenomenon generalizes to all the integrals associated with finite-type cluster algebras--there are a set of ``$u$" variables that provide a perfect, binary realization of the geometry of the polytope, see also Section~\ref{sec:clusterU}.

The $u$-presentation of the string integral is in every way superior to the usual Koba-Nielsen presentation. It is completely gauge-invariant, while the Koba-Nielsen formula has an $\SL(2)$ gauge redundancy. Furthermore, the convergence properties as well as the structure of the poles, and in particular the crucial factorization of the integral on massless poles, even for finite $\alpha^\prime$, is made completely obvious in the $u$-presentation, but takes a bit more analysis to see in the usual Koba-Nielsen form. We will discuss cluster-generalization of the $u$-space, called the cluster configuration space (see Section \ref{sec:clusterU}), and associated generalized particle and string amplitudes in \cite{20193, 20201}. 

Note that the most basic phenomenon we needed for the $u$-variables to all be independent, is that the big polyhedron ${\cal B}$ is a simplex. This is a very special requirement, but can be satisfied more widely than in our favorite illustrative examples. For instance, let's return to our $n=5$ integrand, and  slightly modify the polynomials $p_{13},p_{14},p_{24}$ without changing the shape of their Newton polytope. For instance, we can just change $p_{13} = 1+ y_2$ to $p_{13} = 1 + a y_2$ for a general positive constant $a$. We still have 5 independent $u$'s that we denote as $\hat{u}_{ij}$'s, and indeed the expressions for the $\hat{u}_{ij}$ expressed in terms of $y_{2}, y_3$ and the $p_{i j}$'s are unchanged and the $\B$ will still be a simplex. But the magical ``binary" property is deformed for $a \neq 1$. For instance we find that 
\begin{equation}
{\hat u}_{13} + {\hat u}_{24} \hat{u}_{25} = \frac{1 + 2 y_2 + a y_2^2}{(1 + (a+1) y_2 + a y_2^2)}
\end{equation}
As $a\to 1$, the right hand side becomes exactly equal to one, but not for general $a$. Note interesting that the right-hand side is ``almost" equal to one, in the sense that it is given by a ratio of polynomials $P/Q$, whose Newton polytopes are identical, so that in all extremes of the domain of integration the ratio is indeed equal to one\footnote{Equivalently, we can say that the piecewise-linear functions $\Trop(P)$ and $\Trop(Q)$ are equal.}. Thus, we see that the $u$-variables can exists in a more general setting than (generalized) associahedra, but there is something {\it extra} special about the {\it perfectly} binary character of the $u$ equations in the case of cluster polytopes. 

\subsection{Example: $\G_+(3,6)/T$ vs. $D_4$}

Let us further illustrate these ideas by giving the $u$-variable description of the stringy canonical form for the $D_4$ cluster polytope, which can be specialized to give the form for the $\G_+(3,6)/T$ polytope. 

We follow the same steps described above in finding the ${D}_4$ polytope and the $u$ variables: we choose a positive parametrization of $\G_+(3,6)/T$ and perform the Minkowski sum of the summands associated with minors and the two cross-products. Doing this yields a simple polytope with 16 facets, exactly as needed for the big polyhedron ${\B}$ to be a simplex. We then invert to solve for the 16 $u$-variables. The integral is then given as 
\begin{equation}
{\I}_{{D}_4} = (\alpha^{\prime})^4 \int_{\R_{>0}^{4}} \omega \prod_{J=1}^{16} u_J^{\alpha^\prime X_J}
\end{equation}
where $\omega$ is the canonical form for $\G_+(3,6)/T$ and the 16 $u_I$ are given as

\begin{eqnarray}
&&u_{13} = \frac{(2\times3, 4\times5, 6 \times 1)}{(245)(136)},  u_{24}=\frac{(246)(345)}{(245)(346)},  
u_{31} = \frac{(1\times 2, 3 \times 4, 5 \times 6)}{(125)(346)},  u_{42} = \frac{(126)(135)}{(125)(136)} \nonumber\\
&&u_1 = \frac{(135)(456)}{(145)(356)},  u_2 = \frac{(2 \times 3, 4\times 5, 6\times1)}{(146)(235)},  u_3=\frac{(123)(246)}{(124)(236)}, u_4=\frac{(1\times 2, 3 \times 4, 5 \times 6)}{(134)(256)}  \nonumber\\
&&\tilde{u}_1 = \frac{(135)(234)}{(134)(235)},  \tilde{u}_2=\frac{(2\times 3, 4\times5, 6\times1)}{(145)(236)},
 \tilde{u}_3=\frac{(156)(246)}{(146)(256)},  \tilde{u}_4=\frac{(1\times2, 3\times4, 5\times6)}{(124)(356)} \nonumber\\
&&u_{12} = \frac{(235)(145)(136)}{(135)(2 \times 3, 4 \times 5, 6 \times 1)},  u_{23} = \frac{(236)(146)(245)}{(246) (2 \times 3 ,4 \times 5 ,6 \times 1)},\nonumber\\
&&u_{34} = \frac{(124)(256)(346)}{(246)(1\times2, 3\times 4, 5\times 6)},  u_{41} = \frac{(134)(356)(125)}{(135)(1\times2, 3\times4, 5 \times 6)} %\\\nonumber
\end{eqnarray}
The labels $u_i, \tilde{u}_i, u_{ij}$ for $i, j=1,\ldots, 4$ have been given for ease of comparison with the ${D}_4$ cluster algebra, which manifest a cyclic rotation $i \to i+1$; the reader unfamiliar with cluster algebras can just ignore this and treat them as dummy indices for the sixteen $u$'s. 

The $u$-variables further remarkably satisfy a set of 16 non-linear equations,  giving a ``binary" representation of the ${D}_4$ cluster polytope \cite{20193}:
\begin{align}
\begin{split}
u_1 + \tilde{u}_4 \tilde{u}_3 \tilde{u}_2 u_{23} u_{34} u_{24} = 1\,, \quad {\rm + cyclic}  \\ 
\tilde{u}_1 + u_4 u_3 u_2 u_{23} u_{34} u_{24} = 1, \quad {\rm + cyclic} \\ 
u_{42} + u_{13} u_{31} u_{23} u_{34} u_3 \tilde{u}_3 = 1, \quad {\rm + cyclic} \\ 
u_{12} + u_{23} u_{34}^2 u_{41} u_{24} u_{31} u_3 \tilde{u}_3 u_4 \tilde{u}_4=1\,, \quad {\rm + cyclic}
\end{split}
\end{align}
These equations capture all pairs of facets of the polytope that intersect non-trivially (called {\it compatible}). For instance, from the first equation, we learn that the exponent variable $X_1$ is incompatible with $\tilde{X}_3, \tilde{X}_2, X_{23}, X_{34}, X_{24}$ but compatible with the rest of the $X$ variables. 

From here we can immediately write the expression for $\alpha^\prime \to 0$ limit of the integral (the canonical function of the ${D}_4$ polytope): since the polytope is simple, it is just given by $\sum_{a,b,c,d} \frac{1}{X_a X_b X_c X_d}$ for all quadruplets of mutually compatible $(X_a, X_b,X_c,X_d)$, giving a sum of 50 terms, one for each vertex of the polytope:
\begin{eqnarray}
&&\frac{1}{X_1 X_2 X_3 X_4} + \frac{1}{X_{13} X_{31} X_1 X_3}+\frac{1}{X_{24} X_{42} X_2 X_4}+
\left(\frac{1}{X_{31} X_3 X_4 X_1} + {\rm cyclic} \right) \nonumber\\ 
&&+ \left(\frac{1}{X_1 X_2 X_{12} X_{13}}
+ \frac{1}{X_1 X_2 X_{12} X_{42}} + {\rm cyclic}\right)  + (X_i \leftrightarrow \tilde{X}_i)\nonumber\\ 
&&+\frac{1}{X_1 \tilde{X}_1} \left(\frac{1}{X_{31} X_{41}} + \frac{1}{X_{31} X_{13}} + \frac{1}{X_{12} X_{42}} + \frac{1}{X_{41} X_{42}} + \frac{1}{X_{12} X_{13}} \right) + {\rm cyclic}
\end{eqnarray}

From here, it is easy to get the canonical form for the polytope ${\cal P}(3,6)$ from the $\G_+(3,6)/T$ integral. Note that the power of $(1 \times 2, 3\times 4, 5 \times 6)$ in the integral is given by $X_{31} + X_4 + \tilde{X}_4 - X_{41} - X_{34}$, while that of $(2 \times 3, 4\times 5, 6 \times 1)$ is given by $X_{13} + X_2 + \tilde{X}_2 -X_{12} - X_{23}$. Thus to obtain the form for $\G_+(3,6)/T$, we simply take the above expression for ${D}_4$, and substitute $X_{31} \to X_{41} + X_{34} - X_4 - \tilde{X}_4$ and $X_{13} \to X_{12} + X_{23} - X_2 - \tilde{X}_2$. 

Note that the polytope ${\cal P}(3,6)$ is smaller than the ${D}_4$ polytope, with $48$ vertices and $98$ edges in 
place of the $100$ edges and $50$ vertices of ${D}_4$. Furthermore, while the ${D}_4$ polytope is simple, the polytope ${\cal P}(3,6)$ is not: two pairs of vertices connected by an edge in ${D}_4$ have been contracted to a point. See~\cite{BCL} for further related discussion.

 \subsection{Big polyhedron from tropical functions}

We consider a general integral of the form
\begin{equation}\label{eq:int}
\I(S) := (\alpha')^d \int_{\R_{>0}^d} \prod_{i=1}^d \frac{d x_i}{x_i} \prod_{J=1}^r p_J(\x)^{-S_J}
 \end{equation}
 where $p_J(\x)$ are subtraction-free Laurent polynomials.   Let $\PP = \PP_1+\PP_2 + \cdots +\PP_r$ be the Minkowski sum of the Newton polytopes $\PP_J = \New[p_J(\x)]$, and we assume that $\PP$ is full-dimensional.  Let $\N(\PP)$ denote the normal fan of $\PP$ and let $r_1,\ldots,r_N$ denote the lattice generators of the rays of $\N(\PP)$, where $N$ is also equal to the number of facets of $\PP$.  %We denote by $r_1,\ldots,r_s$ the lattice generators of these rays.
 
 Let us consider the vector space 
 \be
V_{\{p_J(\x)\}}= V_{\{p\}}:= \left\{\sum_{J=1}^r -S_J \Trop(p_J(\x)) \mid S_J \in \R \right\}
 \ee
 of piecewise-linear functions that are linear combinations of the functions $\Trop(p_J(\x))$.  Since each $f(\X) \in  V_{\{p\}}$ is linear when restricted to a maximal cone $C_v \in \N(\PP)$, the function $f$ is uniquely determined by its value $f(r_A)$ on each of the ray generators $r_A$.  Thus $V_{\{p\}} \subseteq \R^{N}$ is naturally embedded in a vector space of dimension $N$.  Let us call the set $\{p_J(\x)\}$ of Laurent polynomials a complete set for $\N$, or simply \emph{complete}, if $V_{\{p\}} = \R^{N}$, and \emph{irredundant} if $r = \dim(V_{\{p\}})$.  (If $\{p_J(\x)\}$ is redundant, we can always replace $\{p_J(\x)\}$ by an irredundant subset without changing $V_{\{p\}}$.)
 
 Now define the {\bf big polyhedron} as the cone of nonnegative tropical functions
 \be
\CC_{\{p_J(\x)\}}= \CC_{\{p\}}:= \left\{f \in V_{\{p\}} \mid f \geq 0 \right\} = V_{\{p\}} \cap \R_{\geq 0}^{N}.
 \ee
 The cone $\CC_{\{p\}}$ is exactly the cone of functions $\Trop(R(\x))$ for nearly convergent integrands $R(\x)$, as in Section \ref{ssec:trop}.  When $\{p_J(\x)\}$ is complete, the cone $\CC_{\{p\}}$ is simply the simplicial cone $\R_{\geq 0}^{N}$.  For each generator $r_A$, let us denote by $\delta_A$ the piecewise-linear function determined by
\be
\delta_A (r_B) = \begin{cases} 1 & \mbox{if $A= B$,} \\
0 & \mbox{otherwise.}
\end{cases}
\ee
The functions $\delta_A$ form a basis for $\CC_{\{p\}}$ when $\{p_J(\x)\}$ is complete.  In this case, we denote by $u_A$  the product of $p_J(\x)$-s satisfying $\Trop(u_A) = \delta_A$.  (In our examples, $u_A$ is a rational function, but this is not true in all cases.)
 
 \subsection{Simple polytopes and complete integrands}
 Let $\N$ be the normal fan of a (full-dimensional) lattice polytope.  We now prove that a complete set of subtraction-free Laurent polynomials $\{p_J(\x)\}$ can be found for $\N$ {\it exactly} when $\N$ is the normal fan of a simple polytope.  Suppose first that $\N$ is not simplicial, so that it has a maximal cone $C_v$ which has extremal rays with lattice generators $r_{A_1},\ldots,r_{A_c}$, where $c > d$.  A function $f$ is linear on $C_v$ if and only if the values $f(r_{A_1}),\ldots,f(r_{A_c})$ satisfy the same linear relations that $r_{A_1},\ldots,r_{A_c}$ do.  This shows that $V_{\{p\}}$ can never be equal to $\R^{N}$.

Conversely, let us suppose that $\N$ is simplicial.  If $\{p_J(\x)\}$ is not complete, then all vectors $g \in V_{\{p\}}$ satisfy some linear relation, say $L \cdot g = 0$.  Let $Q$ be a lattice polytope with normal fan equal to $\N(Q) = \N$.  After replacing $Q$ by $kQ$ for some large integer $k$, we can find a lattice polytope $Q'$ such that all but one chosen facet of $Q'$ and $Q$ are in the same position.  Denote by $\Trop(Q)$ the tropicalization of some monic Laurent polynomial $q(\x)$ satisfying $Q = \New[q(\x)]$.  If $L \cdot \Trop(Q) = 0$, it is easy to find such a deformation $Q'$ such that $L \cdot \Trop(Q') \neq 0$.  Thus, including $q'(\x)$ in our integrand will increase the dimension of $V_{\{p\}}$.  Repeating, this allows us to construct a complete set $\{p_J(\x)\}$.
 
We know of some natural examples of complete integrands.  First, any lattice polygon in $\R^2$ is simple, and thus we can always find a complete (and irredundant) integrand.  Next, as explained in Section~\ref{sec:u}, the open-string integrals are complete and the functions $u_A$ are the $u_{ij}$ given in \eqref{eq:uij}.  We shall further show in \cite{20201} that for any $\Phi$, the cluster string integral $\I_\Phi$ is complete and irredundant.  Here, we work out two examples, one cluster, and one not.
 
 \begin{example}[Cluster string integral for $B_2$]
 We consider the integral \eqref{eq:IB2}.
% \be
%\I_{B_2=C_2} = (\alpha')^2 \int_{\mathbb{R}_{>0}^2} \frac{d x~d y}{x~y} x^{\alpha' X} y^{\alpha' Y} (1{+}x)^{-\alpha' c_1} (1{+}y)^{-\alpha' c_2} (1{+}x{+}x y)^{-\alpha' c_3} (1{+}x{+}2 x y{+}x y^2)^{-\alpha' c_4}\,.
%\ee
As mentioned in Section~\ref{sec:cluster}, the Minkowski sum $\PP$ is the hexagon drawn in Figure \ref{fig:hexagon}.  
The tropicalizations of the six rational functions $x$, $y$, $1/(1+x)$, $1/(1+y)$, $1/(1+x+xy)$, and $1/(1+x+2xy+xy^2)$ are:
 \begin{align} \nonumber
 f_1 &= X & f_2& = Y &
 f_3 &= -\min(0,X) &
 \\ f_4 &=-\min(0,Y) &
  f_5 &= -\min(0,X,X+Y) & f_6& = -\min(0,X,X+2Y). 
  \end{align}
 The lattice generators of the rays of $\N$ are the following six vectors:
  \begin{align}
\nonumber
 r_1 &= (1,0) & r_2& = (0,1) & 
 r_3 &= (-1,0) &\\ r_4 &=(0,-1) &
  r_5 &= (1,-1) & r_6& =(2,-1).
  \end{align}
  The $6 \times 6$ matrix $M= (f_i(r_j))$ has determinant 1 and thus our integral is complete and irredundant as claimed.  $M$ and its inverse are given by
 \be
M =
\left[
\begin{array}{cccccc}
 1 & 0 & -1 & 0 & 1 & 2 \\
 0 & 1 & 0 & -1 & -1 & -1 \\
 0 & 0 & 1 & 0 & 0 & 0 \\
 0 & 0 & 0 & 1 & 1 & 1 \\
 0 & 0 & 1 & 1 & 0 & 0 \\
 0 & 0 & 1 & 2 & 1 & 0 \\
\end{array}
\right] \qquad \qquad M^{-1} =
\left[
\begin{array}{cccccc}
 1 & 0 & 0 & -2 & 0 & 1 \\
 0 & 1 & 0 & 1 & 0 & 0 \\
 0 & 0 & 1 & 0 & 0 & 0 \\
 0 & 0 & -1 & 0 & 1 & 0 \\
 0 & 0 & 1 & 0 & -2 & 1 \\
 0 & 0 & 0 & 1 & 1 & -1 \\
\end{array}
\right].
\ee
Reading the columns of $M$ we obtain the facet inequalities of the hexagon in Figure \ref{fig:hexagon}:
\begin{align}
\nonumber
X & \geq 0 & Y &\geq 0 & c_1+c_3+c_4- X &\geq 0  \\
 c_2+c_3+2c_4 - Y &\geq 0  &
X-Y+c_2+c_4 &\geq 0 & 2X-Y+c_2 &\geq 0
\end{align}
Reading the rows of $M^{-1}$ we obtain the $u$-variables:
\begin{align}
\nonumber
u_1 &= \frac{x(1+y)^2}{1+x+2xy+xy^2} & u_2&=\frac{y}{1+y} & u_3&=\frac{1}{1+x} \\
u_4&=\frac{1+x}{1+x+xy} & u_5&=\frac{(1+x+xy)^2}{(1+x)(1+x+2xy+xy^2)} & u_6 &=\frac{1+x+2xy+xy^2}{(1+y)(1+x+xy)}
\end{align}
Thus for example $\Trop(u_1) = X + 2\min(0,Y)-\min(0,X,X+2Y) = \delta_1$ takes value 1 on $r_1$ and 0 on the other $r_A$-s.
A direct calculation shows that we have
\be \label{eq:uB2}
u_1+u_3u_4^2 u_5 =1 \qquad \text{ and } \qquad u_2 + u_4u_5u_6 = 1
\ee
and the same identity cyclically shifting indices by two.  These equations are the $B_2 = C_2$-analogues of \eqref{eq:u+u} and will be established for general cluster string integrals in \cite{20193,20201}.
\end{example}
 \begin{example}[Heptagon] 
We give a non-cluster example that is complete and irredundant.  Consider the integral
 \be
\I =  (\alpha')^2 \int_{\R_{>0}^2} \frac{dx}{x} \frac{dy}{y} x^A y^B (1+x)^C (1+y)^D (1+x+y)^E (1+x+xy)^F (1+y+xy)^G.
 \ee
 The normal fan $\N$ is drawn in Figure \ref{fig:heptagon}.
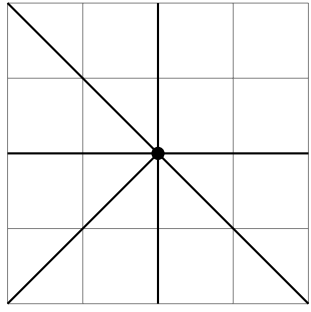
\begin{figure}
\begin{center}
\begin{tikzpicture}
\draw [help lines, step=1cm] (-2,-2) grid (2,2);
\node[fill=black,circle, inner sep=0pt,minimum size=5pt] at (0,0) {};
\draw[thick] (0,0) --(2,0);
\draw[thick] (0,0) --(-2,0);
\draw[thick] (0,0) --(0,-2);
\draw[thick] (0,0) --(0,2);
\draw[thick] (0,0) --(2,-2);
\draw[thick] (0,0) --(-2,-2);
\draw[thick] (0,0) --(-2,2);
\end{tikzpicture}
\caption{The normal fan of the heptagon.}
\label{fig:heptagon}
\end{center}
\end{figure}
 The tropicalizations of the seven polynomials are the following seven functions:
 \begin{align}
 \nonumber
 f_1 &= X & f_2& = Y &
 f_3 &= \min(0,X) & f_4 &=\min(0,Y) \\
  f_5 &= \min(0,X,Y) & f_6& = \min(0,X,X+Y) &
 f_7 &= \min(0,Y,X+Y). 
 \end{align}
 The lattice generators of the rays of $\N$ are the following vectors:
  \begin{align}
  \nonumber
 r_1 &= (1,0) & r_2& = (0,1) & 
 r_3 &= (-1,1) & r_4 &=(-1,0) \\
  r_5 &= (-1,-1) & r_6& =(0,-1)&
 r_7 &= (1,-1).
  \end{align}
  The $7 \times 7$ matrix 
 \be 
\left(f_i(r_j)\right) =\left[
\begin{array}{ccccccc}
 1 & 0 & -1 & -1 & -1 & 0 & 1 \\
 0 & 1 & 1 & 0 & -1 & -1 & -1 \\
 0 & 0 & -1 & -1 & -1 & 0 & 0 \\
 0 & 0 & 0 & 0 & -1 & -1 & -1 \\
 0 & 0 & -1 & -1 & -1 & -1 & -1 \\
 0 & 0 & -1 & -1 & -2 & -1 & 0 \\
 0 & 0 & 0 & -1 & -2 & -1 & -1 \\
\end{array}
\right]
 \ee
is non-singular with determinant $-1$ and thus our $\{p_J(\x)\}$ is complete and irredundant.  Inverting this matrix and reading the rows, the rational functions $u_A$, $A=1,2,\ldots,7$ are given by the following subtraction-free rational functions:
    \begin{align}\nonumber
 u_1 &= \tfrac{x(1+y)}{1+x+xy} & u_2& = \tfrac{y(1+x)}{1+y+xy} & 
 u_3 &= \tfrac{1+y+xy}{(1+x)(1+y)}& u_4 &=\tfrac{(1+x)(1+y)^2}{(1+x+y)(1+y+xy)}\\
 u_5 &= \tfrac{1+x+y}{(1+x)(1+y)} & u_6& =\tfrac{(1+x)^2(1+y)}{(1+x+y)(1+x+xy)}&
 u_7 &= \tfrac{1+x+xy}{(1+x)(1+y)}.
  \end{align}
  Thus for example $\Trop(u_3) = \min(0,Y,X+Y)-\min(0,X)-\min(0,Y)= \delta_3$ takes the value $1$ on $r_3$ and $0$ on all other $r_A$.
 \end{example}
% 
% {{1, 0, 0, 1, 0, -1, 0}, 
% {0, 1, 1, 0, 0, 0, -1},
%  {0, 0, -1, -1, 0, 0, 1},
%   {0, 0, 1, 2, -1, 0, -1}, 
%   {0, 0, -1, -1, 1, 0, 0}, 
%   {0, 0, 2, 1, -1, -1, 0}, 
%  {0, 0, -1, -1, 0, 1, 0}}
%\subsection{Cluster string integrals revisited}\label{sec:cluster2}
%We shall show in \cite{20201} that for any $\Phi$, the cluster string integral $\I_\Phi$ is complete and irredundant.  ...
%
%In section \ref{sec:big} , we will see how special these stringy canonical forms for finite-type cluster algebra are: not only we have $N=d{+}m$ and they correspond to simple polytopes, but by considering the natural ``big polytope" we find dual variables which manifest residues of the integral at finite $\alpha'$. The properties of these integrals are as nice as the open-string amplitudes with factorizations associated with finite-type cluster algebra.

 \section{Tropical compactification}\label{sec:10}
The moduli space $\M_{0,n}(\R)$ of $n$-points on the Riemann sphere has a well-known (Deligne-Knudsen-Mumford) compactification $\oM_{0,n}(\R)$, which is smooth with normal-crossing divisors, the latter being an algebro-geometric avatar of the fact that the associahedron is a simple polytope.  The complement $\oM_{0,n}(\R)\setminus \M_{0,n}(\R)$ consists of $2^{n-1}{-}n{-}1$ boundary components, of which only $n(n{-}3)/2$ (those that touch $\M_{0,n}^+$ in codimension one) are associated to the facets of the associahedron $\A_{n-3}$.  The union of $\M_{0,n}(\R)$ and these $n(n{-}3)/2$ is a space $\M'_{0,n}(\R)$ satisfying
\be
\M_{0,n}(\R) \subsetneq \M'_{0,n}(\R) \subsetneq \oM_{0,n}(\R).
\ee
The space $\M'_{0,n}(\R)$ has a stratification with the same combinatorics as that of the associahedron $\A_{n{-}3}$, and was studied in \cite{Brown:2009qja}.  In this section, we construct two spaces $U^\circ \subset U$, associated to any stringy canonical form which are analogues of the two spaces $\M_{0,n} \subsetneq \M'_{0,n}$.  In particular, we obtain purely ``synthetic" constructions of $\M_{0,n}$ and $\M'_{0,n}$ with only the string integral $\bI_n$ as input.

\subsection{$U$, $U^\circ$ and $U_{\geq 0}$}
We consider a general integral of the form \eqref{eq:int}.
%\begin{equation}\label{eq:int}
%\I(S):= \int_{\R_{>0}^d} \prod_{i=1}^d \frac{d x_i}{x_i} \prod_{J=1}^r p_J(\x)^{-S_J}
% \end{equation}
% where $p_J(\x)$ are subtraction-free Laurent polynomials. 
Let $\PP = \PP_1+\PP_2 + \cdots +\PP_r$ be the Minkowski sum of the Newton polytopes, and we assume $\PP$ to be full-dimensional.  The integrand $R(\x) =  \prod_J p_J(\x)^{-S_J}$ is called {\it convergent} (respectively, {\it nearly convergent}) if $\Trop(R(\x))$ is positive (respectively, nonnegative), and it is called {\it integral} if $R(\x)$ belongs to the ring of rational functions $\C(\x)$.
 For example, if $p(\x) = x^2$, then $
 R(\x) = p(\x)^{1/2}$ would still be integral.  We now define two subrings $\C[U] \subset \C[U^\circ]$ inside the ring of rational functions $\C(\x)$: 
\begin{align}
\begin{split}
\C[U] &:= \C[R(\x) \mid \text{$R(\x)$ is integral and nearly convergent}] \subset \C(\x)\\
\C[U^\circ] &:= \C[R(\x) \mid \text{$R(\x)$ is integral}] \subset \C(\x).
\end{split}
\end{align}
And let $U = \Spec(\C[U])$ and $U^\circ:= \Spec(\C[U^\circ])$ be the corresponding affine schemes.

%\subsection{Compactification}
For simplicity, we now suppose that $p_J(\x) = x_J$ for $J=1,2,\ldots,d$.  Then we have
$
\C[U^\circ] = \C[U][x_1^{\pm 1},\ldots,x_d^{\pm1} ]
$.
In particular, $U^\circ$ is an open subset of the torus $(\C^\times)^d$ whose coordinates are $x_1,\ldots,x_d$, and both $U$ and $U^\circ$ are irreducible.  Since we assumed $p_J(\x)$ to be subtraction-free, any monomial $R(\x)$ is defined on the positive part $\R_{>0}^d \subset (\C^\times)^d$.  We conclude that 
\be
\R_{>0}^d \subset U^\circ(\R) \subset U(\R).
\ee
Let $U_{\geq 0}$ denote the closure of $\R_{>0}^d$ in $U(\R)$.  We call $U_{\geq 0}$ the {\it tropical compactification}.  As we shall show in Section \ref{sec:Utoric}, the affine variety $U$ is an affine open subset of the projective toric variety $X_\PP$ and $U_{\geq 0}$ is diffeomorphic to the polytope $\PP$.  The rings $\C[U]$ and $\C[U^\circ]$ are finitely-generated integral domains.  %(see Section \ref{sec:Utoric}).

\begin{example}\label{ex:stringU}
Let $\I = \bI_n$ be the open-string integral.  As explained in Section \ref{sec:big}, the nearly convergent integrands in this case form a simplicial cone, with generators given by the $u_{ij}$ of \eqref{eq:uij}.  In this case, we have $\C[U] = \C[u_{i\,j}]$ and $\C[U^\circ] = \C[u_{i\,j}^{\pm 1}]$, and the two spaces $U$ and $U^\circ$ are $\M'_{0,n}$ and $\M_{0,n}$ respectively.  The tropical compactification $U_{\geq 0}$ is equal to the closure of $\M_{0,n}^+$ in $\oM_{0,n}(\R)$ and is diffeomorphic to an associahedron $\A_{n-3}$.  The ring $\C[U]$ has the presentation 
\be
\C[U] = \C[u_{i\,j}]/(u_{i\,j} + \prod_{(k\,l) \, \text{crossing} \, (i\,j)} u_{k\,l} - 1)
\ee
where the ideal is generated by the relations \eqref{eq:u+u}.  The ring $\C[U]$ is defined using a larger set of ideal generators in \cite{Brown:2009qja}.  See \cite{20201} for further discussion on this point.
\end{example}

\subsection{Stratification}\label{sec:Ustrata}
The space $U_{\geq 0}$ has a stratification 
\be \label{eq:Ustrata}
U_{\geq 0} = \bigsqcup_{F \text{ a face of } \PP} U_F
\ee
indexed by faces $F$ of $\PP$.  The strata $U_F$ are defined as follows.  Recall the cones $C_F$ of the  normal fan $\N(\PP)$ defined in Section~\ref{ssec:decomp}.  A point $u$ belongs to $U_F$ if it satisfies the following condition for any nearly convergent monomial $R(\x)$,
\be
R(u) \text{ is }  \begin{cases} 0& \mbox{if $\Trop(R(\x)) >0$ on the relative interior of the cone $C_F$} \\
> 0 & \mbox{if $\Trop(R(\x)) = 0$ on the relative interior of the cone $C_F$}.
\end{cases}
\ee
%For any convergent monomial $M$, if $\Trop(M) > 0$ on the interior of the cone $C_F$ where $v \in F$, then $M$ takes a constant nonzero value on $U(F)$.
%\item If $\Trop(M) > 0$ on the interior of all the maximal cones $C_v$ where $v \in F$, then $M$ vanishes on $U(F)$.
%\item If $\Trop(M) = 0$ on the interior of some maximal cones $C_v$ for $v \in F$ and $\Trop(M) > 0$ on the interior of some maximal cones $C_v$ for $v \in F$, then $M$ takes varying values on $U(F)$.
Here, we consider $\PP$ to be a face of $\PP$ with $C_\PP = \{0\}$, and we have $U_P =U_{>0}=\R_{>0}^d$.  We can think of the points in $U_F$ as follows.  Let $\bla = (\lambda_1,\ldots,\lambda_d)$ be an integer point in the relative interior of $C_F$.  Then $u = \lim_{t \to \infty}  (t^{\lambda_1},\ldots,t^{\lambda_d}) \in U_{\geq 0}$ belongs to $U_F$.  In the other direction, suppose that we are given a point in $u \in U_{\geq 0}$.  Then we can find an analytic curve $\gamma: [1,\infty) \to U_{>0}$ such that $\lim_{t \to \infty} \gamma(t) = u$.  Then $\gamma(t) = (c_1 t^{\lambda_1} + O((1/t)^{1-\lambda_1 }),\ldots, c_d t^{\lambda_d} + O((1/t)^{1-\lambda_d }))$ where $c_i > 0$ and if $\bla=(\lambda_1,\ldots,\lambda_d)$ lies in the relative interior of $C_F$, we have $u \in U_F$.  Another description of this same stratification will be explained in Section \ref{sec:Utoric}.

\subsection{The cluster configuration space}\label{sec:clusterU}
Let us now take $\I = \I_\Phi$ to be the cluster string integral \eqref{cluster}.  We obtain two spaces $\M'_\Phi = U$ and $\M_\Phi = U^\circ$ that we call {\it cluster configuration spaces}, generalizing the construction of $\M'_{0,n}$ and $\M_{0,n}$ of Example \ref{ex:stringU}.  These spaces will be studied in \cite{20201}, but here we list some of the main properties.  Both $\M_\Phi$ and $\M'_\Phi$ are affine algebraic varieties, with the latter being a partial compactification of the former.  The space $\M'_\Phi$ is equipped with a stratification that is indexed by faces of the generalized associahedron $\PP(\Phi)$: the top-dimensional open stratum is simply $\M_\Phi$ and there are $N(\Phi)$ codimension one strata, indexed by the facets of $\PP(\Phi)$.  The space $\M'_\Phi$ is smooth, and the boundary stratification is simple normal-crossing.

The ring $\C[\M'_\Phi] = \C[U]$ is generated by distinguished variables $u_\gamma$, for $\gamma \in \Gamma$ (in bijection with the cluster variables $x_\gamma$ of $\A(\Phi)$).  The $u_\gamma$ take values in $[0,1]$ on $U_{\geq 0}$ and satisfy relations of the form
\be \label{eq:uu1}
u_\gamma + \prod_\delta u_\delta^{a_{\gamma \delta}} = 1
\ee
generalizing \eqref{eq:u+u} and \eqref{eq:uB2}.  Here, $a_{\gamma \delta} \in \Z_{\geq 0}$ is the {\it compatibility degree}.  The stratification \eqref{eq:Ustrata} follows from \eqref{eq:uu1}: if $u_\gamma =0$ then $u_\delta= 1$ for all $\delta\in \Gamma$ such that $a_{\gamma\delta} \neq 0$.  Thus, if we have both $u_\gamma = 0$ and $u_{\gamma'} = 0$, then $\gamma$ and $\gamma'$ must index {\it compatible} cluster variables.  A face $F \subset \PP(\Phi)$ of the generalized associahedron corresponds to a set $\{\gamma_1,\gamma_2,\ldots,\gamma_c\}$ of compatible cluster variables; the stratum $U_F$ is then given by $u_{\gamma_1} = u_{\gamma_2} = \cdots = u_{\gamma_c} = 0$.

\subsection{Proof of properties of $U$ and $U_{\geq 0}$}

%\subsubsection{$U$ is an affine open in a projective toric variety}
\label{sec:Utoric}

Associated to any lattice polytope $\Q$ is a projective toric variety $X_\Q$ that depends only on the normal fan of $\Q$.  In particular, $X_\Q = X_{r\Q}$ for any integer $r > 0$. A lattice polytope $\Q$ is called {\it very ample} if for sufficiently large integers $r > 0$, every lattice point in $r\Q$ is a sum of $r$ (not necessarily distinct) lattice points in $\Q$.  When $\Q$ is very ample with lattice points $\v_1,\v_2,\ldots,\v_r$, we can describe $X_\Q$ as the closure of the image of the map
\be
(\C^\times)^d \to \PPP^{r-1} \qquad \text{given by} \qquad (x_1,\ldots,x_d) \mapsto [\x^{\v_1}: \x^{\v_2}: \cdots: \x^{\v_r}]
\ee
inside $\PPP^{r-1}$.
It is known that if $\Q$ is any lattice polytope, then $k\Q$ is very ample for some $k$.  Thus by replacing $\Q$ by $k\Q$, any projective toric variety can be described as the closure of a monomial map in the above way.  

Now suppose that $\Q$ is very ample and fix $Q(\x)$ to be a subtraction-free Laurent polynomial with Newton polytope equal to $\Q$.  The equation $Q(\x) = 0$ cuts out a hypersurface in $X_\Q$, which is given by the intersection of $X_\Q$ and a hyperplane $H_Q$ inside $\PPP^{r-1}$.  Let us denote by $X'_\Q := X_\Q \setminus H_Q$ the open subset of $X_\Q$ where $Q(\x)$ does not vanish.  The variety $X'_\Q$ is affine: it is a closed subvariety of the affine space $\C^{r-1} \subset \PPP^{r-1}$ given by $H_Q \neq 0$.  If $y_1,\ldots,y_r$ denote the homogeneous coordinates of $\PPP^{r-1}$, the coordinate ring of the affine space $\{H_Q \neq 0\} \simeq \C^{r-1}$ is given by $\C[y_1/H_Q,\ldots,y_r/H_Q]$ (these generators satisfy a single linear relation).  Thus the coordinate ring of $X'_Q$ is given by 
\begin{equation}\label{eq:generators}
\C[X'_\Q] = \C\left[\frac{\x^{\v_1}}{Q(\x)},\ldots,\frac{\x^{\v_r}}{Q(\x)}\right] \subset \C(\x),
\end{equation}
where $\v_1,\v_2,\ldots,\v_r$ are the lattice points in $\Q$.

Let $p(\x)$ be the product of all the Laurent polynomials $p_J(\x)$.  Its Newton polytope is $\PP$.  Let $r$ be such that $r\PP$ is very ample.  We claim that
\be \label{eq:CU}
\C[U] = \C\left [\frac{\x^\uu}{p(\x)^r} \mid \uu \text{ is a lattice point in } r\PP\right].
\ee
Since $\x^\uu/p(\x)^r$ is clearly a nearly convergent monomial, the RHS of \eqref{eq:CU} is contained in the LHS.  Now suppose that $R(\x) = A(\x)/B(\x)$ is any nearly convergent integrand.  The denominator $B(\x)$ divides $p(\x)^{kr}$ for sufficiently large $k$, so we may write $R(\x) = \tA(\x)/p(\x)^{kr}$ for some polynomial $\tA(\x)$.  Let $\x^\uu$ be some monomial appearing in $\tA(\x)$.  Then $\uu$ is a lattice point in $kr\PP$, and since $r\PP$ is very ample, we have $\uu= \uu_1+\uu_2+\cdots +\uu_k$ where $\uu_i$ are lattice points in $r\PP$.  Thus 
$\x^\uu/p(\x)^{kr} = (\x^{\uu_1}/p(\x)^{r}) ( \x^{\uu_2}/p(\x)^{r}) \cdots( \x^{\uu_k}/p(\x)^{kr})$ belongs to the RHS, and summing over monomials in $\tA(\x)$, we obtain the equality of \eqref{eq:CU}.

Setting $\Q = r\PP$, we conclude from \eqref{eq:generators} that $U$ is isomorphic to the affine open subvariety $X'_\PP$ of $X_\PP = X_{r\PP}$ given by $p(\x)^r \neq 0$.  Furthermore, $U^\circ \subset U$ is the intersection $X'_\PP \cap T$ where $T$ denotes the open subtorus in $X_\PP$.  We have shown that $U$ is an affine open subvariety of the toric variety $X_\PP$.

The nonnegative part $X_{\PP,\geq 0}$ of $X_\PP$ is the closure of $\R^d_{>0} \subset T$ inside $X_\PP$.  Since $p(\x)$ is subtraction-free with Newton polytope equal to $\PP$, the function $p(\x)$ does not vanish on $X_{\PP, \geq 0}$.  Thus
under the isomorphism $U \simeq X'_\PP$, the tropical compactification $U_{\geq 0}$ is identified with the positive part $X_{\PP,\geq 0}$.  It is well known (see~\cite{Fulton, Arkani-Hamed:2017tmz}) that $X_{\PP, \geq 0}$ is diffeomorphic to the polytope $\PP$, and is in particular compact.  The faces of $\PP$ correspond to the strata $U_F$ described in Section \ref{sec:Ustrata}.  We remark that $X_\PP$ is smooth, and its torus orbit closure stratification is simple normal-crossing, exactly when the normal fan $\N(\PP)$ is ``smooth".  This is not the case for all polytopes, and indeed, $U$ is not always smooth, nor is the boundary stratification always simple normal-crossing.

\section{Conclusion and Discussion}

We have proposed for any (rationally-realizable) polytope, a class of integrals which provide a new intrinsic definition of it's canonical form, naturally deformed by a ``string scale" $\alpha'$, with many properties reminiscent of string amplitudes. Both the convergence properties and the leading order in $\alpha'$ expansion of the stringy canonical form  \eqref{eq:stringy} are controlled by the Minkowski sum of the Newton polytopes of its regulating polynomials. The integral shares properties of ordinary string amplitudes at finite $\alpha'$, and the $\alpha'\to 0$ limit gives the canonical form of the polytope we are interested in. This is equivalently computed as the volume of the dual polytope, which is a halfspace cut out by the tropicalization of the integrand. The ``scattering equations" from $\alpha'\to \infty$ provide a diffeomorphism from the integration domain to the polytope, and thus a pushforward formula for its canonical form.  Among the new integrals we propose are the cluster string integrals that will be studied further in \cite{20193, 20201}, and the Grassmannian string integrals \cite{ALS}.

There are numerous unanswered questions and new avenues for explorations exposed by our preliminary investigations. To begin with, while we provide a geometric understanding for the leading order of any stringy integral, there is an obvious question about higher orders in the Laurent expansion in $\alpha'$. As familiar from string amplitudes, the higher order terms in the expansion involve {\it periods} and are transcendental rather than rational numbers. Unlike the leading order, they depend on coefficients in the integrands and cannot be determined by convex geometry alone: we expect that these higher orders have both a combinatorial component and a transcendental component. It would be fascinating to understand how the number-theoretic structure in the $\alpha'$-expansion of the integral is encoded in the polynomials. For example, for what polynomials do we have multiple zeta values only (similar to the case of string amplitudes)? Of particular interests are the cluster string integrals, and preliminary results indicate that this is the case at least for the type $B_d$ and $C_d$ cluster string integrals. One can study even simpler situations, such as \eqref{eq:main}  (just one polynomial and no $\prod x^{\alpha' X}$ factors), then the integral is simply a function of $\alpha'$ (one can set $S=1$) and at each order in the $\alpha'$ expansion we have a {\it transcendental number} that depends on the polynomial. It is an intriguing question to determine these numbers.

Our investigations of closed stringy integrals, \eqref{eq:closed_int}, as well as open-stringy integrals with shifts, \eqref{eq:openshift}, are rather preliminary. It would be instructive to write general open- and closed-stringy integrals in this way. Indeed even for the usual string integrals over ${\cal M}_{0, n}$ this rewriting is useful. For example, \eqref{eq:closed_int}  and \eqref{eq:openshift} also apply to open and closed-string integrals on ${\cal M}_{0,n}$ where we integrate a form (or both forms for closed-string case) that is not the Parke-Taylor form,  {\it e.g.} those for Cayley polytopes~\cite{Gao:2017dek, He:2018pue}. 

Another fascinating question is if we can understand the geometric structures underlying the open superstring amplitudes with $n$ gluons, which are linear combinations of our open-string integrals. Also since closed string amplitudes can be obtained as single-valued projection of open-string ones at all order in $\alpha'$~\cite{Schlotterer:2018zce, Brown:2018omk}, it would be interesting to study possible generalizations to any stringy integrals, as we have seen that they agree at leading order (trivial case of the projection).

We have shown that the most basic and fundamental phenomena related to scattering equations are easily conceptually understood in the most general setting of stringy canonical forms, but there is still a great deal to be understood about the special cases of interest in physics.  While the CHY formula computing the leading order of open- and closed-string amplitudes is a special case of this general phenomenon, there is still something even more special about the string integral: the number of solutions of the scattering equations for the ``real" string integrals--given by $(n{-}3)!$--are strikingly smaller than the number obtained from more generic realizations of the associahedron. We conjecture that this number is in fact the smallest possible for any stringy canonical forms reproducing ABHY associahedron (bi-adjoint $\phi^3$ amplitudes). It is plausible that the same is true for cluster string integrals, and it would be interesting to prove this. We leave the discussion of topological properties for cluster configuration space to~\cite{20193, 20201}, and only listed the number of saddle points (or the dimension of twisted (co-) homology group) for certain cases (including Grassmannian cases). Moreover, the pullback of the pushforward formula to a subspace gives CHY-like formulas, which is closely related to intersection theory~\cite{Mizera:2017rqa}. The latter has been successfully applied to the study of Feynman integrals ({\it c.f.}~\cite{Mastrolia:2018uzb, Frellesvig:2019uqt}), and there are likely applications of stringy canonical forms in that setting.

Another obvious set of questions concerns the behavior of stringy integrals as meromorphic functions at finite $\alpha'$. We have understood the residues of the integral for any massless pole at finite $\alpha'$. However, it remains an important open question to determine the locations and residues of all the infinite poles.  All the poles for real stringy integrals are of the form $X=m$ for $m\in \Z$ (where $X=0$ is a massless pole corresponding to a facet $F$ of $\PP$)~\cite{Mellin, berkesch2014}, and it would be interesting to study the residues at these ``massive" poles. For the closed-stringy cases, even the locations of the poles need to be investigated. We have presented recurrence relations for (open) stringy integrals (there are similar relations for closed stringy integrals), and it would be very interesting to study them further, and perhaps even solve them in closed form. These relations are not only important conceptually, but they provide an extremely useful way for evaluating the integrals, both analytically and numerically.

Finally returning to the underlying physical motivations of this work, what is the physical meaning of stringy canonical forms, beyond string amplitudes?  Given that the cluster integrals for type $B$, $C$ and $D$ factorize nicely at finite $\alpha'$, and produce the integrands for one-loop tadpole diagrams and one-loop planar $\phi^3$ diagrams as leading order respectively, it would be fascinating to see if they have any direct relationship to ordinary string amplitudes at genus one.  Other potential applications include  stringy integrals for cosmological polytopes~\cite{Arkani-Hamed:2017fdk} (which are naturally associated with an especially simple set of polynomials for polytopes in the Minkowski sum), and also those for unbounded polyhedra for bi-color amplitudes with particles in (anti-)fundamental representations~\cite{Herderschee:2019wtl}. It would also be interesting to study the stringy integrals applied to Stokes polytopes for $\phi^4$ interactions~\cite{Banerjee:2018tun} and find a physical interpretation at finite $\alpha^\prime$. 

Finally, we have defined stringy canonical forms for polytopes, but some of the most interesting examples of positive geometries and canonical forms go beyond polytopes, as with the amplituhedron. It is natural to ask how to write string-like integrals with leading order given by the canonical form of a positive geometry in this most general setting.  A fascinating possibility would be a ``stringy integral" or $\alpha'$ deformation for the ${\cal N}=4$ SYM amplitude, starting at tree level; the latter is naturally written as a differential form which can be obtained as pushforward from G$_+(2,n)$ to the momentum space~\cite{He:2018okq}. 
A natural conjecture is to write an integral over G$_+(2,n)$, with regulators whose saddle-point equations correspond to the four-dimensional scattering equations~\cite{Roiban:2004yf}. 
%even beyond ({\it e.g.} ${\cal M}_{g,n}$)?

\section*{Acknowledgements}
We would like to thank Hugh Thomas, Giulio Salvatori, Marcus Spradlin, Zhenjie Li and Chi Zhang for interesting discussions, and especially to Erik Panzer for useful comments and references. We also thank a referee for an exceptionally detailed report on the first version, which helped greatly improving the quality of the manuscript. S.H. thanks the Institute for Advanced Study, Princeton and the Center for Mathematical Sciences and Applications, Harvard for hospitality during various stages of the work. SH's research is supported in part by the Thousand Young Talents program, the Key Research Program of Frontier Sciences of CAS under Grant No. QYZDBSSW-SYS014, Peng Huanwu center under Grant No. 11747601 and National Natural Science Foundation of China under Grant No. 11935013.  T.L. was supported by NSF DMS-1464693 and by a von Neumann Fellowship from the Institute for
Advanced Study.

\appendix

\section{Canonical forms of polytopes}\label{sec:canform}

In this Appendix, we show how to compute the canonical form in various ways for a quadrilateral, which serves as a simple toy model to illustrate the main ideas of~\cite{Arkani-Hamed:2017tmz} (more details can be found in Section 7 of the reference). 

Let's first recall the definition and a few simplest examples of the canonical forms and positive geometries.  The canonical form of a $d$-dim positive geometry ${\PP}$ (for example, a polytope) is defined to be the unique rational $d$-form $\Omega(\PP)$ whose only singularities are logarithmic singularities on the boundaries of $\PP$; each boundary component of $\PP$, $\partial \PP$, is a positive geometry of dimension $d{-}1$ and the recursive definition requires the residue to be 
\be
{\rm Res}_{\partial {\cal P}} \Omega({\cal P})=\Omega (\partial P)\,.
\ee
The $0$-dimensional positive geometries are points and the canonical $0$-form is either the function $1$ or $-1$ (depending on the orientation).  A simple example of a positive geometry is the line segment (or interval) $[a,b] \subset \mathbb{RP}^1$. The canonical form is 
\be
\Omega([a, b])=\frac {d x}{x-a} - \frac {d x}{x-b}=\frac{(b-a) d x}{(b-x) (x-a)}=d\log \frac{x-a}{x-b}\,, 
\ee
which (only) has logarithmic singularities at $x=a$ and $x=b$, with residues $\pm 1$; the latter is the canonical form of a point ($0$-form) depending on the orientation. For a projective $d$-simplex, $\Delta \subset \mathbb{RP}^d$ which can be cut out by $d{+}1$ inequalities as 
\be
\Delta=\{ Y \in \mathbb{RP}^d~|~Y \cdot W_i \geq 0~{\rm for}~i=1,2,\ldots, d{+}1\}\footnote{These inequalities are interpreted as follows: we have $Y \in \Delta$ if there exists a vector $\tilde Y \in \R^{d+1} \setminus \{0\}$ representing $Y \in \mathbb{RP}^d$ satisfying $\tilde Y \cdot W_i \geq 0$ for $i=1,2,...,d+1$ simultaneously.} 
\ee
where the $d{+}1$ dual vectors $W_1, \ldots, W_{d{+}1}$ are the facets of the simplex. It is straightforward to see that the canonical form can be written as
\be
\Omega(\Delta)=\frac{\langle W_1 W_2 \cdots W_{d{+}1}\rangle~\langle Y d^d Y\rangle}{d!~(Y \cdot W_1) (Y\cdot W_2) \cdots (Y\cdot W_{d{+}1})}\,.
\ee
Equivalently we can write the form in terms of the $d{+}1$ vertices of $\Delta$, which we denote as $Z_1, Z_2, \ldots, Z_{d{+}1}$. If the facet $W_i$ contains the vertices $Z_{i{+}1}, Z_{i{+}2}, \ldots, Z_{i{+}d}$ , then we find that, up to an overall sign that depends on $d$, the canonical form $\Omega(\Delta)$ reads
\be
[1,2,\ldots, d{+}1]:=\frac{ \langle Z_1 Z_2 \cdots Z_{d{+}1} \rangle^d~\langle Y d^d Y\rangle}{d!~\langle Y Z_1 \cdots Z_d \rangle \langle Y Z_2 \cdots Z_{d{+}1}\rangle \cdots \langle Y Z_{d{+}1} \cdots Z_{d{-}1}\rangle}\,.
\ee
Here we have introduced a notation for the canonical form of the simplex with vertices $Z_1, \ldots, Z_{d{+}1}$. For convenience we will also define {\it e.g.} $\langle 1 2 \cdots d{+}1 \rangle= \langle Z_1 Z_2 \cdots Z_{d{+}1} \rangle$. 

Now let's turn to the example of a (projective) quadrilateral, which we denote as ${\A}={\cal A}(Z_1, Z_2, Z_3, Z_4)$, where $Z_i \in \mathbb{R}^3$ for $i=1,2,3,4$ are the vertices, and we have $\langle i\,j\,k\rangle >0$ for $i<j<k$. The quickest way to compute its canonical form is by triangulating the quadrilateral. For example, $\A$ can be triangulated into two triangles using one of the diagonals, say, one with vertices $Z_1, Z_2, Z_3$ and the other with vertices $Z_3, Z_4, Z_1$. The canonical form for the quadrilateral is given by the sum $\Omega({\cal A})=[1,2,3]+[3,4,1]$ of the canonical forms of the two triangles,
\be
\Omega({\cal A})= \frac {\langle Y d^2 Y\rangle} 2 \left(\frac{\langle 1 2 3\rangle^2}{\langle Y 1 2\rangle \langle Y 2 3\rangle  \langle Y 3 1 \rangle} + \frac{\langle  3 4 1 \rangle^2}{\langle Y 3 4 \rangle \langle Y 4 1 \rangle\langle Y 1 3 \rangle}\right)\,,
\ee
where the pole $\langle Y 1 3 \rangle$ is spurious and should be cancelled in the sum. By putting the denominator together we see that indeed it is cancelled, and we find
\be
\frac{\langle Y d^2 Y\rangle \langle Y (12)\cap (34) (23) \cap (41)\rangle}{2 \langle Y 1 2 \rangle \langle Y2 3 \rangle \langle Y 3 4 \rangle \langle Y 4 1 \rangle }\,,
\ee
where in the numerator we have the intersection of two lines $(12) \cap (34):= Z_1 \langle 2 3 4 \rangle- Z_2 \langle 1 3 4 \rangle$ and similarly for $(23) \cap (41)$.  Indeed, the numerator is the unique combination to kill potentially non-vanishing residue at spurious singularities at the two intersections, {\it i.e.} $\langle Y12\rangle=\langle Y 34 \rangle=0$, and $\langle Y23 \rangle=\langle Y 41 \rangle=0$, where the form should not have a pole.

Moreover, in terms of the dual vectors, $W_1=(12), \ldots, W_4=(41)$ (equivalently, the edges of the quadrilateral), the denominator factors are $(Y\cdot W_1) (Y \cdot W_2) ( Y \cdot W_3) ( Y\cdot W_4)$ and the numerator reads $(W_1-W_3) \times (W_2-W_4)$ (where $\times$ denotes the cross-product). It is interesting that by going to affine space with $Y$ at infinity as $Y=(1,0,0)$ and $W_i=(1, w_i)$, we find that the canonical function becomes the familiar formula for the volume of the dual quadrilateral with vertices $w_1, w_2, w_3, w_4$. 

Another way to compute the canonical form is via a pushforward based on the {\it Newton-polytope map}~\cite{Arkani-Hamed:2017tmz}.  This map is a rational map from a simplex to a convex (projective) polytope ${\A}$, and it restricts to a diffeomorphism on the interiors ${\rm Int}(\Delta) \to {\rm Int}({\A})$.  Let's denote the vertices of ${\cal A}$ by $Z_1, \ldots, Z_n$, and let $z_1, \ldots, z_n \in \mathbb{Z}^{d{+}1}$ be an integer matrix with the same oriented matroid as $Z_1, \ldots, Z_n$, {\it i.e.} $\langle Z_{i_0} \cdots Z_{i_d}\rangle$ and $\langle z_{i_0} \cdots z_{i_d} \rangle$ have the same sign. Assuming $z_i=(1, z_{1i}, \ldots, z_{d i})$, the Newton-polytope map is the rational map $\Phi$ given by
\be \label{eq:Newton}
\Phi(X)=\sum_{i=1}^n x_1^{z_{1, i}} x_2^{z_{2, i}} \cdots x_d^{z_{d, i}} Z_i\,.
\ee
Here, the polytope with integer vertices $z_i$ is called the Newton polytope, which is $\New[p(\x)]$ for any polynomial of the form $p({\x})=\sum_{i=1}^n a_i x_1^{z_{1, i}} \cdots x_d^{z_{d, i}}$ with $a_i\neq 0$. For the square, we may take the Newton polytope to be the square with vertices $(0,0), (1,0), (1,1), (0,1)$ (we denote coordinates as $x,y$) and the Newton polytope map is given by
\be\label{eq:mapquad}
Y=\Phi(x,y)=Z_1+ x Z_2 + x y Z_3 + y Z_4%=(1+x+ x y + y, {\bf z}_1 + x {\bf z}_2 + x y {\bf z}_3 + y {\bf z}_4)
\,, 
\ee
and it is easy to see that it indeed provides a diffeomorphism from $\mathbb{R}^2_{> 0}$ to the interior of the quadrilateral $\A(Z_1,Z_2,Z_3,Z_4)$.  As shown in~\cite{Arkani-Hamed:2017tmz}, this implies that the pushforward of the form $\frac {d x d y}{x y}$ is the canonical form $\Omega(\A)$:
\be
\Phi_* \left(\frac {d x d y}{x y}\right)=d\log \frac{\langle Y 1 2 \rangle}{\langle Y 3 4\rangle} d\log \frac{\langle Y 2 3 \rangle}{\langle  Y 4 1 \rangle}=\Omega({\A})\,.
\ee
This identity can be checked directly by choosing a parametrization of $Z_i$ and $Y$, where the pushforward involves summing over the two roots from solving the $2$ equations from \eqref{eq:mapquad}.

%\begin{thebibliography}{xxx}
%\bibitem[ABHY]{Arkani-Hamed:2017mur}
%\bibitem[ABL]{Arkani-Hamed:2017tmz}
%\bibitem[AHLT1]{clusterletter} Letter on binary geometries
%\bibitem[AHLT2]{clusterconfig} Cluster configuration spaces of finite type
%\bibitem[ALS]{ALS} Arkani-Hamed, Lam, Spradlin, Non-perturbative geometries for planar N = 4 SYM amplitudes
%\bibitem[Bro]{Brown-thesis}
%\bibitem[CFZ]{CFZ} Chapoton-Fomin-Zelevinsky
%\bibitem[FZ]{FZ} Fomin Zelevinsky
%\bibitem[Ful]{Fulton} Fulton, Toric varieties
%\bibitem[SW]{SW} Speyer Williams
%\end{thebibliography}

\bibliographystyle{utphys}
\bibliography{refs}
\end{document}